\documentclass[12pt]{report}
\usepackage[a4paper,margin=1in,footskip=0.5in]{geometry}
\usepackage{amsmath,amssymb,amsthm}
\usepackage{color}
\usepackage{outlines}

\usepackage{graphicx}
\usepackage{caption}
\usepackage{subcaption}
\usepackage{amsmath}
\newcommand\numberthis{\addtocounter{equation}{1}\tag{\theequation}}
\usepackage[utf8]{inputenc}
\usepackage[english]{babel}
\usepackage[utf8]{inputenc}
\usepackage{physics}
\usepackage{natbib}
\usepackage{tikz}
\usetikzlibrary{quantikz}
\usepackage{float} 

\usepackage{blindtext}

\DeclareMathOperator{\Log}{log}
\newcommand{\kt}{\rangle}
\newcommand{\br}{\langle}
\newcommand{\mbf}[1]{\mathbf{#1}}
\usepackage{hyperref}

\begin{document}

\begin{titlepage}
    \begin{center}
        \vspace*{1cm}
            
        \Huge
        \textbf{A STUDY OF CHAOS AND RANDOMNESS IN QUANTUM SYSTEMS}
            
        \vspace{0.5cm}

        \vspace{1.5cm}
            
        \textbf{Sreeram PG}
            
        \vfill
            
        A thesis presented for the degree of\\
        Doctor of Philosophy
            
        \vspace{0.8cm}
            
        \includegraphics[width=0.4\textwidth]{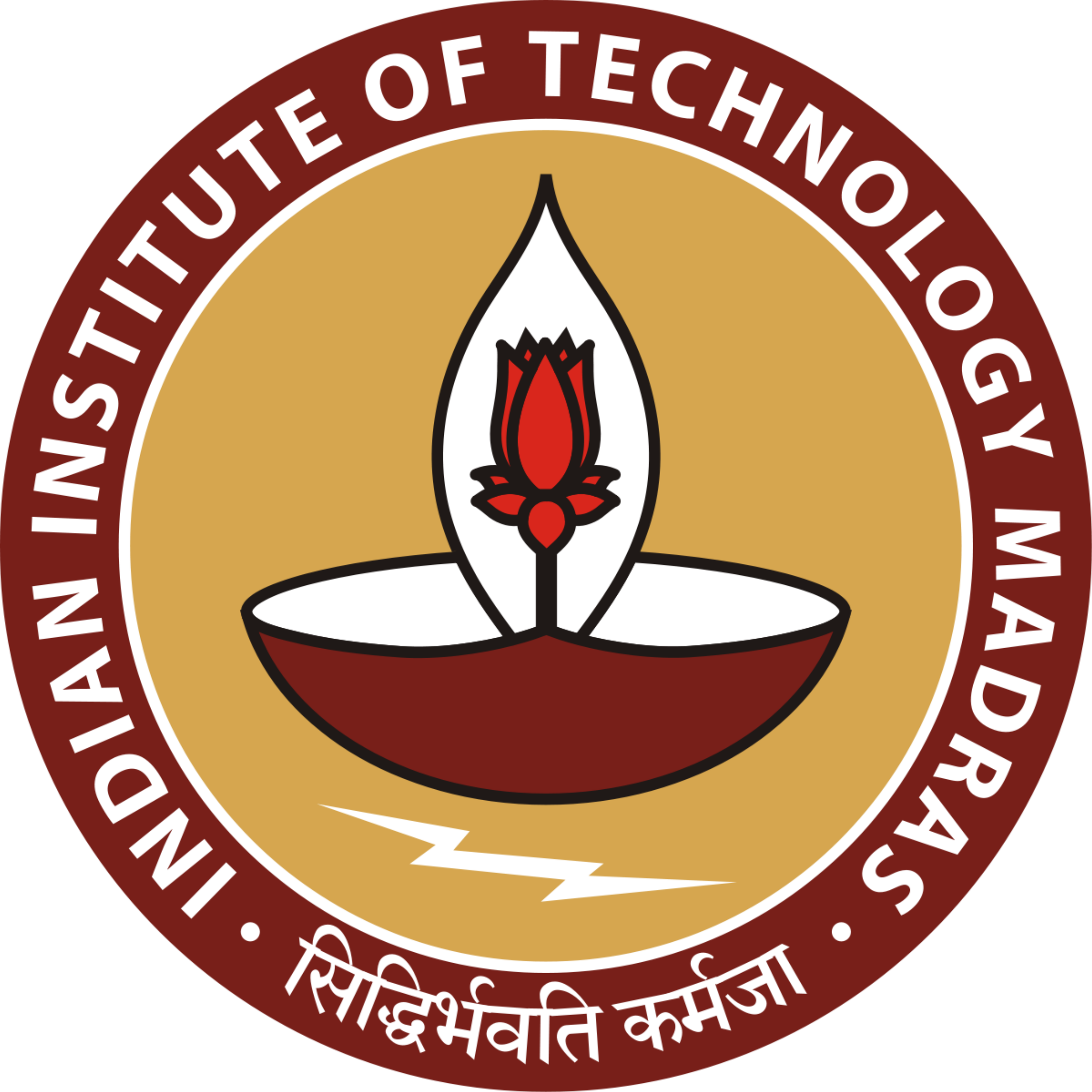}
            
        \Large
        Department of Physics\\
        IIT Madras\\
        India\\

    \end{center}
\end{titlepage}

\tableofcontents
\newpage
\chapter*{Abstract}
How classical chaos emerges from the underlying quantum world is a fundamental problem in physics. The origin of this question is in the correspondence principle, which states that quantum mechanics, the physical theory of microscopic objects, should reproduce classical (Newton's) laws in the macroscopic limit. Classical chaos arises due to non-linear dynamics, whereas quantum mechanics, driven by unitary evolution, is linear. The question that still remains is - what are the footprints of classical chaos in the quantum world? One can understand the quantum signatures of classical chaos by studying a quantum system whose classical analogue is chaotic. In this work, we use the quantum kicked top model of few qubits in the deep quantum regime to investigate signatures that can be considered as a precursor to chaos in the classical limit. In particular, we study out-of-time-ordered correlators (OTOCs) and Loschmidt echo, the two well-known dynamical diagnostics of chaos. We find vestiges of classical chaos even in such a deep quantum regime. 
There is an intimate connection between chaos and complexity which can be tied together using information theory. Classically chaotic evolution causes localized probability distributions in the phase space to spread out due to ergodicity and mixing. Quantum mechanically, chaotic maps take fiducial wave packets to pseudo-random states in the Hilbert space, which typically have high entropy. The consequences of such a "pseudo-random" evolution have been extensively studied in terms of dynamical generation of entanglement and other quantum correlations, randomized benchmarking, information scrambling and operator spreading, and so on. 

Another arena where one can study the effects of chaos and randomness is quantum state tomography. Quantum state reconstruction is a nontrivial problem in itself because of the inherent quantum uncertainty. The traditional approach, employing projective measurements to find out information about the state, requires a large number of measurements on identical copies of the system for an accurate estimate of the state. A continuous weak measurement protocol performed on an ensemble of the system can get around this. We study quantum tomography from a continuous measurement record obtained by measuring expectation values of a set of Hermitian operators generated by a unitary evolution of an initial observable. For this purpose, we consider the application of a random unitary, diagonal in a fixed basis at each time step.  We quantify the information gain in tomography using Fisher information of the measurement record and the Shannon entropy associated with the eigenvalues of the covariance matrix of the estimation. Surprisingly, high fidelity of reconstruction is obtained using random unitaries diagonal in a fixed basis even though the measurement record is not informationally complete. We then compare this with the information generated and fidelities obtained by applying a different Haar random unitary at each time step. We give an upper bound on the maximal information obtained in tomography and show that a covariance matrix taken from the Wishart-Laguerre ensemble of random matrices and the associated Marchenko-Pastur distribution saturates this bound. We find that physically, this corresponds to applying a different Haar random unitary at each time step. We show that repeated application of random diagonal unitaries gives a covariance matrix in tomographic estimation corresponding to a new ensemble of random matrices.
We show that the information gain obtained from  Porter-Thomas distribution comes close to that obtained from diagonal-in-a-basis evolution.

As another contribution of this thesis, we have harnessed the power of randomness inherent in the maximally mixed state to give an efficient quantum algorithm to measure OTOCs. The protocol achieves an exponential speedup over the best known classical algorithm, provided the OTOC operator to be estimated admits an efficient gate decomposition. We also discuss a scheme to obtain information about the eigenvalue spectrum and the spectral density of OTOCs. For this purpose, we adapt the deterministic quantum computation using one clean qubit (DQC1) circuit. This protocol also helps benchmark unitary gates, which is important from the quantum computation and control perspective.

Lastly, a chapter of this thesis is devoted to studying the phenomena of measure concentration on a higher dimensional sphere, also known as Levy's Lemma. This lemma finds numerous applications in quantum information theory - from entanglement theory to foundations of statistical mechanics. In its original form, Levy's lemma deals with properties of concentration of measure on a hypersphere when points on the sphere are chosen in a uniformly random manner. We generalize Levy's lemma and obtain the concentration of measure inequalities when the points on the hypersphere can be chosen with respect to any Lipschitz probability distribution. This generalizes the applications of Levy's Lemma in quantum information theory. In particular, we discuss the applications of our results to bipartite entanglement and Fannes' inequality. 

\chapter{Introduction}

\label{chap:intro}

{The universe as we know it is quantum mechanical, yet classical mechanics gives an excellent description of the macroscopic world around us. The question of quantum-to-classical transition, that is, how classical mechanics arises from the underlying quantum framework has been intriguing. Intimately related is the question of the origin of classical chaos with roots in non-linear phenomena, from the underlying quantum world whose evolution is given by linear and unitary dynamics. The study of quantum chaos straddles both the theories of classical and quantum mechanics, providing an understanding of the phenomena like the emergent unpredictability, complexity, thermalization in closed quantum systems, and scrambling of information. However, this has not come easy. Classically, chaos is characterized by the extreme sensitivity of the dynamics to initial conditions. Two dynamical evolutions starting from nearby initial points diverge exponentially, leading to very different outcomes. We may not know the system's initial condition perfectly because of measurement errors and resolution limits. Then our prediction about its future is going to be inaccurate. This is why classical chaos was aptly described in one line by {Lorenz} as ``Chaos: When the present determines the future, but the approximate present does not approximately determine the future" \citep{danforth2013chaos}. However, in quantum mechanics, initial conditions are not described by points in the phase space because of the uncertainty principle. A state vector describes a quantum state in a Hilbert space. Since Schr\"{o}dinger equation is linear, there is no scope for exponential separation in the dynamics starting from two nearby initial conditions. In fact, the evolution is unitary, and the initial overlap between two state vectors is preserved throughout. Then how is classical chaos reflected in quantum systems?}

{It is worth mentioning that this quest is not merely an effort to ``fix" a definition of chaos in quantum mechanics. {Erhenfest's} correspondence gives timescales to which quantum expectation values of observables are expected to follow classical trajectories \citep{ehrenfest1927bemerkung}. This characteristic timescale, known as the ``break time", is exponentially small for chaotic systems as compared to regular/non-chaotic systems. Therefore, understanding the dynamics of quantum wave packets as they stretch and fold, causing the interference effects to become significant for the expectation values to depart from classical trajectories, has profound dynamical consequences. This led Zurek to make predictions about how Hyperion, a satellite of Jupiter, will depart from its classical trajectory in a matter of decades \citep{zurek1998decoherence}. The role of decoherence in preventing this and restoring classical dynamics is a closely related and important issue.} 

{The study of quantum signatures of classical chaos involves a broad spectrum of overlapping research directions. One studies Hamiltonian systems whose classical analogues are chaotic to unravel the correspondence between quantum and classical theories. One way to see the signatures of chaos is to look at the statistical properties of the Hamiltonian spectra. The distribution of the difference between nearest eigenvalues (level spacing distributions) behaves differently for chaotic and integrable dynamics. If the classical analogue of the quantum system is chaotic, then the level spacing distribution follows that of a suitable random matrix ensemble \citep{bohigas1984characterization}. On the other hand, if classical dynamics is integrable, the level spacings follow a Poissonian distribution \citep{berry1977level}}.
{Random matrix theory has been used to characterize and quantify some properties of chaos in quantum systems. Eugene Wigner pioneered the use of random matrices to model large nuclei \citep{wigner55}. He found that such heavy nuclei are so complicated that their Hamiltonians act like  random matrices. Some spectral properties of the system simply follow from the random matrix. This  carries over to the chaotic Hamiltonians too.}

{The study of level statistics, semiclassical approximations of the spectrum, and connections to the periodic orbit theory have been of focus traditionally \citep{berry1977level, bohigas1971spacing,Haake,delande1986quantum,  atas2013joint, bhosale2018scaling,, tekur2018exact, tekur2018higher}. However, several ``dynamical" signatures of chaos are being vigorously pursued with important consequences to quantum information processing as well as fundamental physics.
	These include quantum-to-classical transition, classical emergence of chaos via decoherence/weak continuous measurement, to recent trends in studying the dynamical generation of entanglement/quantum correlations and information scrambling using out-of-time-ordered correlators \citep{larkin,Zurek/Paz,MillerSarkar, oz02-2, Lakshminarayan,BandyopadhyayArul2002,Bandyopadhyay04,Tanaka-2002,LakSub2003,ArulSub2005, bhattacharya2003continuous,Ghose2004, habib2006emergence,trail2008entanglement,LombardiMatzkin2011,PhysRevLett.112.014102, madhok2015signatures, Madhok2018_corr, swingle2016measuring,Swingle-2018, chaos1, PhysRevX.7.031011,Hashimoto-2017,Cotler-2018,Carlos-2019, Maciej-2019,Meenu-2019,Pappalardi-2020,   yan2020information, santos2020speck}}.

{Studying the properties of stationary states (level statistics) and localization of eigenstates can give us a great deal of insight about the system. However, the actual dynamics often have surprises, like decoherence and information scrambling. Traditionally, there has been an emphasis on the ``energy domain" since one could not control quantum systems in the past. So one rarely could do an experiment in which the system was prepared at time t=0 in a pure state and then allowed to be evolved and measured. That is a much more recent development with atomic, molecular, and optical physics, and progress in quantum technologies like superconducting qubits playing a huge part.
	The study of quantum systems is fuelled by recent experiments where questions like the ability to control quantum systems \citep{Chaudhary, poggi2020quantifying}, thermalization in closed quantum systems \citep{Neill16} and quantum simulations of chaotic and non-integrable Hamiltonians \citep{sieberer2019digital, PhysRevX.7.031011} are of prime importance.} 
	
	{Signatures of  chaos
	based on statistical spectral properties, such as
	nearest-neighbor spacing statistics \citep{Haake},  are of little value for understanding systems with small Hilbert spaces, since statistical analysis may result in misleading conclusions for small sample sizes. Alternatively, one can study the dynamics of correlation functions in the system. In principle, the effective Planck constant $h_{\text{eff}}$ in such systems is large and hence quantum-classical correspondence time scales, such as
	the Ehrenfest time of $E_f\sim \log(1/h_{\text{eff}})/\lambda_C$, where $\lambda_C$ is the classical Lyapunov exponent
	are very short. The quantum correlators have a classical correspondence only in this short period \cite{shepelyansky1983some}.}

This thesis studies the signatures of chaos in the deep quantum regime. We use quantities such as out-of-time-ordered correlators~(OTOCs), Loschmidt echo, and paradigms like quantum state tomography to unravel the same. Quantum chaos also has interesting connections with quantum control. For a three-qubit system, it was established that the degree of control over a subsystem is inversely related to the level of chaos in the full system \citep{mirkin2021quantum}.
Later in this thesis, we find out the residual signatures of chaos in the behavior of OTOCs and Loschmidt echo for three and four qubit systems. We also find analytical expressions for these quantities, which is rare for chaotic systems. Another arena where we study the effects of chaos is in quantum state tomography. We discover a basis-dependent signature of chaos in the rate of information gain in tomography.

{Closely related is the problem of estimating chaos quantifiers. We give a computation protocol utilizing only a single pure qubit that achieves exponential speedup over any known classical algorithm in measuring OTOCs in quantum systems. Lastly, we study the concentration of measure phenomena in higher-dimensional Hilbert spaces, extend and generalize the existing results, and discuss their application to quantum information.}  

\section{Dynamical signatures of chaos}

Alternative to the static signatures, there are dynamical signatures of chaos that can be explored using tools such as Loschmidt echo and out-of-time-ordered correlators~(OTOCs).  Loschmidt echo captures the sensitivity of the dynamics to perturbations \citep{hahn1950spin,peres1984stability, jalabert2001environment}, while OTOC is a measure of quantum information scrambling \citep{Swingle-2018, shenker2014black}. These two quantities are closely related, and both employ imperfect time-reversal of the quantum system. Their relationship has been established recently \citep{yan2020information}. 
In quantum mechanics, time-irreversibility arises from the non-controllability of the Hamiltonian \citep{peres1984stability}. This contrasts with classical mechanics, where irreversibility arises because of the  mixing and coarse-graining.  No system is completely isolated from the environment. The lack of perfect knowledge about the Hamiltonian affects a chaotic system much more than a regular one.


The Loschmidt echo arose out of the ``reversibility paradox", which goes back to the dispute between {Boltzmann} and {Loschmidt} in the 1870s. The fact that macroscopic systems follow an arrow of time despite the underlying physical laws being perfectly reversible caused a lot of trouble to scientists. Loschmidt thought that the reversal of dynamics should be achievable by reversing the sign of velocities. However, Boltzmann refuted it soon after, noting that it is practically impossible \citep{stephen1966brush}. Reversing the dynamics of a system would require perfect knowledge of its initial conditions. 

Let us say our system is a large number of gas molecules in a container, expanding into its environment. To reverse this dynamics, one would require the initial knowledge of all the molecules perfectly. Even a slight difference in the initial knowledge of one of the molecules will lead to the disequalization of time very quickly. Not just that, no system is truly isolated in real life. There are always interactions with the environment, which has a large number of degrees of freedom. Imperfect knowledge along with environmental interactions establish an arrow of time for the system. These imperfections can be modeled as a perturbation to the system Hamiltonian. Then the difference in the overlap between the forward evolution and the backward evolution of the system is called Loschmidt echo. It is defined as $F(\tau)=\lvert\bra{\psi_0} e^{iH'\tau/\hbar}e^{-iH\tau/\hbar}\ket{\psi_0}\rvert^2,$ where $\ket{\psi_0}$ is the state undergoing evolution for a time $\tau.$ 

Echo acts as a measure of sensitivity to perturbations. In the absence of perturbations, Loschmidt echo attains the maximum  value of one. The presence of imperfections leads to decay from there. This  immediately suggests that the Loschmidt echo must be closely connected with the decoherence phenomena. The coupling of quantum systems with the environment causes decoherence. It is the process by which the quantum behavior is lost from the system. The connection between echo dynamics and decoherence has been established and the former has been used to quantify the latter \citep{zurek2001sub,cucchietti2003decoherence,cucchietti2004universality}. 

Loschmidt echo is employed in various  fields of physics, including quantum chaos, quantum computation and quantum information, elastic waves, quantum phase transition, statistical mechanics of small systems, etc. It was first implemented experimentally in NMR systems \citep{hahn1950spin}. The time-reversal was achieved using a radio frequency pulse. Perfect reversal is hampered by pulse imperfections and environmental interactions, which act as perturbations.

Echo decay can be used to diagnose chaos in quantum systems. The behavior of the Loschmidt echo depends on the initial state, the underlying dynamics, and the perturbation applied. For single-particle quantum systems whose classical limit is chaotic, the Loschmidt echo dynamics is well-understood \citep{shepelyansky1983some,jalabert2001environment,jacquod2001golden,cerruti2003uniform}. Its typical behavior with respect to time and perturbation strength are as follows. At short times, echo decays parabolically, followed by an asymptotic decay at intermediate times. The type of asymptotic decay is dependent on the perturbation strength. For small perturbations, the decay is Gaussian, whereas, for stronger perturbations, the echo slumps exponentially. The functional form of exponential decay is further dependent on whether the perturbation is global or local. This regime is followed by a saturation region at long times.

OTOC on the other hand captures the scrambling of quantum information in an extended quantum system. Quantum information scrambling is the spreading of information throughout the system via correlations. This process leads to a loss of memory of the system about its initial conditions, similar to classical chaos. Scrambling is captured in the commutator $ C(t)=-\langle[W(t),V(0)]^2\rangle$ between two Hermitian or unitary operators $V$ and $W$. Here the time evolution is determined by the system Hamiltonian  $H$ as, $W(t)=\mathrm{exp}({iHt}) W(0) \mathrm{exp}({-iHt})$. The expectation value of the squared commutator is obtained usually with respect to the thermal state. If the two operators are local and separated, initially the commutator is zero, since the operators are acting in independent subspaces. The time at which the commutator becomes appreciably divergent from zero is termed  scrambling time. If the operators $W, V$  are unitary, then $C(t)=2- 2 \mathrm{Re}[\langle W(t)V(0)W(t)V(0) \rangle]$. The quantity 
$F(t)=\langle W(t)V(0)W(t)V(0) \rangle$ is called an out-of time-ordered correlator since the terms inside are irregularly ordered in time. 
  At early time, $F(t)$ goes like $F(t)= 1-\epsilon \mathrm{exp}(-\lambda_Q t)$, where $\lambda_Q$ is the quantum analogue of the classical Lyapunov exponent \citep{shenker2014black}.  For large quantum numbers, $\lambda_Q$ reflects the classical Lyapunov exponent.
  
The  OTOC can be seen as a variant of the echo itself. $W(t)= \mathrm{exp}({iHt}) W(0) \mathrm{exp}({-iHt})$ can be seen as a unitary perturbation $W(0)$ preventing the cancellation of forward and backward evolutions. As a result, $W(t)$ grows in complex ways, leading to an exponential growth of the commutator. 

Although quantum signatures of chaos are traditionally explored in large systems in the semiclassical limit, the search for any residual signatures in the deep quantum regime has engaged researchers. Do statistical signatures of chaos exist in small quantum systems? Statistical mechanics holds on the presupposition that all possible configurations are equally probable for the system. Chaotic dynamics in classical systems ensures ergodicity in dynamics. However, in quantum systems, linear evolution disallows chaotic motion. Surprisingly, researchers have found ergodicity and thermalization even in  three qubit system modeled as a kicked top \citep{Neill16}. Kicked tops are Hamiltonian systems, whose classical limit can be chaotic depending on the parameter value. These are important class of systems which can act as a test bed for asking fundamental questions related to chaos and thermalization. One reason is that, despite being nonintegrable and chaotic in the classical limit, such systems have been exactly solved \citep{dogra19}. Furthermore,
 kicked top systems are experimentally realizable using superconducting qubits and  NMR systems  \citep{Neill16,MaheshUdayExpt-2019}. 

The effects of chaos can also be seen in much more natural and common correlations such as entanglement and discord \citep{MillerSarkar,Ghose2004,Wang2004, norris2007chaos,madhok2015signatures,Neill16,RuebeckArjendu2017, Madhok2018_corr,bhosale2017signatures,Bhosale-2018, tomsovic2018eigenstate, MaheshUdayExpt-2019,dogra19, Meenu-2019, pulikkottil2020entanglement}. Their dynamical behavior distinguishes chaotic and regular regimes and demonstrates correlations with classical phase space. It is indeed remarkable that entanglement and discord, which are very quantum phenomena, hold signatures of classical chaos.

Quantum chaos  also has interesting  connections with quantum control. For a three-qubit system, it was established that the degree of control over a subsystem is inversely related with the level of chaos in the full system \citep{mirkin2021quantum}.
Later in this thesis, we find out the residual signatures of chaos in the behavior of OTOCs and Loschmidt echo for three and four qubit systems. We also find analytical expressions for these quantities, which is rare for chaotic systems.

\section{Chaos and quantum tomography}

Another place where chaos manifests is in the rate of information gain during quantum state tomography. Tomography is about estimating the initial condition. Classically, chaos leads to exponential separation of trajectories starting from two nearby initial conditions, which helps in determining the starting point. Can we find its analogue in quantum state tomography? It turns out we can, as shown in~\citep{PhysRevLett.112.014102}.

In quantum tomography, one makes a series of measurements on a collection of identically prepared systems to gain information about the system. An estimate of the state is obtained by inverting the statistics of measurement records. For the assessment to be a good one, the measurements have to be tomographically complete. That is, the observables measured must span all of the operator space. The straight forward method to carry out tomography is to do projective measurements and then use the Born rule to invert the spectrum. Since projective measurements also destroy the state,  only one observable can be measured after each preparation, and very many replicas of the initial state are required for its reconstruction. 

One can choose to perform weak measurements instead, which prevent the state from collapsing. If the measurements are weak enough to exert minimal disturbance on the states, a good reconstruction fidelity can be achieved with a finite number of copies of the state. There is an added advantage of being able to optimize the measurements for the required fidelity in the estimate. {Silberfab} \emph{et~al.} proposed a weak collective measurement scheme on the identically prepared and collectively evolved ensemble, which will be of interest to us in this thesis~ \citep{PhysRevLett.95.030402}. Collective measurements are known to be much more effective and optimal than local measurements \citep{massar2005optimal,vidal1999optimal,gisin1999spin,bagan2006separable,hou2018deterministic}. The reason for this phenomenon is connected to the lower mutual information when measurements are performed locally \citep{bennett1999quantum}. In collective measurements,  the measurement apparatus interacts with the entire ensemble of states as a single quantum system rather than individually. This helps in reducing the effect of measurement induced disturbance on the system.  The measurement backaction gets distributed among the ensemble, leading to negligible effect on any individual system. 

In 2006, {Greg A Smith} \textit{et al.} performed  the protocol described above of weak measurements on a cloud of Cs atoms, coupled to an off-resonant optical probe \citep{PhysRevLett.97.180403}. Here the Stokes vector of the optical probe gets correlated to the atomic spin. Measuring the polarization of the probe weakly measures a collective spin observable. This is a standard Von-Neumann type measurement, where one couples the system with a meter and strongly measures the latter. Information completeness is achieved by keeping the Cs cloud in a time varying magnetic field. The varying magnetic field couples with the spin system, changing the observable measured each time.

Quantum tomography is useful in estimating the correctness of various control protocols. Estimating the quantum state gives information on its dynamics and spread across the Hilbert space during evolution and helps characterize quantum chaos. State tomography is also an essential tool in estimating quantum channels in process tomography.  

In this thesis, we adapt the weak continuous measurement protocol by  Silberfarb \textit{et al.} \citep{PhysRevLett.95.030402} and use a restricted class of unitaries, diagonal in a given basis, to drive the dynamics.  We quantify its performance and compare it with Haar-random dynamics. This is fascinating from a fundamental perspective since the diagonal-in-a-basis unitaries form a proper subspace of random unitaries, leading to incomplete measurements. It is of practical importance if they achieve good fidelity since constructing the diagonal-in-a-basis unitaries is easier in a laboratory, needing fewer resources/component gates. Diagonal gates are fault-tolerantly realized experimentally in super- and semi-conducting systems \citep{aliferis2009fault, nakata2014diagonal}. They are also robust to environmental decoherence than non-diagonal circuits { \citep{buscemi2007quantum}}. Furthermore, the diagonal gates have better computational power than classical computers \citep{PhysRevLett.112.140505, bremner2010classical} and  are also very efficient in generating entanglement \citep{Arul}.
  After studying their performance in state reconstruction, we establish certain random matrix connections of measurement correlations. We also pose an interesting question regarding the role of eigenvalues and eigenvectors of random unitaries in the ergodicity of the dynamics. We use the rate of information gain to decouple and establish their roles in chaos. 
 \section{ Deterministic quantum computation using one clean qubit (DQC1) for chaos estimation}
 In the standard model of quantum computation involving pure states, entanglement was identified as the resource required for exponential speedup over classical computation. Multipartite entanglement, with a large number of qubits taking part, is necessary for this speedup \citep{jozsa2003role}, with the amount of entanglement growing in accord with the size of the computation \citep{vidal2003efficient}. Thus the power of regular quantum computation stems from quantum entanglement and the controllability over pure states.
 
 However,  researchers later identified that entanglement is not essential for exponential speedup over classical computation \citep{braunstein1999separability,biham2004quantum,kenigsberg2006quantum,meyer2000sophisticated}.
  Ideally for quantum computation, we desire the quantum states to not have unsolicited interactions with their environment.  But in real life, there are always interactions and states are usually mixed and not pure. These can hamper our control over the states.  In 1998, {Knill} and {Laflamme} proposed a restricted model of computation where the  system can be in a completely mixed state, and still achieve exponential speedup over any known classical algorithm~\citep{knill}.
  
  In this model, there is only one pure qubit acting as the control. Even though entanglement was not discerned, non-classical correlations quantified by discord are found to be non-zero in this model~\citep{PhysRevLett.100.050502}. These non-classical correlations other than entanglement are the resource enabling the exponential speedup in DQC1~\citep{PhysRevLett.100.050502}. The presence of discord at the output states of DQC1 was experimentally confirmed in \citep{lanyon2008experimental}, and an analytical expression for geometric discord at the output for arbitrary dimension has been put forth in \citep{passante2012measuring}. This suggests using discord as a universal measure of quantum computational resources.   DQC1 algorithm uses only a single quantum bit, yet it is much more powerful than classical computation, which is surprising. Therefore DQC1 is famously called the ``power of one qubit model". 
  
  Computation using only a single pure state, although restricted in applications, can be used to solve some useful problems out of hand for classical computers. Estimating the trace of a unitary and approximating Jones polynomial are among them \citep{shor2007estimating}. In this thesis, we use the DQC1 model to compute OTOCs and estimate their eigenvalue spectrum. The highlight of our method is that it performs exponentially faster than any known classical protocol.
  The eigenvalue spectrum of OTOCs shows different distributions, depending on the underlying chaos \citep{spectrum2, spectrum1}. Another important application of DQC1 we find is in benchmarking quantum circuits. We design a circuit to estimate the gate fidelities of unitary gates. Charecterizing the performance of quantum gates is very important from the perspective of quantum computing and quantum control.

\section{Concentration of measure in Hilbert spaces}

In the quantum state tomography section, we shall come across random quantum states. All of them behave in a typical fashion in tomography, which helps in making statistical inferences. Such typicality is associated with higher dimensional Hilbert spaces, whose origin is in the phenomenon of concentration of measure~\citep{milman2009asymptotic}. Typical properties are shared by most states. The concentration of measure phenomena has deep implications for the foundations of statistical physics. The traditional postulate of  ``equal apriori probabilities", which is unprovable, could be replaced with a typicality relation \citep{popescu2006entanglement}.

Consider that the system $(H_S)$ and environment $(H_E)$ make up the universe $(H_R=  H_S \otimes H_E)$. One typically assumes that the environment is much bigger compared to the system. Following the ``equal apriori probabilities" postulate,  if the universe is in  a maximally mixed state $(\Omega_R= I/R)$, the state of the system can be obtained by tracing out the environment $(\Omega_S= \mathrm{Tr}_E \Omega_R)$, called the canonical state. Now even if we assume that the universe is in a random pure state, one obtains that the state of the system is very close to $\Omega_S$ almost always. Thus one can do away with the assumption of ``equal apriori probabilities", and replace it with the mathematically derivable ``apparently equal apriori probability principle" \citep{popescu2006entanglement}.
The main ingredient in deriving this principle is Levy's lemma \citep{milman2009asymptotic, ledoux2001concentration}.

Looking at the measure concentration phenomena from the geometry of a higher dimensional hypersphere, Levy's lemma is stated as follows:  Given a Lipschitz continuous function $f: S^{n-1} \rightarrow \mathbb{R}$ defined on a large dimensional hypersphere $S^{n-1}$, and a point $x \in S^{n-1}$ chosen uniformly at random. Then the measure $\mu$ of points on the hypersphere is given by
\begin{equation}
\mu\left\lbrace x \in S^{n-1}: |f(x)- \mathbb{E}{f(x)} |\geq \epsilon \right\rbrace \leq 2 \mathrm{exp} \left(\frac{- kn \epsilon^2}{\eta^2} \right) \label{one},
\end{equation}

where $\mathbb{E}{f(x)}$ is the mean value of $f(x),$
and $k$ is a positive constant.   $\eta$ is called the Lipschitz constant of the function $f$.  It is defined as $\eta= \mathrm{sup}|\bigtriangledown f|$. Simply put,  what this lemma communicates can be described as  follows. We ask a question - what is the probability that a point taken at random on a higher dimensional sphere lies in a narrow belt surrounding any particular equator? According to Levy's Lemma, as the dimension of the sphere increases, this probability approaches one.

Typicality relations induced by the concentration of measure are prevalent in quantum information theory. For example, quantum correlations such as entanglement \citep{popescu2006entanglement}, discord \citep{ferraro2010almost}, coherence \citep{zhang2017average} all show typical behavior in higher dimensional Hilbert space. The concentration of measure applies to quantum channels too. The distance between almost all quantum channels is mostly a constant value \citep{nechita2018almost}.
In chapter \ref{chap:chap6}, we extend Levy's lemma to apply to a more general situation and explore its consequences in entanglement in higher dimensional Hilbert spaces. 

\section{Organization of the thesis}
 
 This thesis is organized as follows. The next chapter  contains preliminary information for readers unfamiliar with the precincts of the thesis. Chapter \ref{chap:chap3} discusses the signatures of chaos in small quantum systems investigated in \citep{sreeram2021out}. Chapter \ref{chap:chap4} is mostly comprised of our work in random state tomography~\citep{sreeram2021quantum}, along with a brief discussion on coherent state tomography  \citep{sahu2022revisiting}. Later, we describe our study employing the DQC1 protocol \citep{pg2021exponential} in Chapt. \ref{chap:chap5}. Finally, we turn toward the ``concentration of measure phenomena" in Chapt. \ref{chap:chap6} before concluding in Chapt. \ref{chap:chap7}. We also have an appendix \ref{appendixA} where we discuss some of the effects of concentration of measure in biological evolution.

 \chapter{BACKGROUND}
\label{chap:background}

This chapter gives essential background information about the concepts and tools used in the thesis. It may be helpful for the reader not familiar with the area to first read this chapter before proceeding.

\section{Loschmidt echo and OTOC}
  
In classical dynamical systems, chaos is marked by extreme sensitivity to initial conditions. Quantum mechanics, in the large quantum number limit, reduces to classical mechanics, according to the correspondence principle. Therefore one expects to discover the signatures of chaos in the quantum regime. However, since Schr\"{o}dinger evolution is unitary, sensitivity to starting points does not arise. Therefore one has to look elsewhere for the signatures of chaos, if at all they exist. Such vestiges have been identified  in the sensitivity to perturbation of the Hamiltonian \citep{Haake}, entropy production \citep{Zurek/Paz} and entanglement \citep{furuya1998quantum}. 

Loschmidt echo captures the sensitivity of quantum dynamics to perturbations \citep{gorin2006dynamics,goussev2012loschmidt}. Suppose an initial state $\ket{\psi_0}$ is evolved forward in time by a Hamiltonian $H.$ A perfect time-reversal operation will get back the initial state. That would correspond to applying Hamiltonian $-H$ and evolving from time $\tau$ to $2\tau$. However, such a perfect time reversal is not possible to achieve in reality. Let us say the backward evolution is achieved by the application of a perturbed Hamiltonian $-H'.$ Then, the Loschmidt echo is defined as
\begin{equation}
F(\tau)=\lvert\bra{\psi_0} e^{iH'\tau/\hbar}e^{-iH\tau/\hbar}\ket{\psi_0}\rvert^2.
\end{equation}
{Erwin Hahn}, in 1950, provided the first experimental implementation of time-reversal and hence Loschmidt echo \citep{hahn1950spin}. Loschmidt echo is used as a tool to quantify the decoherence effect due to interaction with the environment. The behavior of the Loschmidt echo  is quite complex and it depends on the initial state, the underlying dynamics, and the perturbation applied. For quantum systems whose classical limit is chaotic, the Loschmidt echo dynamics is well-understood \citep{shepelyansky1983some,jalabert2001environment,jacquod2001golden,cerruti2003uniform}. At short times, echo decays parabolically, followed by an asymptotic decay at intermediate times. The type of asymptotic decay is dependent on the perturbation strength. For small perturbations, the decay is Gaussian, whereas, for stronger perturbations, the echo slumps exponentially. This regime is followed by a saturation region at long times.

There is another closely related quantifier of quantum chaos, called out-of-time-ordered correlator~(OTOC), defined as 
\begin{equation}
C_{W,V}(\tau)=-\langle[W(x, \tau), V(y, 0)]^2\rangle,
\end{equation}
where $V$ and $W$ are local unitary or Hermitian observables. The first index denotes the position coordinate. The expectation value is usually taken with respect to the thermal state at a particular inverse temperature $\beta$.
\begin{figure}
    \centering
    \includegraphics[width=1\linewidth]{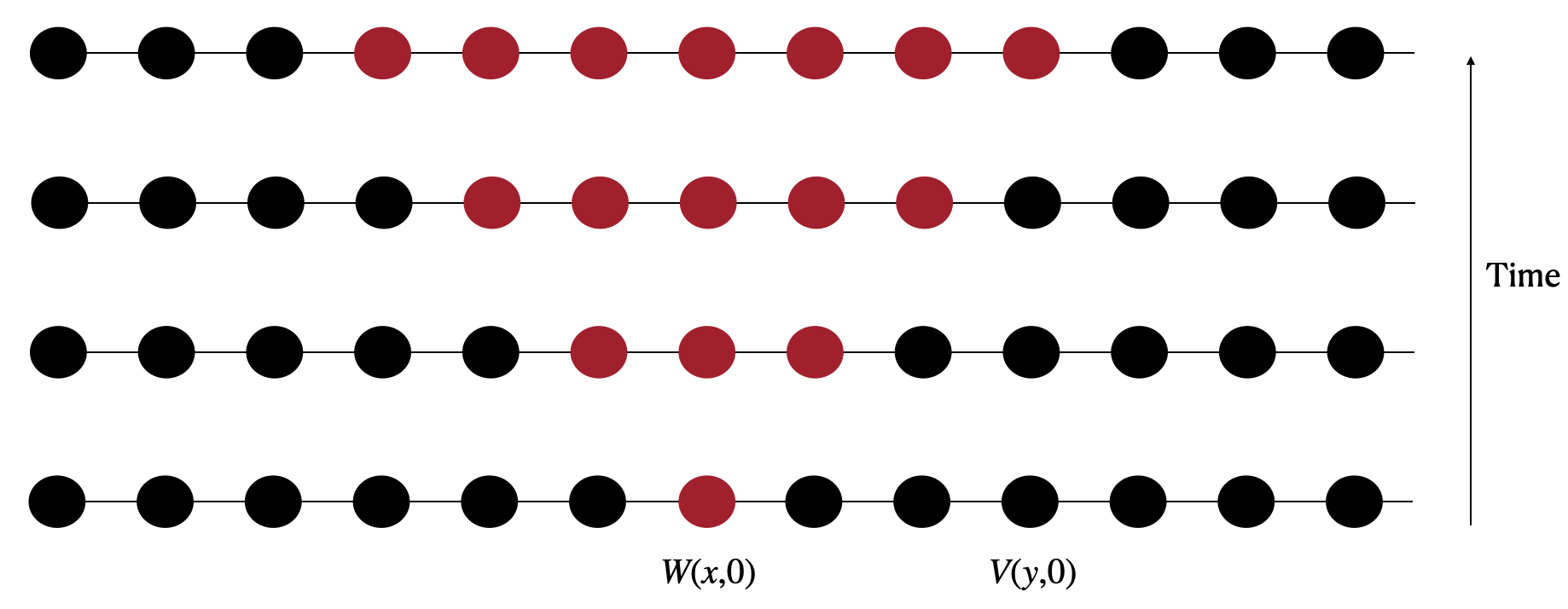}
    \caption{In an extended spin system, the operators $W$ and $V$ initially act on disjoint subsystems and they commute. Only one of these operators~($W$) evolve in time according to the unitary~($U$) generated from the global system Hamiltonian. The  As time progresses along the vertical axis, $ W(x, \tau)=U^{\dagger}(\tau)W(x, 0)U(\tau)$  grows in its support~(shown in maroon), and extends the spin subsystem where the operator $V$ acts. At this point the commutator $[W(x,\tau),V(y,0)] \neq 0.$ }
    \label{fig:otoc_growth}
\end{figure}
There have been many efforts to connect both quantities. Recently, {Bin Yan} \emph{et~al.} established OTOC as a thermal average of the Loschmidt echo \citep{yan2020information}. There were already experimental protocols to measure OTOC using Loschmidt echo type sequences \citep{swingle2016measuring,garttner2017measuring, scrambling2, sanchez2020perturbation}. One can see the relation of OTOC with chaos by taking the operators to be $x(\tau)$ and $p(\tau)$ as given by

\begin{equation}
  C(\tau)= -\langle[x(\tau),p(0)]^2\rangle.  
\end{equation}
 In the large quantum number limit, quantum commutator reduces to a classical commutator, and 
\begin{equation}
C(\tau)=-\langle \left(i\hbar \lbrace x(\tau),p(0) \rbrace \right)^2\rangle=\hbar^2 \left(\frac{\delta x(\tau)}{\delta x(0)}\right)^2. \label{otoc}
\end{equation} 
In a classically chaotic system, dynamics is very sensitive to initial conditions, and $x(\tau)$ diverges exponentially, $\left(\frac{\delta x(\tau)}{\delta x(0)} \right) \sim e^{\lambda_C \tau}$, where $\lambda_C$ is called the classical Lyapunov exponent.   Therefore in the semiclassical limit, one can see from Eq.~(\ref{otoc}) that the OTOC  captures the exponential growth  $C(\tau)\sim \hbar^2 e^{2\lambda_Q \tau}$, with a quantum Lyapunov exponent $\lambda_Q,$ reflective of the classical one ($2\lambda_Q =\lambda_C$).  Thus, OTOCs work as quantifiers of quantum chaos. However, OTOCs show such growth only till Ehrenfest time, and then they saturate. They are useful tools in studying the dynamics of quantum information and thermalization. 

Let us consider an extended spin chain as shown in Fig. \ref{fig:otoc_growth}, with two local operators acting at two different sites separated from each other. Initially, their commutator is zero, since the operators are acting on independent Hilbert spaces. However, as time evolves, the dynamics establishes correlations between spins, and the operators spread over increasing degrees of freedom in the system. Then the two operators become non-commuting as they no longer act in independent spaces. Their commutator grows until Ehrenfest time and then saturates. Saturation is after the wave function has spread over all the available Hilbert space, and there is no more room left to further  scramble the information. Chaotic quantum dynamics thermalizes an isolated, many-body quantum system. Suppose two pure orthogonal states of the chain had different initial expectation values of a local operator.  After thermalization, they turn out to have the same expectation value equal to the expectation with respect to the thermal state.
\section{Weak Measurement Tomograpghy}
We follow an excellent pedagogical review on quantum measurements by Svensson \citep{svensson2013pedagogical} to give a background on weak continuous measurements.
The total system under consideration consists of the object system and the pointer/meter. Let us assume that the object system and the meter are initially in an uncoupled state. Their time evolution is governed by the total Hamiltonian $H_\tau =H_S+ H_M +H_{int}.$ Here $H_S$ and $H_M$ denote the Hamiltonian of system and meter respectively. $H_{int}$ is the interaction Hamiltonian between the system and the meter. 
 Let the initial state of the total system be $\sigma_0 \otimes \mu_0,$ where the state $\sigma_0$ belongs to the object system and $\mu_0$ to the pointer. 
Both the system and the meter interact via a time evolution operator $\mathcal{U}$, generated by the total Hamiltonian, $H_\tau$. The evolution is given by
\begin{equation}
\mathcal{U} \tau_0 \mathcal{U}^\dagger= \mathcal{U} \sigma_0 \otimes \mu_0 \mathcal{U}^\dagger,
\end{equation}
where $\mathcal{U}= \mathrm{exp} \left(- \frac{i}{\hbar} \int \mathrm{d}t H_\tau \right)$.

Let the initial state of the probe~(meter) be $|m^{(0)}\rangle.$ The meter and system undergoes a collective unitary evolution, and let  $|m^{(i)}\rangle$ be the state of the meter at time $t$. The Hilbert space $H_M$ of the meter is spanned by a complete, orthogonal set of basis states $\lbrace \ket{q} \rbrace$, which are eigenvectors of an observable $\hat Q$, called the pointer observable/pointer variable. Expanding  the meter state in terms of a continuous pointer variable $\hat{Q},$ with pointer states $|q \rangle$,
\begin{equation}
|m^{(0)}\rangle= \int \mathrm{d}q |q \rangle \langle q|m^{(0)}\rangle= \int \mathrm{d}q|q\rangle \psi_0(q)
\end{equation}
\begin{equation}
|m^{(i)}\rangle= \int \mathrm{d}q |q \rangle \langle q|m^{(i)}\rangle= \int \mathrm{d}q|q\rangle \psi_i(q). \label{post_meas}
\end{equation}
Let us say that initially, the pointer of the meter is centered around $q=0$ so that $\langle \hat Q\rangle_0=0$. That is a convenient choice because the difference in the pointer variable is what characterizes a measurement. Let us also choose the initial wave function of the meter to be a Gaussian, centered at zero. 

\begin{equation}
\psi_0(q)= \dfrac{1}{(2 \pi \Delta^2)^{1/4}} \mathrm{exp}\left(\frac{-q^2}{4 \Delta^2} \right),
\end{equation}
where $\Delta$ is the spread of the Gaussian probability distribution. In a collective weak measurement, the collective observable say $\mathcal{O}_c= \sum \mathcal{O}^j$, where $\mathcal{O}^j$ acts on the $j^{th}$ subsystem. 
The interaction Hamiltonian  $H_{int} = \gamma \mathcal{O}_c \otimes {P}$ captures the coupling of the object system with the meter. The variable $P$ is chosen which is conjugate to the pointer variable $ Q$, and $\gamma$ is a  coupling constant. The measurement takes place during a short time interval $\delta \tau_u$, so that
\begin{equation}
\int \mathrm{d}\tau H_{int} = \int \mathrm{d}\tau \gamma \mathcal{O}_c \otimes {P}= \gamma \mathcal{O}_c \otimes {P} \delta \tau_u.
\end{equation}
The combination $\gamma \delta \tau_u = g$ is an effective coupling constant. Measurements depend only on the interaction term of the Hamiltonian. Therefore we also assume for simplicity that $H_S$ and $H_M$ are zero.
Then
\begin{align}
\mathcal{U} &= \mathrm{exp} \left( \frac{-i}{\hbar}\int \mathrm{d}\tau H_\tau \right) \\ &= \mathrm{exp} \left( \frac{-i}{\hbar}\int \mathrm{d}\tau H_{int} \right) \\&=
\mathrm{exp} \left( \frac{-i}{\hbar}g \mathcal{O}_c \otimes {P} \right).
\end{align}
To understand how the coupled evolution changes the meter variable, assume that the system starts in one of the eigenstates $|o_i\rangle$ of the collective observable $\mathcal{O}_c.$ The pre-measurement interaction with the meter is given by

\begin{align}
|o_{i}\rangle \otimes |m^{(i)}\rangle &= \mathcal{U}|o_i\rangle \otimes |m^{(0)}\rangle \\&=
\mathrm{exp} \left( \frac{-i}{\hbar}g \mathcal{O}_c \otimes {P} \right) |o_{i}\rangle \otimes |m^{(0)}\rangle \\&= |o_i\rangle \otimes \mathrm{exp} \left( \frac{-i}{\hbar}g o_i  {P} \right) |m^{(0)}\rangle \\&= |o_i\rangle \otimes \mathrm{exp} \left( \frac{-i}{\hbar}g o_i  {P} \right) \int \mathrm{d}q |q\rangle \psi_0(q) \\&= |o_i\rangle \otimes \int \mathrm{d}q |q\rangle \psi_0(q- go_i). \label{pre_meas}
\end{align}
Then the meter state after evolution, $|m^{(i)}\rangle = \int \mathrm{d}q |q\rangle \psi_0(q- go_i)$, which implies that $\psi_i(q) = \psi_0(q-go_i)$, from Eq.~(\ref{pre_meas}).
So the initial pointer state of the meter has been transformed into a superposition of shifted initial states. The shift is proportional to the eigenvalue of the object system variable.
Until now, a measurement  has not been performed. The initial system state would generally be a linear combination of eigenstates. Now a measurement of the meter can be made.
If the meter is projected onto a particular outcome, the rest is an operator acting on the system Hilbert space, called a Kraus operator.
\begin{align}
M_q&= \langle q| \mathrm{exp} \left( \frac{-i}{\hbar}g \mathcal{O}_c \otimes {P} \right)|m_0\rangle\\&= \mathrm{exp} \left( \frac{-i}{\hbar}g \mathcal{O}_c \otimes {P} \right) \psi_0(q) \\ &= \sum_i \psi_0(q-go_i) |o_i\rangle \langle o_i| \\&= \sum_i \dfrac{1}{(2 \pi \sigma^2)^{1/4}} \mathrm{exp}\left(\frac{-(q-go_i)^2}{4 \sigma^2} \right) |o_i\rangle \langle o_i|.
\end{align}
Let $\rho_0$ denote the initial state of the object system. Then the  post measurement state is given by  \begin{equation}
  \rho_q'= \frac{M_q \rho_0 M_q^\dagger}{P(q/\rho_0)}.  \label{post_measure}
\end{equation}
The corresponding POVM element $ E_q = M_q ^{\dagger} M_q$ is given by
\begin{equation}
E_q =\sum_i \dfrac{1}{(2 \pi \sigma^2)^{1/2}} \mathrm{exp}\left(\frac{-(q-go_i)^2}{2 \sigma^2} \right) |o_i\rangle \langle o_i|.
\end{equation}
Note that    $\underset{\sigma \rightarrow0}{\mathrm{Lim}}E_q= |go_i =q\rangle \langle go_i=q|$. For a finite $\sigma$ the measurement has finite strength. As $\sigma \gg 0$, the measurement is very weak. The probability for the outcome $q$ is given by $P(q/\rho_0)= \mathrm{Tr}(E_q \rho_0).$  In the weak measurement regime, when $\sigma \gg o_i$ holds for all eigenvalues,  probability function in Eq.~(\ref{prob})  can be rewritten as follows.
  \begin{align}
  \mathrm{Prob}(q) &=  \dfrac{1}{(2 \pi \sigma^2)^{1/2}}\sum_i |\alpha_i|^2 \mathrm{exp}\left(\frac{-(q-go_i)^2}{2 \sigma^2} \right) \\
  &\approx \dfrac{1}{(2 \pi \sigma^2)^{1/2}} \mathrm{exp}\left(\frac{-(q-\sum_i |\alpha_i|^2 go_i)^2}{2 \sigma^2} \right) \label{approx0}\\
  &= \dfrac{1}{(2 \pi \sigma^2)^{1/2}} \mathrm{exp}\left(\frac{-(q- g \langle o\rangle)^2}{2 \sigma^2} \right). 
  \end{align}
  Equation~(\ref{approx0}) is obtained by Taylor expanding the exponential function up to first-order around $q=0$ \citep{vaidman1996weak}.
 $\sigma^2$ is called shot noise of the probe. There is also another noise arising due to the fundamental uncertainty in the outcomes of the projective measurements. It is called projection noise. If projection noise is large, that affects the state of the object system each time we measure the probe, called the back-action. It is easy to observe this in the above equations~(\ref{pre_meas} ,\ref{post_measure}) - different outcomes $\ket{q}$ at the probe lead to different post-measurement states of the meter, which affects further measurements. When the fluctuation caused by shot noise is much larger than the projection noise, we can neglect the back-action and assume that the system state remains reasonably unaffected during measurements. The measurement record can be written down as
\begin{equation}
    M(\tau)= \mathrm{Tr}(\mathcal{O}_0 \rho(\tau)) +\sigma w(\tau),
\end{equation}
where $w(\tau)$ is a Weiner process with mean zero and variance one. 
To make tomographically complete measurements, we need to evolve $\rho_o$ in such a way that it spans all of the available state  Hilbert space. Alternatively in the Heisenberg picture, the operators we measure have to span the operator space. We make use of Haar random unitaries to achieve this.   Let $d$ be the dimension of the Hilbert space; then Haar measure is the uniform measure on the set of unitary matrices, $U(d)$. To generate a Haar random unitary, we follow \citep{mezzadri2006generate}. Any matrix belonging to the general linear group $\mathrm{GL}(d,\mathcal{C})$ can be decomposed as a product of a unitary matrix $(Q)$, and an upper triangular matrix $(R)$. Since $R$ is invertible, we get 
\begin{equation}Q=ZR^{-1}.
\end{equation} If both the real and complex parts of each of the entries of the matrix $Z$ are distributed according to an independent and identical standard normal distribution, then $Z$ is said to be in the Ginibre ensemble. In that case, $Q$ is distributed according to the Haar measure. However, the problem is that $Q$ is not unique, since for any diagonal  unitary $\Lambda$,
\begin{equation}
QR=(Q \Lambda) (\Lambda^{\dagger} R) = Q'R' \label{qr}.
\end{equation}
Note that $Q'$ is still a unitary, and   $R'$ is still an upper triangular matrix. To make the decomposition unique, one demands that the diagonal entries of $R$ be real and positive. Then  the set of $\Lambda$ matrices leading to  equivalent transformations  in Eq.~(\ref{qr}) reduces to include  just the  identity matrix \citep{ozols2009generate}. The unitary $Q$ is  Haar distributed and unique. 

Later in this thesis, we will be looking at a restricted class of unitaries, which do not  lead to tomographically complete measurements. They are the set of  unitaries diagonal in a particular randomly chosen basis.  We choose the random basis by rotating the computational basis by a Haar random unitary. A unitary diagonal in this basis  looks like
\begin{equation}
\mathcal{U}_{\mathrm{diag}}=
\begin{pmatrix}
e^{i\phi_1}&0&0&...\\
0&e^{i\phi_2}&0&...\\
0&0&e^{i\phi_3}&...\\
...&...&...&e^{i \phi_d}
\end{pmatrix}.
\end{equation}
The phases multiplying the diagonal elements are chosen uniformly at random between $0$ and $2\pi$. We will see that they perform quite well despite missing out on some information.

We also compare the covariance matrices generated by Haar random unitaries and diagonal-in-a-basis unitaries with certain random matrix ensembles and find very good correlations. Random matrices, which have random variables as entries, could be used to study the statistical properties of big complex systems. An appropriate ensemble of random matrices has to be chosen, reflecting the system's correlations.
   \section{Random matrices}
  
  Random matrices are matrices formed by random variables as elements. Random matrix theory deals with understanding such matrices' properties (like eigenvalue statistics) \citep{mehta2004random}. These matrices are useful in modeling physical systems, and they are widely used in various fields of physics. {Eugene Wigner} kick-started a whole new field when he used random matrices to model nuclei of heavy atoms in 1955 \citep{wigner1993characteristic}. In 1984, the BGS conjecture established that the eigenvalue statistics of chaotic quantum systems could be described by random matrices \citep{bohigas1984characterization}. Other fields where random matrices are found helpful include quantum optics \citep{aaronson2011computational}, mesoscopic physics \citep{sanchez2004magnetic}, quantum gravity \citep{franchini2009horizon}, quantum chromodynamics \citep{verbaarschot2000random} and superconductivity \citep{bahcall1996random}.
  
  Dyson, while studying spectral properties of many-body quantum systems, introduced ensembles of random matrices called Gaussian ensemble \citep{dyson1962statistical}. Such statistical ensembles have proven to be useful in systems exhibiting chaos in the classical limit. Appropriate ensembles of random matrices can closely resemble the spectral statistics of fully chaotic quantum systems. Hamiltonians of such systems can be modeled using random Hermitian matrices. If the Hamiltonian is time-reversal invariant, then the Gaussian orthogonal ensemble (GOE) or the Gaussian symplectic ensemble (GSE) is the relevant sets of matrices. On the other hand, if the dynamics is not time-reversal invariant, then the Gaussian unitary ensemble (GUE) models the system.
  
  Suppose one constructs a square matrix $H$, with each element being chosen from a standard Gaussian distribution. Now to make it Hermitian,  symmetrize the matrix $H_s= (H+H^T)/2$. Then $H_s$ is said to be an element of the Gaussian orthogonal ensemble. To produce elements of a Gaussian unitary ensemble, one simply needs to replace the real entries with complex ones, where both the real and complex parts are chosen according to a Gaussian distribution. For GSE, one uses quaternions. 
  
  Modifying these Gaussian ensembles to form measures on the unitary space, one arrives at circular orthogonal matrices (COE) and circular unitary ensemble (CUE)  \citep{dyson1962statistical}. The CUE consists of unitary matrices distributed according to the Haar measure on the unitary group $U_N.$ The set of all symmetric unitary matrices makes up the circular orthogonal ensemble. The COE describes unitary evolutions with time-reversal symmetry.
  
\textbf{Wishart ensemble:}  In this thesis, we will be using an important ensemble called the Wishart ensemble \citep{wishart1928generalised}. Wishart matrices were the first random matrices to have been studied in the literature.   Wishart matrices~$(\mathcal{W})$ are square matrices constructed by multiplying  Gaussian rectangular matrices by their adjoint. \textit{i.e.,}
\begin{equation}
\mathcal{W}= HH^{\dagger}.
\end{equation}
$H$ is an $N\times M$ matrix, with each element chosen from a Gaussian probability density function. Hence $\mathcal{W}$ is an $N \times N$ square matrix. An ensemble of Wishart matrices is also called a Laguerre ensemble. The eigenvalue density distribution of Wishart matrices follows the Marchenko-Pastur distribution.   For a Wishart matrix constructed from $D \times N$ rectangular random matrix with  $D \leq N$, the Marchenko-Pastur density function denoted by $\rho(\lambda)$ is given by
\begin{align}
\rho(\lambda) &= \frac{N}{2 \pi  \lambda} \sqrt{(\lambda-\lambda_-) (\lambda_{+}-\lambda)} \\
\lambda_{\pm}&= \frac{1}{N}\left(1- \left(\frac{D}{N}\right)^{-1/2}\right)^2,
\end{align}
where $\lambda_{-}$, and $\lambda_{-}$ are the minimum and maximum eigenvalues respectively,  $\lambda \in [\lambda_{-}, \lambda_{+}].$ Another important distribution we use in this thesis is the Porter-Thomas distribution. In analysing the  resonance width of large nuclei, { Porter} and {Thomas} found  that  a chi square distribution with a single degree of freedom is consistent with the data \citep{porter1956fluctuations}.  The Porter-Thomas distribution  represents frequency distribution of components of a pure unit vector, chosen uniformly at random in a $d^2$- dimensional real Hilbert space. Let $a_i$ be the $i^{th}$ component of the random real pure state, then  probability for obtaining the $i^{th}$ outcome $p_i= a_i^2$. When the dimension of the Hilbert space $d^2$ is large,  the  $i^{th}$ outcome occurs  $\lambda_i= d^2p_i$ times. The distribution of these frequencies follow the Porter-Thomas distribution given by
\begin{equation}
\rho(\lambda)= \frac{1}{\sqrt{2\pi \lambda}} e^{-\lambda/2}. 
\end{equation}

\section{The kicked top}
  Periodically kicked quantum systems are popular in studying the quantum origins of chaos because of the ease of analysis. The kicked top is one such prototype model \citep{haake1987classical}. Classical Hamiltonian of the kicked top has the angular momentum components $(J_x,J_y,J_z)$ as the dynamical variables. The square of the angular momentum $J^2= J_x^2+J_y^2+J_z^2$ is conserved, and the motion of $\overset{\rightarrow}{J}$ is restricted to the surface of a sphere with radius $J^2.$ The classical Hamiltonian is given by
 \begin{equation}
   H(\overset{\rightarrow}{J},t)= \alpha J_x + \tau J_z^2 \sum_{n=0,\pm 1,\pm 2,..} T \delta(t-nT),
 \end{equation}
where $\alpha$ is the velocity of rotation generated about the $x$ axis, and $\tau$ is the torsion strength. Torsion or the kick is a state-dependent rotation about the $Z$ axis, proportional to $J_z$. The kick takes place at a temporal distance $T$, periodically. A kicked top model can be seen as the collective motion of a set of spin-1/2 particles, each interacting with every other spin with an interaction strength $\tau$ \citep{ghose-pra-2008}. Equivalently it describes the motion of one large spin with angular momentum $J.$  In the quantum limit, one replaces the angular momentum vectors by operators to get the quantum Hamiltonian. The unitary operator obtained by exponentiating the Hamiltonian is given by
\begin{equation}
    U= \mathrm{exp}\left(\frac{-i\tau J_z^2}{\hbar^2(2j+1)}\right) \mathrm{exp}\left(\frac{-i\alpha J_x}{\hbar}\right).
\end{equation}
Evolution of an initial state using powers of this unitary yields a stroboscopic map at discrete times $t=nT$, where $n=\lbrace0,1,2...\rbrace$. Since $J^2$ commutes with $U$, the motion is confined to  $2j+1$ dimensional Hilbert space. To compare the classical and quantum dynamics, one uses the directed angular momentum states called the spin coherent states for analysis since they are the closest to being classical.

 \section{Fisher information}
  
 Let $X$ be a random variable, dependent on an underlying parameter $\theta.$ Then, the Fisher information captures the volume of knowledge about $\theta$ contained in a realization of the random variable $X.$ If the probability density of the realizations denoted by $f(X;\theta)$ is sharply peaked, then a small change in the parameter leads to a large change in the probability density. This means that a given outcome of the random variable contains a lot of information about the underlying parameter. On the other hand, if the density function is flat, then a particular realization of the random variable does not give much information about $\theta.$ Then, it takes a large sample of $X$ outcome data to reach a conclusion about the actual value of the parameter. This urges a quantification of the amount of information gain in terms of the variance with respect to the parameter. The formal definition of the Fisher information is given as follows \citep{fisher1922mathematical, cover2012elements},
 \begin{equation}
     \mathcal{F}(\theta)= \mathrm{E}\left[\left( \left.\dfrac{\partial}{\partial \theta} \mathrm{log}f(X;\theta)\right)^2\right|\theta\right].
 \end{equation}
 When there are multiple parameters $\theta=\lbrace \theta_1,\theta_2,...\theta_n \rbrace$ determining $X,$ the Fisher information becomes a matrix. An element of the matrix is given as
 \begin{equation}
      \mathcal{F}(\theta)_{i,j}= \mathrm{E}\left[\left( \left.\dfrac{\partial}{\partial \theta_i} \mathrm{log}f(X;\theta)\right) \left( \dfrac{\partial}{\partial \theta_j} \mathrm{log}f(X;\theta)\right) \right|\theta \right].
 \end{equation}
 An important relation concerning Fisher information was stated in 1946 by {Herald Cramer} and {C.R Rao}, called the Cramer-Rao bound \citep{cramir1946mathematical,rao1992information}. Consider $\hat{\theta}$, an unbiased estimator of the parameter $\theta.$  $\hat{\theta}$  is a rule/function for calculating the estimate of $\theta$, whose expectation is equal to the true value of the parameter. Then the Cramer-Rao bound states that the inverse of the Fisher information is a lower bound for the error in the estimator,
 \begin{equation}
     \mathrm{Var}(\theta) \geq \frac{1}{\mathcal{F}(\theta)}.
 \end{equation}Thus Fisher information sets a fundamental limitation to the precision of the parameter estimation.
  \section{Shannon and Von Neumann entropy}
 Shannon entropy of a classical random variable \citep{shannon2001mathematical, cover2012elements}  quantifies the average information gained in learning the value of the random variable $X.$ Alternatively, it is the amount of uncertainty about $X$ before the observation has been made. It is written as a function of the probability distribution of the random variable realizations.
 Let $\lbrace p_1,p_2...p_n\rbrace$ be the probabilities for various outcomes of $X,$ then the Shannon entropy of this set is given by
 \begin{equation}
     \mathcal{H}(X)= \mathcal{H}(p_1,...p_n)= - \sum p_x \mathrm{log}p_x.
 \end{equation}
 Shannon entropy attains its maximum value if all the outcomes are equally probable. Instead, if the outcome of the random variable is absolutely certain, then Shannon entropy is zero since there is no uncertainty. 
 The analogue of Shannon entropy in quantum mechanics is called the Von Neumann entropy \citep{von2018mathematical}. In quantum theory,  states are described by density matrices. Entropy of a quantum state $\rho$ is defined as
 \begin{equation}
     S(\rho)= - \mathrm{Tr}(\rho \mathrm{log}\rho).
 \end{equation}
  Let $\lbrace \lambda_1, \lambda_2... \lambda_n \rbrace$ denote the eigenvalues of the density matrix $\rho,$ then the Von Neumann entropy can be written in a fashion similar to Shannon entropy as
  \begin{equation}
      S(\rho)= - \sum \lambda_i \mathrm{log}\lambda_i.
 \end{equation}

  \section{Deterministic quantum computation with one pure qubit}
Quantum computation is beneficial since it can solve certain problems much faster than classical computers. The potency of quantum computers is ascribed to quantum interference, superposition, and quantum parallelism. Unlike classical computers, quantum computers can perform many tasks in parallel. Classically, the basic unit of information is a bit, and it can be either a zero or a one. But the quantum mechanical unit, called a qubit, can be zero and one simultaneously! This is because of the superposition principle. If a classical computer has four bits, it can represent a number from zero to fifteen at a time. However, with four qubits, a quantum computer can represent all of them simultaneously. If we have a quantum processor, it can work out the outputs for all sixteen inputs at once. This is called quantum parallelism.

Another source of speedup is attributed to entanglement. The amount of information contained in entangled qubits grow exponentially with the number of qubits involved. The same number of classical bits contain much less information as the classical information only grows linearly.

Despite all this, a model of quantum computation was proposed, which showed exponential speedup in performing certain restricted tasks over the classical algorithms \citep{knill}. It was called the deterministic quantum computation with one pure qubit  (DQC1 model). Here we illustrate one of the computations that can be performed using this model. 
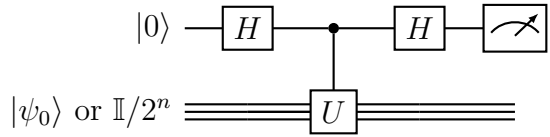
\begin{figure}[!ht]
		\centering
	\begin{quantikz}
		\lstick{$\ket{0}$} & \gate{H} &  \ctrl{1}   & \gate{H} & \meter{} \\
		\lstick{$\ket{\psi_0}$ or $\mathbb{I}/2^{n}$}  & \qwbundle[alternate]{} & \gate{U}  \qwbundle[alternate]{}& \qwbundle[alternate]{} & \qwbundle[alternate]{} 
	\end{quantikz}
	\caption{Circuit for computing the trace of the unitary $U$}
	\label{trace}
\end{figure}
We can compute the trace of the unitary $U$ using the circuit shown in Fig.~\ref{trace}. In this the state of the probe is initiated to the pure state $\ket{0}.$  For the ease of demonstration, assume for now that the state of the system  also starts out as a pure state $\ket{\psi_0}$. Then the initial state of the combined probe-system is $\ket{0} \otimes \ket{\psi_0}$. After the action of the first Hadamard on the probe, the combined state changes to  $\frac{\ket{0+1}}{\sqrt{2}} \otimes \ket{\psi_0}$. In the next step, unitary acts conditionally on the system. After its action, the new state is  $\frac{\ket{0}\otimes \ket{\psi_0}}{\sqrt{2}} +\frac{\ket{1} \otimes \ket{U \psi_0}}{\sqrt{2}}.$ After the  the second Hadamard, the state changes to $\frac{1}{2} \left[(\ket{0}+\ket{1}) \otimes \ket{\psi_0} + (\ket{0}-\ket{1}) \otimes \ket{U\psi_0}\right] = \frac{1}{2}\left[ \ket{0} \otimes (1+U) \ket{\psi_0} + \ket{1}(1-U)\ket{\psi_0}\right].$  Now measuring $\sigma_z$ on the probe, the probability to get $0$ is given by

\begin{align}
\mathrm{Pr}(0) &=\frac{1}{4} \mathrm{Tr} \left[ (1+U) \ket{\psi_0} \bra{\psi_0} (1+U^\dagger) \right] \\
&= \frac{1}{4} \mathrm{Tr}\left[ \ket{\psi_0} \bra{\psi_0} + U\ket{\psi_0}\bra{\psi_0}U^\dagger + U \ket{\psi_0} \bra{\psi_0} +\ket{\psi_0} \bra{\psi_0}U^\dagger \right]\\
&= \frac{1}{2} \left( 1+ \mathrm{Re}(\mathrm{Tr}[ U \ket{\psi_0}\bra{\psi_0}] \right).
\end{align}
Replacing $\ket{\psi_0}$ by the maximally mixed state $\mathbb{I}/2^n$, we get 
\begin{equation}
    \mathrm{Pr}(0)=  \frac{1}{2} \left(1+ \frac{1}{2^n} \mathrm{Re(Tr}[U]) \right).
\end{equation}
 Similarly, the probability to measure $1$ is given by 
 \begin{equation}
     \mathrm{Pr}(1)=  \frac{1}{2} \left(1- \frac{1}{2^n} \mathrm{Re(Tr}[U]) \right).
 \end{equation}
 
   Thus from the above two equations, one can get the real part of the trace of the unitary. By measuring $\sigma_y$ instead, one can get the imaginary part of the trace. Combining both, we have the trace of the unitary. $L$  measurements on the top qubit give us an estimate of the trace with fluctuations of size $1/\sqrt{L}$. Therefore, to achieve an accuracy $\epsilon$, one requires $L \sim 1/\epsilon^2$ implementations of the circuit. If $P_e$ is the probability that the estimate departs from the actual value by an amount $\epsilon$, then one needs to run the experiment $L \sim \log(1/P_e)/\epsilon^2$ times.
This accuracy in the estimate is independent of the size of the unitary matrix and hence performs exponentially faster than what is classically achieved.

 \section{Levy's lemma}
 A sphere of higher dimensions has nontrivial properties. To see this, consider a point on the surface of the hypersphere $S^{n-1}$, denoted  $x= (x_1,x_2,...x_{n-1}).$ Let us look at the set of points lying on a narrow belt of width $\epsilon$ around an equator about a coordinate  $x_j$, 
 \begin{equation}
     S_{j\epsilon}= \lbrace x\in S^n  : |x_j| \leq \epsilon/2 \rbrace.
 \end{equation}
 It turns out that the measure ($\mu$) of points on the surface of the hypersphere, normalized to unity  obey
 \begin{equation}
     \mu(S_{j\epsilon}) \geq 1-\mathrm{exp}(-k n \epsilon^2/2),
 \end{equation}
 where $k>0$ is a constant.
  The above relation holds for any coordinate $x_j$. This means that most of the surface area of the hypersphere is concentrated around the equator. This observation can be generalized to a class of functions called Lipschitz continuous functions.

 \textbf{Lipschitz functions}: Lipschitz continuity \citep{o2006metric} is a strong form of uniform continuity. It limits how much the function's slope can vary in its domain. The definition of Lipschitz continuity is as follows. 
 A function $f$ from the metric space $(X,d_x)$ to the metric space $(Y,d_y)$, where $d_x,d_y$ denote the metrics in the corresponding spaces, is said to be Lipschitz continuous if $\exists k>0$ such that $\forall \lbrace x_1,x_2\rbrace \in X, $  
 \begin{equation}
    d_y(f(x_1),f(x_2)) \leq k d_x(x_1,x_2). 
 \end{equation}
 
 \textbf{Levy's Lemma}: Given a Lipschitz continuous function $f: S^{n-1} \rightarrow \mathbb{R}$ defined on a higher dimensional hypersphere $S^{n-1}$, and a point $x \in S^{n-1}$ chosen uniformly at random. Then the measure of such points ($\mu$) is
\begin{equation}
\mu\left\lbrace x \in S^{n-1}: |f(x)- \mathbb{E}{f(x)} |\geq \epsilon \right\rbrace \leq 2 \mathrm{exp} \left(\frac{- kn \epsilon^2}{\eta^2} \right). \label{one}
\end{equation}
Here $\mathbb{E}{f(x)}$ is the expectation value of the function, $\eta$ is the Lipschitz constant of $f$, given by $\eta= \mathrm{sup}|\bigtriangledown f|$ and $k$ is a positive constant \citep{milman2009asymptotic, ledoux2001concentration}. A derivation of the lemma can be found in the lecture note \citep{gerken2013measure}.

\chapter{Signatures of chaos in small quantum systems}
\label{chap:chap3}
The contemporary interest and progress in quantum information processing have happened along with control over single
or few particle systems that are driving home the novelty of unique quantum phenomena such as
entanglement. It has also opened doors for investigation in the time domain, with exquisite control of individual quantum systems in the laboratory and the ability to drive these systems with designer Hamiltonians that can simulate phenomena as diverse as many-body-localization to ergodicity, chaos 
and thermalization. Two experiments that preserve the coherence and purity of complex many-body
time-evolving states illustrate the richness of this domain \citep{Neill16, Kaufman2016}. 

The first of these \citep{Neill16} involved the study of 3 qubits in a superconducting transmon setup that simulated the quantum kicked
top. Using state tomography they made connections between the onset of chaos and concomitant enhancement
in the entanglement. The second \citep{Kaufman2016} involved a two-dimensional Bose-Einstein condensate of $^{87}$Rb atoms,
implementing effectively a 6-particle Bose-Hubbard Hamiltonian. The study of thermalization via the development
of entanglement in such experiments on isolated quantum systems is of interest in the foundations of statistical mechanics,
and they test the Ergodic Thermalization Hypothesis (ETH) that is currently of great theoretical interest as well. 
Connections between low-dimensional ergodicity and chaos with entanglement, general quantum correlations and state tomography have long been studied, mostly theoretically, (for example in \citep{MillerSarkar,Lakshminarayan,BandyopadhyayArul2002,Tanaka-2002,LakSub2003,Bandyopadhyay04,Ghose2004,ArulSub2005,trail2008entanglement,LombardiMatzkin2011,PhysRevLett.112.014102, madhok2015signatures, Madhok2018_corr,Maciej-2019,Meenu-2019,Pappalardi-2020}), although a cold-atom experiment as early as 2009 \citep{Chaudhary} was a pioneering work in this direction. 

These experiments also beg the question of how statistical properties such as thermalization and semiclassical 
properties such as chaos manifest in such low-dimensional quantum systems. The 3-qubit transmon experiment is based on 
the mapping of the well-studied quantum kicked top to a many-spin Floquet system. However, while traditional
studies of quantum chaos are for large spin $j$ \citep{Haake}, this experiment involved only $j=3/2$ (3 qubits) and the mapped
system is in fact a nearest neighbor transverse field Ising model which is integrable. In any case, the solvability 
of this as well as the 4 qubits or $j=2$ system which involves non-integrable next-nearest-neighbor interactions was demonstrated
in \citep{dogra19}. Such a study did show that it is possible to see some generic features and even some random matrix theory 
properties in such small systems. For example, it showed how with increasing the parameter controlling the non-integrability, entanglement moves
from being bipartite to multipartite, sharing it globally and demonstrating its monogamous nature.

The study of quantum signatures of chaos involves a wide spectrum of overlapping research directions.  
Study of level statistics, semiclassical approximations of the spectrum, and connections to the periodic orbit theory have been the focus traditionally \citep{Haake}. However, several ``dynamical" signatures of chaos are being vigorously pursued with important consequences to quantum information processing as well as fundamental physics.
These include quantum-to-classical transition, classical emergence of chaos via decoherence/weak continuous measurement to recent trends in studying the dynamical generation of entanglement/quantum correlations and information scrambling using out of time-ordered correlators \citep{MillerSarkar,Lakshminarayan,BandyopadhyayArul2002,Tanaka-2002,LakSub2003,Bandyopadhyay04,Ghose2004,ArulSub2005,trail2008entanglement,LombardiMatzkin2011,PhysRevLett.112.014102, madhok2015signatures, Madhok2018_corr,Maciej-2019,Meenu-2019,Pappalardi-2020, bhattacharya2003continuous, habib2006emergence, oz02-2, Zurek/Paz, yan2020information, larkin,swingle2016measuring, chaos1,Cotler-2018,Hashimoto-2017,Swingle-2018,Carlos-2019,  PhysRevX.7.031011}.

Studying the properties of stationary states (level statistics) and localization of eigenstates can give us a great deal of insight, but the actual dynamics often has surprises, e.g., the idea of decoherence, information scrambling, and coherent control. Traditionally, there has been an emphasis on the ``energy domain" since one could not control quantum systems in the past. So one rarely could do an experiment in which the system was prepared at time t=0 in a pure state and then allowed to evolve, and then measured.  That is a much more recent development with atomic, molecular, and optical physics as well as progress in quantum technologies like superconducting qubits playing a huge part.
This is fuelled by recent experiments in quantum information where questions like the ability to control quantum systems \citep{Chaudhary, poggi2020quantifying}, thermalization in closed quantum systems \citep{Neill16} and quantum simulations of chaotic and non-integrable Hamiltonians \citep{sieberer2019digital, PhysRevX.7.031011} are of prime importance.

Starting from \citep{RuebeckArjendu2017} which considered just two qubits ($j=1$) analytically and 3 qubits ($j=3/2$) numerically,
there have been studies that followed the fate of the few qubit kicked top \citep{dogra19,Madhok2018_corr,Bhosale-2018}. 
A recent experiment \citep{MaheshUdayExpt-2019} used NMR to study the 2 qubit version of the kicked top already displaying some semiclassical features, and 
also peculiar quantum ones such as time- and parameter- periodicity \citep{Bhosale-2018}.

The chapter is placed in this context as one that explores how two measures based on the time-evolution fare in ferreting out
non-integrability and chaos out of small quantum systems that are already experimentally realizable, hence the question is how low can we go? These measures are the out-of-time-ordered correlator (OTOC), being intensely studied now in a remarkable variety of contexts, and the Loschmidt echo, which has a longer history of study in low-dimensional chaos. We find that although only very short-time information is available, OTOC of $j=2$ and $j=5/2$ kicked tops already show definite precursors of exponential growth, and many properties of the echo are also shared by large $j$ systems, although the exponential decay may not be apparent, at least in the regimes we have addressed here. Thus the answer to the question seems to be ``pretty low".
We also initiate the study of a kicked top of arbitrary spin $j$, but when the chaos parameter is so ``absurdly large" that the Lyapunov exponent $\lambda_C$ is as large as $\log(1/h_{\text{eff}})$, and the Ehrenfest time is still of order 1! This ``dual" case also manifests for low values of $j$, one does not require is a very large value of the chaoticity parameter for the top. The kicked top Floquet operator, as we shall discuss, can be written as a {\it sum} of just $4$ rotations (for integer $j$), and hence the interactions need not be implemented at all.

An array of quantum signatures of chaos have already been studied. Fidelity decay in quantum systems \citep{ peres1984stability, sc96}, level statistics \citep{berry1977level, bohigas1971spacing}, properties of regular and irregular wave functions \citep{Berry77a, BERRY197926, voros1976semi} and quantum scars  \citep{heller1984bound}, signatures in single particle billiards \citep{McDonald, robnik1985classical}, semiclassical trace formulas \citep{gutzwiller1971periodic} and imprints on quantum correlations and tomography  \citep{ArulSub2005,Bandyopadhyay04,BandyopadhyayArul2002,Chaudhary,Ghose2004,LakSub2003,Lakshminarayan,LombardiMatzkin2011,madhok2015signatures, Madhok2018_corr, trail2008entanglement, MillerSarkar, PhysRevLett.112.014102}.
Recent trends that focus on many-body systems,  include studies involving connections of quantum chaos to OTOCs, entropic uncertainty relations, spread of quantum information throughout the system
with consequences ranging from the foundations of quantum statistical mechanics, quantum phase transitions, and thermalization on the one hand to the spreading of quantum information in many-body systems and black holes on the other hand \citep{swingle2016measuring, hayden2007black, MaldacenaSYK15, chaos1, hartman2013time, shenker2014black, sekino2008fast, Lashkari2013, pawan, IyodaSagawa, PhysRevX.7.031011}.

The OTOC, in their simplest form, captures the growth of the incompatibility between two operators, when one of them is evolved in the Heisenberg picture while the other is stationary \citep{larkin,swingle2016measuring, chaos1,Cotler-2018,Hashimoto-2017,Swingle-2018,Carlos-2019, PhysRevX.7.031011}. 
Incompatibility is analogous to separation in classical phase space.
For two Hermitian observables, OTOC is given by
\begin{equation}\label{eqq}
C_{W,V}(\tau)=-\langle[W(x, \tau), V(y, 0)]^2\rangle, 
\end{equation}
where the local operators $W$ and $V$  act on sites $x$ and $y$ respectively and $W(x, \tau)=U^{\dagger}(\tau)W(x, 0)U(\tau)$ is the Heisenberg evolution of operator $W$ under unitary dynamics $U(\tau)$. The expectation value is taken with respect to the maximally mixed state. In sufficiently chaotic systems, the OTOC  essentially vanishes till the information of the operator perturbation at $x$ reaches $y$, during which phase the operator becomes highly non-local, an occurrence that is dubbed operator scrambling. Thereafter there is a rapid increase of the OTOC before
it saturates in a finite system at which stage the localized information at $x$ is considered to have been scrambled throughout the system, and 
it is not possible to recover it from any local subset. If the rapid increase of the OTOC is exponential $\sim e^{2 \lambda_Q\tau}$, $\lambda_Q$ has
been referred to as a quantum Lyapunov exponent.

If the system has a bound spectrum this implies instability in a finite space and can be taken as a 
definition of quantum chaos. Thinking of systems with a well-defined semiclassical limit, note that simple systems such as the inverted parabolic potential $-x^2$ have trivially exponentially growing OTOC, but are of course not chaotic, but merely unstable. Similarly, there could be naturally isolated unstable orbits in an otherwise integrable system and special operators may still show exponential OTOC growth. Still, the jury is out on the role of OTOC in general and hence studying them in as many scenarios is of interest. Systems with well-defined semiclassical or classical limits are of special interest as it is well understood in what sense they are non-integrable and what the classical Lyapunov exponents are, and there have been several studies on this \citep{Rozenbaum17,arul2,Saraceno-2018,Moudgalya-2018,Klaus-2018,Jalabert-2018,arul3}, including some on the quantum kicked top \citep{arul_vaibhav,sieberer2019digital,yin2020quantum}

Though quantum systems do not show sensitivity to perturbations in initial state vectors, integrable and chaotic quantum systems show remarkably different behavior and sensitivity when the system dynamics itself is perturbed \citep{Peres02, peres1984stability}. One of the concepts used to capture this notion of quantum chaos is the Loschmidt echo that is related to the fidelity between the evolution of a quantum system with exact dynamics and propagation under a slightly perturbed Hamiltonian \citep{prosen2003theory, gorin2006dynamics, PhysRevE.68.036216}. Alternatively, this quantifies the distance between the forward propagation of a system and its time-reversed dynamics under small perturbations.
This is interesting as the question of time-reversal itself and its connections to chaos, both quantum and classical, has been one of the foundational questions in physics.  The debate around the microscopic origins of the second law of thermodynamics from underlying time-reversal invariant classical mechanics leads to interesting paradoxes. For example, could one reverse the momenta of all particles in a system causing the entropy to decrease thereby violating the second law \citep{Peres02}? 
In this chapter, our focus is to study Loschmidt echo for a few qubit kicked top that is exactly solvable \citep{dogra19}.

Loschimidt echo is defined as 
\begin{equation}
F(\tau)=|\langle \phi  |e^{i H'\tau} e^{-i H\tau} |\phi \rangle|^2,
\end{equation}
where $\ket{\phi}$ is the initial state, and $H$ is the Hamiltonian for the forward evolution and $H'$ is the perturbed Hamiltonian representing
imperfect time reversal, {\it i.e.}, the Hamiltonian responsible for backward evolution. The perturbed evolution can be 
due to environmental noise and thus there is an intimate connection between Loschmidt echo and decoherence \citep{Zurek/Paz}
The Loschmidt echo has a rather complex behavior that depends on the state $|\phi\kt$, the nature of the Hamiltonian $H$--whether
it is integrable or not, the degree of chaos if it is not integrable and also on the strength of the perturbation that defines $H'$. In certain regimes,
an exponential decay of the fidelity has been observed with a rate that is the classical Lyapunov exponent.

Recently the question of sensitivity to perturbations is connected to the accuracy and robustness of quantum information processing devices. After all, the quantum device/simulator is a many-body complex quantum system and one needs to benchmark its accuracy \citep{sieberer2019digital, PhysRevE.62.3504, PhysRevE.62.6366}.
How does one trust a quantum simulator that invariably involves a many-body chaotic Hamiltonian with a rapid proliferation of errors, especially near a quantum critical point that is typically characterized by high entanglement/complexity and a large Schmidt rank of the system density matrix \citep{hauke2012can, georgescu2014quantum, deutsch2020harnessing}? While these questions have been under active research for many decades, only recently experiments have reached the level of sophistication and
control where non-integrability, chaos, and thermalization of closed quantum systems are studied by
manipulating individual interacting quantum bits. Another interesting avenue on the applications of Loschmidt echo is the application to quantum-limited metrology and making sensors. Since chaotic systems are sensitive to perturbations, this suggests a way to for high precision metrology \citep{fiderer2018quantum}. 

This chapter is arranged in the following way. In Sec.~(\ref{sec:KTinto}) the kicked top model is described and some of its classical properties are mentioned. A complete solution of the quantum problem for $3$ and $4$ qubit cases is also carried out, in the sense that explicit expressions for the powers of the Floquet operator are given in terms of the Chebyshev polynomials. In Sec.~(\ref{sec:OTOC})
the OTOC is derived for the 3 and 4 qubit kicked tops and their dependence on time and the chaoticity parameter is discussed. The OTOC 
is also compared with that for a larger number of spins, found numerically. The peculiar case when the number of spins is arbitrary by the chaos parameter is very large is also discussed in this section. {In Sec.~(\ref{sec:Echo}), the Loschmidt echo is discussed and we summarise and discuss future directions in  Sec.~(\ref{sec:Conc}).}
\section{The case of kicked top}
\label{sec:KTinto}
The quantum kicked top is  characterized by the angular momentum vector $(J_x,J_y,J_z)$, and the Hamiltonian \citep{KusScharfHaake1987,Haake,Peres02}  is written as
\begin{equation}
\label{Eq:QKT}
H=\frac{\kappa_0}{2j}{J_z}^2 \sum_{n = -\infty}^{ \infty} \delta(t-n\tau)+\frac{p}{\tau} \, {J_y}.
\end{equation}
It consists of rotations and impulsive rotations caused by periodic kicks at regular intervals of time $\tau.$ The time evolution of the top is given by the unitary
\begin{equation}
\mathcal{U} = \exp\left [-i (\kappa_0/2j \hbar) J_z^2 \right]\exp\left[-i (p/\hbar) J_y\right],
\end{equation}
which describes the evolution from one kick to the next. Angle of rotation about the $y$ axis is given by $p$, and $\kappa_0$ is the chaoticity parameter, which is a measure of the twist applied between kicks.
 Here  we take $\hbar=1$ and $p=\pi/2$.
In the limit of very large angular momentum, the classical limit is reached. At $i^{th}$ iteration the classical map   of the unit 
sphere  $X_{i}^2+Y_{i}^2+Z_{i}^2=1$ onto itself 
is given by 
\begin{eqnarray}
X_{i}&=&Z_{i-1}\cos(\kappa_0 X_{i-1})+Y_{i-1}\sin (\kappa_0 X_{i-1}),\nonumber \\
Y_{i}&=&-Z_{i-1}\sin(\kappa_0 X_{i-1})+Y_{i-1}\cos (\kappa_0 X_{i-1}),\nonumber \\
Z_{i}&=&-X_{i-1}.
\end{eqnarray}
where $X_{i},Y_{i},Z_{i}=J_{x,y,z}/j$ 
\begin{figure}
	\centering
	\includegraphics[scale=0.65]{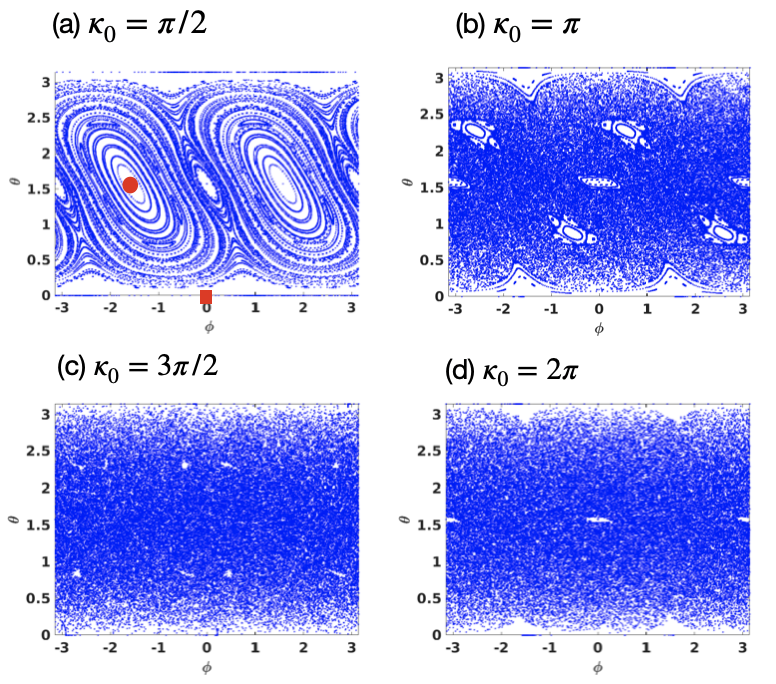}
	\caption{ The figer shows classical phase space of a kicked top at various chaoticities. (a) shows regular phase space at a small $\kappa_0 $ value. (b) and (c) represent mixed phase space with some regular islands appearing amidst the chaotic sea. (d) shows chaotic phase space at $\kappa_0=2\pi$. The point marked with a red square $(\Theta=0, \Phi=0)$, lies on a period-four orbit, while the red circle $(\Theta=\pi/2, \Phi=-\pi/2)$ is at the center of a regular island.}
	\label{fig:classical}
\end{figure}

Dynamics of a particle under these equations are simulated 
numerically for different initial states:
$(X_{0}, Y_{0}, Z_{0})$, and for two 
values of the chaos, $\kappa_0=0.5$ and $2.5$, as shown in Fig.~\ref{fig:classical},
conventionally called regular and mixed phase space structures respectively. At $\kappa_0=0,$ the Hamiltonian is only that of rotation, and the classical map is integrable. 
For $\kappa_0>0$ the phase space contains both regular and chaotic regions. $\kappa_0 >0$, is fully chaotic regime.
The kicked top can be modeled as a many-body system by regarding the spin angular momentum $J$ as made up of multiple spin-1/2 particles. Then  $J_{x,y,z}$ with $\sum_{l=1}^{2j} \sigma^{x,y,z}_l/2$ \citep{Milburn99,Wang2004}.
Then the Floquet operator comprises of  $N = 2j$ qubits. It describes an  ising chain of equal, all-to-all interaction between the qubits, with periodic kicks induced by a delta-pulsed magnetic field in the transverse direction.  
\begin{equation}
\label{uni}
{\mathcal U}=\exp\left(-i \frac{\kappa_0}{4j}  \sum_{ l< l'=1}^{2j} \sigma^z_{l} \sigma^z_{l'}\right)
\exp\left( -i \frac{\pi}{4} \sum_{l=1}^{2j}\sigma^y_l \right).
\end{equation}
Here $\sigma^{x,y,z}_l$ are the spin-1/2 Pauli operators, and an overall phase is neglected.
Ruebeck and others studied the $j=1$ case of just two qubits, and analyzed some very quantum feature, not connected to the classical case \citep{RuebeckArjendu2017}.
 For the two qubit case, many quantum correlation measures were also studied in \citep{Bhosale-2018}. The three-qubit kicked top ising model, with $j=3/2$ is shown to be integrable \citep{Prosen2000,ArulSub2005}.
For larger spins, the model is non-integrable.
We confine to the permutation symmetric subspace of the total space  for our analysis.
\subsection{Solving  the 3 and 4 qubit kicked tops}

The solutions in these cases were discussed first in \citep{dogra19},
where a wide variety of entanglement measures, from entropy to concurrence were studied and compared with
available experimental data. We recount here the essential details of the solutions for the sake of a self-contained narrative.
First, there is the general observation of an ``up-down" or parity symmetry: 
\[ [\mathcal{U},\otimes_{l=1}^{2j} \sigma^y_l]=0, \]
valid for any number of qubits. It is therefore optimal to work with a basis that is both permutation 
symmetric and is adapted to the parity.

For $j=3/2$ or the 3-qubit case, the standard 4-dimensional spin-quartet permutation symmetric space 
$\{|000\kt, |W\kt=(|001\kt+|010\kt+|100\kt)/\sqrt{3},
|\overline{W}\kt =(|110\kt+|101\kt+|011\kt)/\sqrt{3},|111\kt\}$ is spanned by the following set of basis vectors with definite parity.
\begin{eqnarray}
|\phi^{\pm}_1\kt&=&\frac{1}{\sqrt{2}}(|000\kt \mp i | 111 \kt), \label{eq:3qubit_basis_1} \\
|\phi_2^{\pm}\kt&=&\frac{1}{\sqrt{2}} (|W\rangle \pm i |\overline{W}\kt).  \label{eq:3qubit_basis}
\end{eqnarray}
 The Floquet unitary operator in this parity adapted basis makes it block-diagonal.
\begin{equation} 
\label{eq6}
\mathcal{U} = \begin{pmatrix}
\mathcal{U}_{+} & 0 \\ 0 & \mathcal{U}_{-}
\end{pmatrix}, 
\end{equation}
where the blocks $\mathcal{U}_{+}$ and $\mathcal{U}_{-}$ are two-dimensional matrices expressed in the parity adapted basis as
\begin{equation}
\mathcal{U}_{\pm} = \pm  e^{\mp \frac{i \pi}{4}} e^{-i \kappa} \begin{pmatrix}
\frac{i}{2}e^{-2i \kappa} & \mp \frac{\sqrt{3} }{2} e^{-2i \kappa} \\
\pm \frac{\sqrt{3}}{2} e^{2i \kappa} &  -\frac{i}{2}e^{2i \kappa}
\end{pmatrix},
\label{eq:Uplusm}
\end{equation}
corresponding to parity eigenvalue $\pm1$.
Where we have used $\kappa=\kappa_0/6$ for clarity. 

Expressing Eq.~(\ref{eq:Uplusm}) as a rotation ($e^{-i \theta \sigma^{\hat{\eta}}}$)
by an angle `$\theta$' about an arbitrary axis 
($\hat{\eta}=\sin{\alpha} \cos{\beta} \hat{x}
+ \sin{\alpha} \sin{\beta} \hat{y} + \cos{\alpha} \hat{z}$),
and a phase, we obtain,
$ \cos{\theta} =\frac{1}{2} \sin{2\kappa}$, 
$\beta=\pi/2 +2 \kappa$, and 
$\sin{\alpha} = \sqrt{3}/(2\sin{\theta})$.
Thus the time evolution is the propagator which is simply the power $\mathcal{U}^n$
is block-diagonal with blocks $\mathcal{U}_{\pm}^n$, which are 
explicitly given by,
\begin{equation}
\label{eq:Upluspowern}
\mathcal{U}_{\pm}^n = (\pm 1)^n e^{-i n (\pm  \frac{\pi}{4}+\kappa)}
\begin{pmatrix}
\alpha_n &
\mp \beta_n^* \\
\pm \beta_n &  
\alpha_n^*
\end{pmatrix}, 
\end{equation}   
where,
\begin{subequations}
\begin{align}
\label{eq:alpha3qub}
\alpha_n &= T_n(\chi)+\frac{i}{2}\, U_{n-1}(\chi) \cos 2\kappa \quad\\
\label{eq:beta3qub}
\beta_{n} &= (\sqrt{3}/2)\, U_{n-1}(\chi) \,e^{2i \kappa}.
\end{align}
\end{subequations}
Here the Chebyshev polynomials $T_n(\chi)$ and $U_{n-1}(\chi)$ of the first and second kinds are used and are defined as 
\begin{equation}
T_n(\chi)=\cos(n \theta)\;\; U_{n-1}(\chi)=\sin(n \theta)/\sin \theta,
\end{equation}
with 
$\chi=\cos{\theta}=\sin(2\kappa)/2=\sin(\kappa_0/3)/2$. Hence the matrix elements of the time $n$ propagator are explicitly given 
by polynomials of order $n$ in the variable $\sin(\kappa_0/3)$.
We further solve the four qubit kicked top
where each qubit is coupled to every other qubit by the same strength. Hamiltonian for such a system can be easily obtained from Eq.~(\ref{Eq:QKT}),
by substituting $j=2$. 
The four qubit case is fascinating since it is the smallest system having non-nearest-neighbour interactions. In parallel with the three qubit case, here also we are restricted to the subspace invariant under permutations.
The permutation symmetric subspace basis which is parity adapted is given by
\begin{equation}
\label{eq:4QubBasis}
\begin{split}
|\phi_1^{\pm} \rangle &= \frac{1}{\sqrt{2}} (|W\rangle \mp | \overline{W} \rangle),\\
|\phi_2^{\pm} \rangle &= \frac{1}{\sqrt{2}} (|0000\rangle \pm | 1111 \rangle),\\
|\phi_3^{+} \rangle &= \frac{1}{\sqrt{6}} \sum_{\mathcal{P}}|0011\rangle_{\mathcal{P}}
\end{split}
\end{equation}
where $|W\kt =\frac{1}{2}\sum_{\mathcal{P}}|0001\kt_{\mathcal{P}}$, $|\overline{W}\kt =\frac{1}{2}\sum_{\mathcal{P}}|1110\kt_{\mathcal{P}}$, and $\sum_{\mathcal{P}}$ sums over all possible permutations. 
Distinctly, for the 4-qubit case, $|\phi_1^{+}\kt$ is an eigenstate of $\mathcal{U}$ with eigenvalue $-1$ for {\it all} values of the parameter $\kappa_0$. 
Then the permutationally symmetric subspace breaks up into $1\oplus2\oplus2$ subspaces.
The Floquet unitary becomes easier to analyse when written down in this basis, as it becomes block diagonal. Any algebra is easier to perform in this form
The $n^{th}$ power of $ \mathcal{U}$ is given by

\begin{equation}
\label{eq:4qubitUpown}
\mathcal{U}^n = \begin{pmatrix}
\mathcal{U}_0^n & 0_{1\times 2}& 0_{1\times 2} \\ 0_{2\times 1} &   
\mathcal{U}_{+}^n & 0_{2\times 2} \\ 0_{2\times 1} & 0_{2\times 2} &  \mathcal{U}_{-}^n
\end{pmatrix}, 
\end{equation}
This simplifies the problem significantly. 
Here
\begin{equation}
\mathcal{U}_0=\langle \phi_1^{+} |\mathcal{U}|\phi_1^{+} \rangle = -1,
\end{equation}
 a part of the positive-parity subspace.
Block $\mathcal{U}_{+}$ in the $\{\phi_2^{+},\phi_3^{+}\}$ basis is given by
\begin{equation}
\label{eq12}
\mathcal{U}_{+} = -ie^{-\frac{i \kappa }{2}} \left(
\begin{array}{cc}
\frac{i}{2} e^{-i \kappa} & \frac{\sqrt{3}i}{2}  e^{-i \kappa} \\
\frac{\sqrt{3}i}{2}  e^{i \kappa} & -\frac{i}{2} e^{i \kappa} \\
\end{array}
\right),
\end{equation}
while $\mathcal{U}_{-}$ in the  negative parity basis $\{\phi_1^{-},\phi_2^{-}\}$, is
\begin{equation}
\label{eq12}
\mathcal{U}_{-} = e^{-\frac{3 i \kappa }{4}} \left(
\begin{array}{cc}
0 & e^{\frac{3 i \kappa }{4}} \\
-e^{-\frac{3 i \kappa }{4}} & 0 \\
\end{array}
\right),
\end{equation}
where $\kappa=\kappa_0/2$.

 In a manner similar to the case of 3-qubits above, the time $n$ propagator is now
 written compactly in terms of the Chebyshev polynomials. We have 
\begin{eqnarray}
\label{eq:4QubUppown}
\mathcal{U}_{+}^n &=&  e^{-\frac{i n(\pi+\kappa) }{2}} 
\begin{pmatrix}
\alpha_n & i\beta_n^{*} \\  i\beta_n & \alpha_n^{*}
\end{pmatrix},
\end{eqnarray}
where
\begin{subequations}
\begin{align}
\label{eq:4QubitAlphaBeta}
\alpha_n =&  T_{n}(\chi)+\frac{i}{2}U_{n-1}(\chi)\cos{\kappa} \\
\label{eq:beta4qub}
\beta_n  =&  \frac{\sqrt{3}}{2}U_{n-1}(\chi)e^{i\kappa},
\end{align}
\end{subequations}
with $\chi=\sin{\kappa}/2=\sin(\kappa_0/2)/2$.
The negative parity subspace evolution operator is
\begin{equation}
\label{eq:4QubUmpown}
\mathcal{U}_{-}^n = e^{-\frac{ 3in \kappa }{4}} \left(
\begin{array}{cc}
\cos \frac{n\pi}{2} & e^{\frac{3 i \kappa}{4}} \sin \frac{n\pi}{2} \\
-e^{-\frac{3 i \kappa }{4}} \sin \frac{n\pi}{2}  &  \cos \frac{n\pi}{2} \\
\end{array}
\right).
\end{equation}
Although for simplicity we use the same symbols $\alpha_n$ and $\beta_n$ for the propagator matrix entries in the 3 and 4 qubit
cases, they are not the same. However, in either case, we note the important identity that $|\alpha_n|^2+|\beta_n|^2=1$, following
 from the unitarity of the propagators involved, arises from the Pell equation for the Chebyshev polynomials:
\begin{equation}
\label{eq:Pell}
T_n(x)^2+(1-x^2) U_{n-1}^2=1.
\end{equation}
\section{OTOC and the kicked top}
\label{sec:OTOC}

The out-of-time-ordered correlators (OTOC) are closely connected to the growth 
of the incompatibility of observables due to the dynamics. They are currently being studied in a wide variety of contexts from many-body physics to field theories,
quantum gravity, and black holes in a remarkable coming together of many research 
communities. They are thought of as a way to investigate the ``quantum butterfly effect",
which was also the role and motivation for the introduction of the Loschmidt echo. 
Both of these quantities, in systems with a semiclassical limit, have 
regimes where the classical Lyapunov exponent plays a role: as (half) the rate of the exponential growth of OTOC and 
as the rate of exponential decay of the echo. The Lyapunov exponent may be seen more clearly
in the OTOC as the echo has a rather complex dependence on the perturbation used, however recent 
works have pointed out explicit connections between OTOC in an averaged sense and the echo \citep{yan2020information}.

Let $A(0)$ be some observable and let $A(\tau)=\mathcal{U}^{-\tau} A(0) \mathcal{U}^\tau$ be its Heisenberg time evolution.  
We define the OTOC as
\begin{equation}
\label{eq:OTOCdefn}
C_{\rho}(\tau)=-\frac{1}{2}\mathrm{Tr} \left(\rho\, [A(\tau),A(0)]^2 \right).
\end{equation}
where $\rho$ is some state of the system. In particular we deal with the infinite temperature state
$\rho=I/(2j+1)$ denote the corresponding OTOC as $C_{\infty}(\tau)$. 
The phrase ``out-of-time-ordered" is justified for these quantities as the commutator contains 
terms such as $\br A(\tau) A(0) A(\tau) A(0) \kt$ wherein the operators are not monotonically ordered in time.
OTOC have been used an indicator of information scrambling as some initially localized operator or 
``information" in a  many-body system gets entangled with other one-particle operators on other sites and 
leads to a complex state wherein the initial information is practically lost. For nonintegrable chaotic 
systems, especially with a semiclassical limit, the expected exponential growth of the OTOC 
\begin{equation}
\label{eq:QuantumLE}
C_{\rho}(t)\sim e^{2 \lambda_Q \tau}
\end{equation}
has been observed and the quantum Lyapunov exponent $\lambda_Q$ has been found to be close to the
classical one. The exponential growth is observed till the log-time or the Ehrenfest time which 
scales as $\ln(1/h)/\lambda_{C}$ where $h$ is a scaled Planck's constant and $\lambda_C$ the classical
Lyapunov exponent.

The kicked top has been previously used in OTOC studies such as in \citep{arul_vaibhav,sieberer2019digital} and 
variations of it that break the permutation symmetry are beginning to be studied as well as potential models of 
``holography" \citep{yin2020quantum} as well as from the point of view of experimental realizations via NMR for example.
Previous studies of the kicked top OTOC were in the semiclassical limit of large $j$, wherein only numerical
results are accessible. It is of interest to ask how these properties manifest themselves in the solvable highly quantum
regime of small $j$ (small number of qubits $N$) which are accessible to present day experiments. We are limited by short time scales and the exponential 
growths cannot be clearly observed in these cases. Yet it is intriguing to have exactly solvable cases wherein we may 
see such growth in a rudimentary form and study the transition to semiclassical regimes.
Due to our restriction to the permutation symmetric subspace, it is not 
possible to use a single qubit operator and we take the symmetric subspace projection of the
collective spin variable $A(0)= \sum_{i=1}^{2j}\sigma^z_i/2=J_z$ as the observable. 
\subsection{OTOC in 3 qubits: $j=3/2$}

For $j=3/2$, the $N=3$ qubit case, this restriction takes the form of 
$J_z=(3/2)|000\kt \br 000|-(3/2) |111\kt \br 111| +(1/2) |W\kt \br W|-(1/2) |\overline{W}\kt \br \overline{W}|$.
Using the basis in Eq.~(\ref{eq:3qubit_basis_1},\ref{eq:3qubit_basis}) in  which the time evolution further block-diagonalizes and
noting that $J_z|\phi_1^{\pm}\kt = (3/2) |\phi_1^{\mp}\kt$, $J_z|\phi_2^{\pm}\kt = (1/2) |\phi_2^{\mp}\kt$,
we use
\begin{equation}
\label{eq:Jz}
J_z=\begin{pmatrix}
0_{2 \times 2 } &S\\S&0_{2 \times 2}
\end{pmatrix},
\;\; S=\frac{1}{2} \begin{pmatrix}
3 &0 \\0&1
\end{pmatrix}.
\end{equation}
This leads to 
\begin{equation}
J_z(n)=\mathcal{U}^{-n} J_z \mathcal{U}^n=\begin{pmatrix}
0 & \mathcal{U}_+^{-n} S \mathcal{U}_{-}^n\\
\mathcal{U}_-^{-n} S \mathcal{U}_{+}^n &0
\end{pmatrix}.
\end{equation}
Considering the case of the infinite temperature OTOC $C_{\infty}(n)$, we
separate it as
\begin{equation}
\label{eq:C2minusC4}
C_{\infty}(n)=C_2(n)-C_4(n),
\end{equation}
where $C_2(n)=\mathrm{Tr} [J_z^2(n) J_z(0)^2]/4$ is the two-point correlator and \\$C_4(n)=\mathrm{Tr}[ J_z(n) J_z(0) J_z(n)J_z(0)]/4$
is the four-point correlator which is out-of-time ordered. 
This leads to
\begin{subequations}
\label{eq:3qubitC2C4}
\begin{align}
C_2(n)&=\frac{1}{4}\left[ \mathrm{Tr}(\mathcal{U}_+^{-n} S^2 \mathcal{U}_+^{n} S^2)+\mathrm{Tr}(\mathcal{U}_-^{-n} S^2 \mathcal{U}_-^{n} S^2)\right]\\
C_4(n)&=\frac{1}{4}\left[ \mathrm{Tr}(\mathcal{U}_+^{-n} S \mathcal{U}_-^{n} S)^2+\mathrm{Tr}(\mathcal{U}_-^{-n} S \mathcal{U}_+^{n} S)^2\right].
\end{align}
\end{subequations}
Plugging in the elements of $\mathcal{U}_{\pm}^n$ from Eq.~(\ref{eq:Upluspowern}) and simplifications lead to 
\begin{equation}
\label{eq:Cinfty3qub}
\begin{split}
C_2(n)&=\frac{1}{16}\left( 41-32 |\beta_n|^2 \right)\\
C_4(n)&=(-1)^n \frac{1}{16} \left(41 -160 |\beta_n|^2+128 |\beta_n|^4 \right),
\end{split}
\end{equation}
where $\beta_n$ is given by Eq.~(\ref{eq:beta3qub}), and hence 
\begin{equation}
|\beta_n|^2=\frac{3}{4} U_{n-1}^2\left[ \frac{1}{2}\sin\left(\frac{\kappa_0}{3}\right) \right].
\end{equation} 
For small $\kappa_0$ when the dynamics is near-integrable these give
\begin{equation}
C_{\infty}(n)\approx \left\{ \begin{split}\frac{1}{6} n^2 \kappa_0^2 -\frac{13}{2592} n^4 \kappa_0^4 &\;\; n \; \text{even}\\
\frac{5}{8}+\frac{1}{288} (n^2-1)^2 \kappa_0^4 & \;\; n \; \text{odd} \end{split} \right.
\end{equation}
This shows a marked odd-even behaviour with the even time OTOC increasingly quadratically with time
at the lowest order. The odd-even effect is quite easily understood as for very small $\kappa_0$ the dynamics is essentially
one of rotation about the $y$ axis by $\pi/2$ and hence the $J_z$ operator with a concentration in the
$z$ direction is rotated practically to its negative at times $2 \,\text{mod}\, 4$ and to itself at times $0\, \text{mod}\, 4$ 
and hence almost commutes, but at times $1\, \text{mod}\, 4$ or $3 \, \text{mod}\, 4$ is 
concentrated on the $y$ and $-y$ directions and maximally fails to commute. Indeed the constant term $5/8$ is nothing but 
$-\mathrm{Tr}[J_y,J_z]^2/4=\mathrm{Tr} J_x^2/4$. A quadratic growth has also been observed in the Hadamard quantum walk \citep{SivaAL-2019} and we may expect 
a general power-law growth of the OTOC to be a general integrable and 
near-integrable feature \citep{Rozenbaum17,Prakash-2019} that we see in this small and solvable system exactly.

Now we turn attention to fixed and small times but for arbitrary values of the parameter $\kappa_0$. It follows from Eq.~(\ref{eq:C2minusC4}) and Eq.~(\ref{eq:Cinfty3qub}) that $C_{\infty}(1)=5/8$ irrespective of the value of $\kappa_0$, as $U_0(x)=1$. This shows no interesting dynamical
behaviour and the OTOC have a diffusive time scale over which the properties depend on the observable chosen as well. The next time steps 
already are of interest:
\begin{equation}
\begin{split}
C_{\infty}(2)&=6 \sin^2(\kappa_0/3)\left(1-\frac{3}{4} \sin^2(\kappa_0/3)\right)\\
C_{\infty}(3)&=\frac{5}{8}+18 \sin^4(\kappa_0/3)\left(1-\frac{1}{2}\sin^2(\kappa_0/3)\right)^2,
\end{split}
\end{equation}
$C_{\infty}(n)$ being a polynomial of order $4(n-1)$  in $\chi=\sin(\kappa_0/3)/2$.
\begin{figure}[h]
	\centering
	\includegraphics[width=0.7\linewidth]{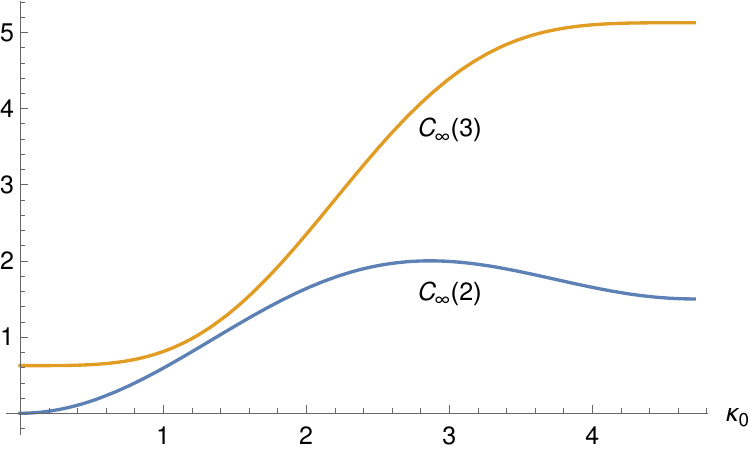}
	\caption{The OTOC for the 3 qubit kicked top at times $2$ and $3$ as a function of the chaos parameter $\kappa_0$. In all figures the observable used is $J_z$. Note the difference in the behavior 
	around $\kappa_0=0$, the near-integrable regime and also that the increase is monotonic at time $3$, and reaches a maximum at $\kappa_0=3 \pi/2$, when the top is essentially already fully chaotic.}
	\label{fig:OTOC2and3}
\end{figure}
The curves for $C_{\infty}(2)$ and $C_{\infty}(3)$ are shown in Fig.~\ref{fig:OTOC2and3} for convenience and 
we see that they increase with $\kappa_0$  and $C_{\infty}(3)$ is monotonically increasing over the entire range
of interest $\kappa_0 \in [0, 3 \pi/2]$, reaching the maximum value at $\kappa_0=3 \pi/2$.
\begin{figure}[h]
	\centering
	\includegraphics[width=0.5\linewidth]{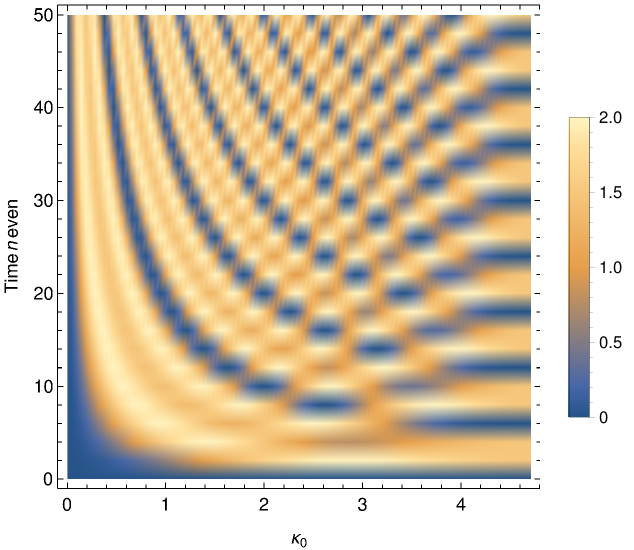}
	\includegraphics[width=0.5\linewidth]{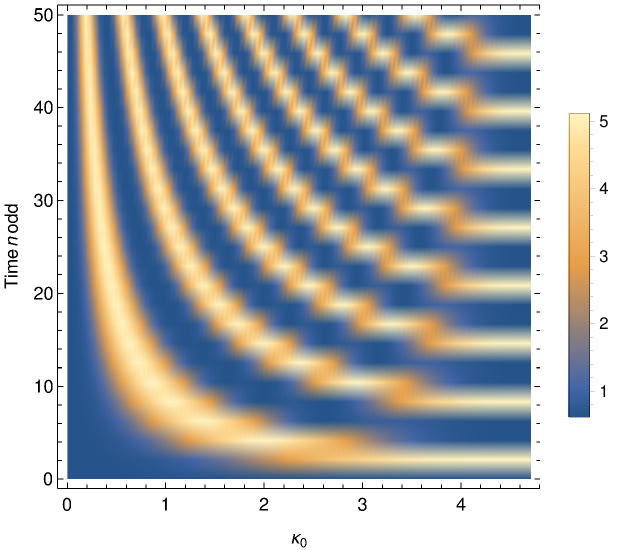}
	\caption{A density plot of the OTOC as function of time and the parameter $\kappa_0$ for 3 qubits, it is separated for 
	even and odd times for reasons explained in the text.}
	\label{fig:OTOC3evenodd}
\end{figure}
\begin{figure}[h]
	\centering
	\includegraphics[width=0.9\linewidth]{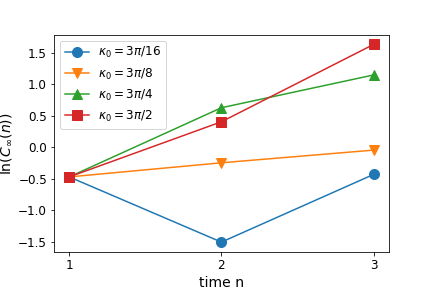}
	\caption{The OTOC is shown in linear-log scale for a few values of $\kappa_0$ for the 3 qubit kicked-top. The dynamics is predominantly chaotic at $\kappa_0=3 \pi/2$ and
	is reflected in what appears to be a near linear OTOC growth on a lin-log plot implying exponential growth.}
	\label{fig:OTOCLogLin}
\end{figure}
A more global view is provided in Fig.~\ref{fig:OTOC3evenodd} where the OTOC for $j=3/2$ shown 
as a function of the time, split into even and odd ones, and the parameter $\kappa_0$. There would
be a periodicity beyond the value of $\kappa_0=3 \pi/2$, which provides an interesting boundary.
Exactly at this point, the classical dynamics is fairly chaotic and we do see a sharp increase in the OTOC
values for short times even in this small $j$ value. 

To give an indication of the growth, $\mathrm{log}[C_{\infty}(n)]$ is 
plotted in Fig.~\ref{fig:OTOCLogLin} for $1\leq n \leq 3$. This has just three points, but the trend is clear and 
we may even interpret this as signs of the exponential growth of the OTOC that one expects in chaotic systems.
\begin{figure}[h]
	\centering
	\includegraphics[width=0.7\linewidth]{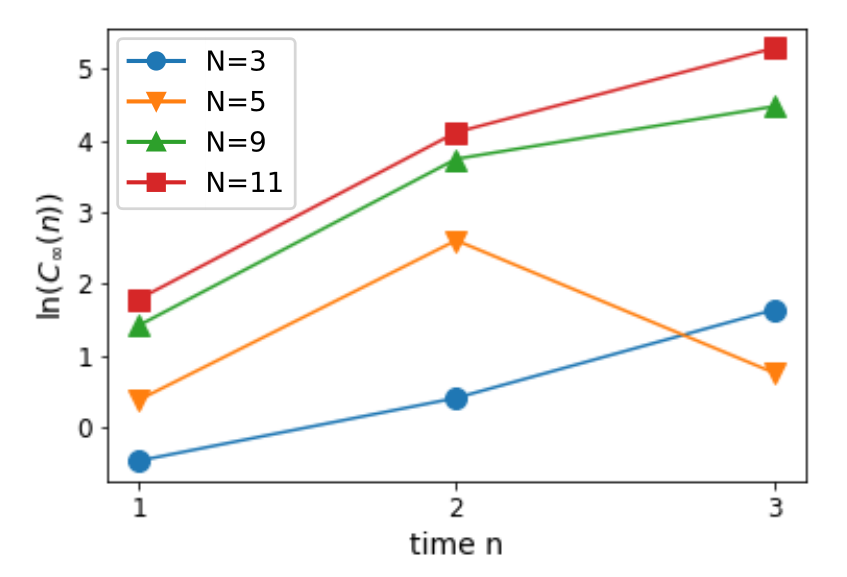}
	\caption{The OTOC in the linear-log scale, when $\kappa_0=3 \pi/2$ and the $N$, the number of qubits,  is increased. The slope at $N=5$ already
	is well-saturated to those corresponding to larger $N$ values.}
	\label{fig:OTOC3qubkappa3piby2}
\end{figure}
To compare this with higher values of $j$, we show in Fig.~\ref{fig:OTOC3qubkappa3piby2} the case
for some larger values of the spin $j$, but with $\kappa_0=3 \pi/2$ in all the cases. We do see an increase and saturation 
in the slope with increasing $j$ values. It is interesting to observe from the same figure that with $j=5/2$ ($N =5$ qubit kicked top) the OTOC slope has already 
saturated and hence at this value of the parameter, while $j=3/2$ is too low, $j=5/2$ may be just enough.
To explore this further we turn to the other solvable case of $j=2$ and compare it with
higher values of $j$, as well as study the peculiar case of $\kappa_0=\pi j$ for arbitrary $j$.
\subsection{OTOC in 4 qubits, $j=2$, and the peculiar case of $\kappa_0=\pi j$ for arbitrary $j$.}

The $4$ qubit case we reiterate can be qualitatively different from the case of $3$ as it has next-nearest neighbor interactions and is
a rudimentary non-integrable model. The calculations do not pose a serious problem as the unitary time evolution is still block-diagonalized
into utmost $2-$dimensional spaces, see Eq.~(\ref{eq:4qubitUpown}). the equations get a little bit more involved, but nevertheless can be
exactly solved, especially with the help of computer algebra. Skipping the details, we present the final results again separating the 
cases of different time parities. For time $n$ even we get
\begin{equation}
C_{\infty}(n)=\frac{1}{5}[ 34 -16\, |\beta_n|^2 -32 \,\text{Re} \left( \alpha_n^2 e^{in \kappa_0/4}\right)  
 -2 \cos(3 n \kappa_0/4) ],
\end{equation}
and for odd $n$, 
\begin{equation}
C_{\infty}(n)=\frac{1}{5}[ 25 -16\, |\beta_n|^2 -16 (-1)^{(n-1)/2} \, \text{Im}\left(\alpha_n e^{in \kappa_0/2}\right)].
\end{equation}
Here the $\alpha_n$ and $\beta_n$ involve the Chebyshev polynomials and are from Eq.~(\ref{eq:4QubitAlphaBeta}).
It follows that $C_{\infty}(1)=1$ irrespective of $\kappa_0$. 
Expressions for short times maybe explicitly extracted and for $n=2,3$ are
\begin{equation*}
\begin{split}
C_{\infty}(2)=&\frac{1}{5}\left(28-30\cos(\kappa_0/2)+6 \cos(\kappa_0)-4 \cos(3 \kappa_0/2)\right)\\
C_{\infty}(3)=&\frac{1}{10}\left(37-36 \cos(\kappa_0) +9 \cos(2 \kappa_0)\right).
\end{split}
\end{equation*}
While $C_{\infty}(2)$ is a monotonically increasing function for $0\leq \kappa_0 \leq 2 \pi$ and is a maximum at $\kappa_0=2\pi$,
$C_{\infty}(3)$ vanishes at this point having a maximum at $\kappa_0=\pi$. These special values of $\kappa_0$ correspond to $\pi j$ and $\pi j/2$. Notice that for $j=3/2$, $C(3)$ was a maximum at $\kappa_0=\pi j$ (see Fig.~\ref{fig:OTOC2and3}), this difference between half-integer angular momenta and integer ones persists, and such features have also been noticed in entanglement before \citep{dogra19}. 

\begin{figure}[h]
	\centering
	\includegraphics[width=0.7\linewidth]{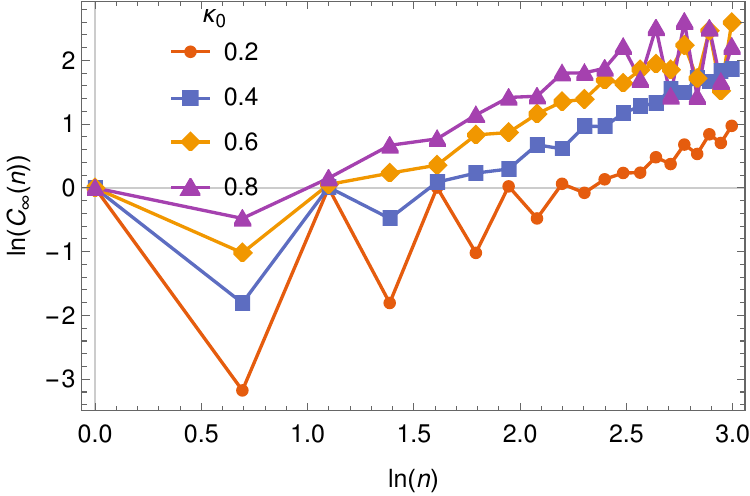}
	\caption{The 4 qubit OTOC growth in log-log scale for values of $\kappa_0$ when the dynamics is near-integrable. The growths
	are consistent with power-laws, taking into account the odd-even features in time.}
	\label{fig:OTOC4Qub_smallk}
\end{figure}
\begin{figure}[h]
	\centering
	\includegraphics[width=0.7\linewidth]{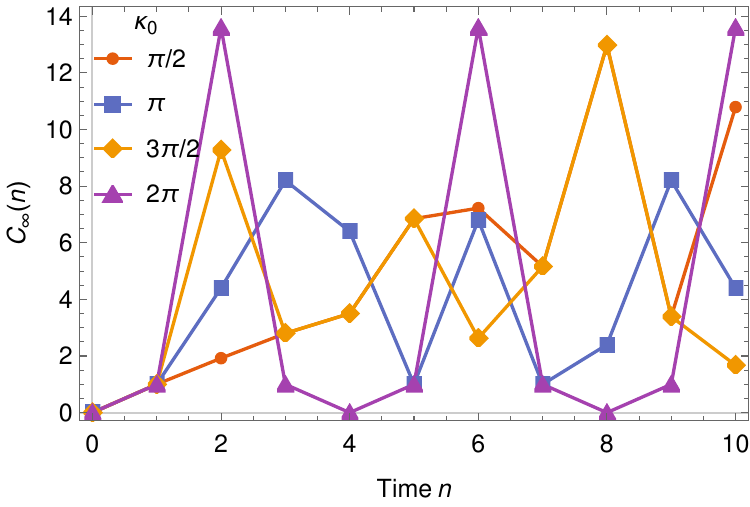}
	\caption{The 4 qubit OTOC for larger values of $\kappa_0$, the large growth at $\kappa_0=2 \pi$ is to be noted along with its periodicity.}
	\label{fig:OTOC4Qub_largek}
\end{figure}

For relatively small values of $\kappa_0$, when the classical system is near-integrable there is modest 
OTOC growth mostly governed by power laws as shown in Fig.~\ref{fig:OTOC4Qub_smallk}. At large values of $\kappa_0$, the
OTOC grows rapidly, as seen in Fig.~\ref{fig:OTOC4Qub_largek} and then oscillates in an apparently irregular manner. 
\begin{figure}[h]
	\centering
	\includegraphics[width=0.7\linewidth]{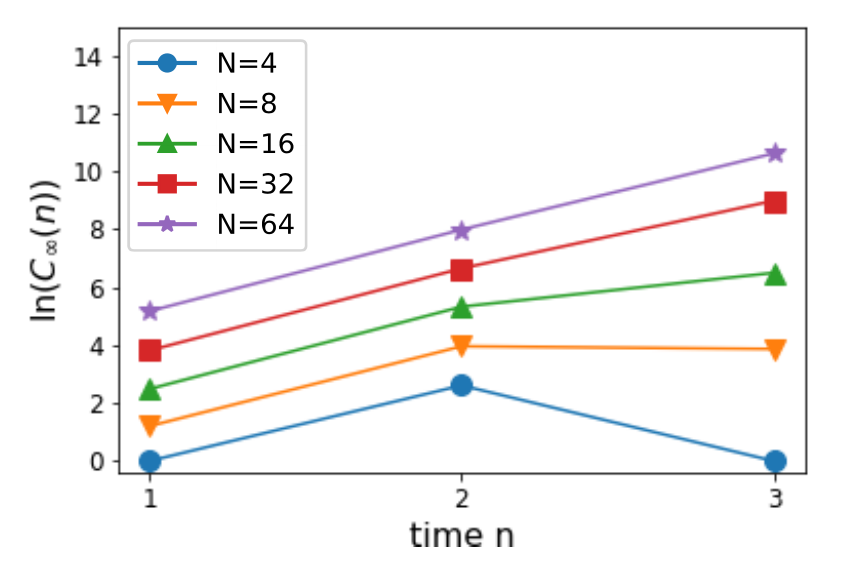}
	\caption{The $N = 4$, 8 qubit OTOC at $\kappa_0=2 \pi$ compared with that of larger number of qubits, showing how the initial growth spurt is already reflecting the semiclassical Lyapunov exponent.}
	\label{fig:OTOC4Qub_pis}
\end{figure}
Of special interest again is $\kappa_0=2 \pi$, beyond which there is a symmetric behavior equivalent to a smaller value of $\kappa_0$ and hence certainly not reflecting any semi-classical property. For this case, it is amusing that the initial growth between $C_{\infty}(1)=1$ and $C_{\infty}(2)=68/5$, which is all that is there, in the sense that there is time-symmetry and periodicity beyond, already reflects the large $j$ growth of OTOC at $\kappa_0=2 \pi$.
The classical dynamics is highly chaotic at this parameter value and we may expect purely exponential growth of the OTOC. This is shown in Fig.~\ref{fig:OTOC4Qub_pis}, where we only plot the first 3 time steps. Using the first 2 steps of the case $j=2$, we may be bold enough to 
find the quantum Lyapunov exponent of Eq.~(\ref{eq:QuantumLE}) as $0.5 \log(68/5) \sim 1.3$ and compare with the classical value of $\lambda_{C}=\log(\kappa_0)-1\sim 0.84$. We note of course that the classical exponent comes from an infinite time average and the kicked top, unlike the baker's or the cat map, is not a uniformly hyperbolic system. Thus it can hardly be expected that finite-time quantum properties from a particular observable reflect this
number exactly and we see that even for large $j$ the slope is not significantly changed towards the classical value. Thus it seems plausible that with only 4 qubits one can observe the exponential growth of the OTOC due to quantum chaos.

As the extreme case of $\kappa_0=\pi j$ registers the largest growth of the OTOC for the 3 and 4 qubit systems studied above, it is natural to
investigate this for an arbitrary value of $j$. In this case the Floquet unitary operator
\begin{equation}
\mathcal{U}=e^{-i \pi J_z^2/2}e^{-i \pi J_y/2}
\end{equation}
enjoys many special properties, that we intend to investigate in detail elsewhere. For integer $j$ values it is a {\it sum} of 4
pure rotations and in general, for integer $j$, we note that when $\kappa_0=\pi r/s$ where $r$ and $s$ are
relatively prime integers, 
\begin{align}
\mathcal{U}_{r,s}=&e^{-i r \pi J_z^2/2s}e^{-i \pi J_y/2}\\
&=\sum_{l=0}^{2s-1}a_l(r,s) e^{-i \pi l J_z/s}e^{-i \pi J_y/2}
\end{align}
where 
\begin{equation}
a_l(r,s)=\frac{1}{2s}\sum_{m=0}^{2s-1}e^{-i \pi m l/s}e^{-i \pi r m^2/s}
\end{equation}
are Gauss sums. A similar sum over $4s$ terms applies for half-integer $j$ values. We record them as possible
routes to implementing the kicked top experimentally when $\kappa_0$ is some rational multiple of $\pi$, as the
torsion is replaced by a sum of rotations. For the case of $j=2$, or $r=1$, $s=2$, we note that $\mathcal{U}^8=I$, where $I$ is identity. These maps remind one of the cat maps, whose quantum mechanics is exactly periodic.

For large value of $j$ we notice that the quantum-classical correspondence time, the Ehrenfest or log-time is $\sim \log(2j+1)/\lambda_C=\log(2j+1)/\log(\pi j) \sim 1$. Thus we are at the true border of the correspondence and do not  expect to see classical effects for times beyond a few steps, however large $j$ may be, and indeed we find that only $n=1,2$ are unique and of interest. We find remarkably simple expressions for these:
\begin{equation}
\label{eq:PIJC}
\begin{split}
C_{\infty}(1)&=\frac{1}{6}j(j+1)\\
C_{\infty}(2)&=\frac{2}{15}j(j+1)(3 j^2+3j-1),
\end{split}
\end{equation}
they being related to squares and $4^{\text{th}}$ powers of integers. It reassuringly returns $1$ and $68/5$ for the case $j=2$ which we have discussed above.
This results in the quantum Lyapunov exponent of $\log(C_{\infty}(2)/C_{\infty}(1))\sim \log (j)+0.3$ which is to be compared with the classical one 
$\log(\pi j)-1\sim \log(j)+0.14$. Thus the principal growth of the two Lyapunov exponents are identical and we emphasize that this is in itself
quite a remarkable fact. Thus while this extreme case is highly special it does reflect the large classical chaos that underlies the system. Analysis for $\kappa_0$ other fractions of $\pi j$ are therefore of interest.
\section{Loschmidt echo and the kicked top}
\label{sec:Echo}

Loschmidt echo, as discussed above, is a quantifier of quantum chaos based on the overlap of a given state with itself when evolved by a perturbed and an exact Hamiltonian. In general, Loschmidt echo depends on the state undergoing evolution,
 nature and magnitude of perturbation, and degree of chaos.  
 To make the echo state independent, one can look at the decay by considering an average over initial states from Haar measure for finite dimensional systems, $\overline{F_d}(\kappa_0,\kappa_0',n)=\int d\ket{\psi_0} F_d(\kappa_0,\kappa_0',n,\ket{\psi_0})  $ and \citep{garcia2016lyapunov, zanardi2004purity} 
 \begin{equation}
\overline{F_d}(\kappa_0,\kappa_0',n)=\frac{1}{d(d+1)}(d+ \lvert\mathrm{Tr}[\mathcal{U}^{-n}(\kappa_0) \mathcal{U}^{n}(\kappa_0')]\rvert^2)
 \end{equation}
 where $d$ is the dimension of the Hilbert space of the states. Essentially, the echo depends on the quantity $\lvert\mathrm{Tr}[\mathcal{U}^{-n}(\kappa_0) \mathcal{U}^{n}(\kappa_0')]\rvert^2 $, which can be calculated easily to obtain, for the three qubit kicked top,
 \begin{equation}
 \overline{F_3}(\kappa_0,\kappa_0',n)=\frac{1}{5}(1+\lvert\alpha_n \tilde{\alpha^*_n}+\beta_n \tilde{\beta^*_n}+\beta^*_n \tilde{\beta_n}+\alpha^*_n \tilde{\alpha_n}\rvert^2)
 \end{equation}
 where $\tilde{\alpha_n}$ and $\tilde{\beta_n}$ are $\alpha_n(\kappa_0')$ and $\beta_n(\kappa_0')$ respectively and $\kappa_0'=\kappa_0+ \delta \kappa_0$. Here,  $\delta\kappa_0$ is the strength of perturbation.
 For the four qubit top, this gives 
 
 	\begin{multline}
 	\overline{F_4}(\kappa_0,\kappa_0',n)=\frac{1}{30}\Big(5+\vert1+e^{in\delta \kappa_0/4}[\alpha_n \tilde{\alpha^*_n}+\beta_n \tilde{\beta^*_n}+\beta^*_n \tilde{\beta_n}+ \\ \alpha^*_n \tilde{\alpha_n} +2e^{3in\delta \kappa_0/{8}}(\cos^2(n\pi/2)+\sin^2 (n\pi/2)\cos(3\delta\kappa_0/8)]\vert^2 \Big)
 	\end{multline}
 
  Therefore, we have the exact expressions for the Loschmidt echo for the cases at hand and explore. 
 
 \begin{figure}[!htbp]
 	\centering
 	\includegraphics[width=0.7\linewidth]{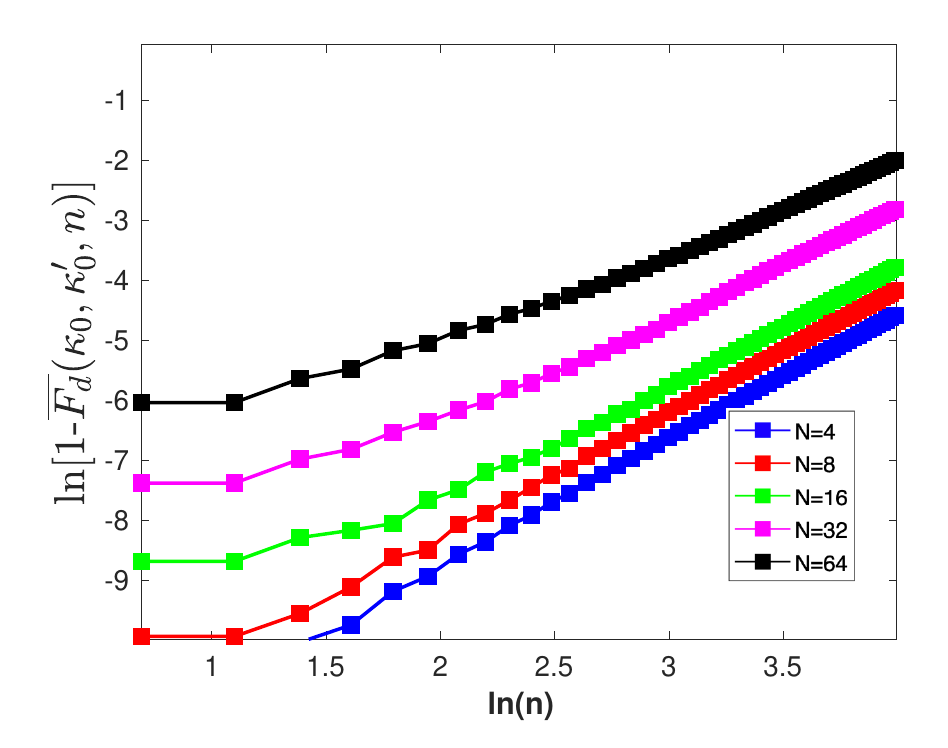}
 	\caption{The quadratic fall for the Loschmidt echo is shown with on a  log-log scale, when $\kappa_0= 2\pi$ for a few $N$ values including $N=4$. The perturbation strength is 0.01.}
 	\label{LE3}
 \end{figure}

  \begin{figure}[!htbp]
 	\centering
 	\includegraphics[width=0.7\linewidth]{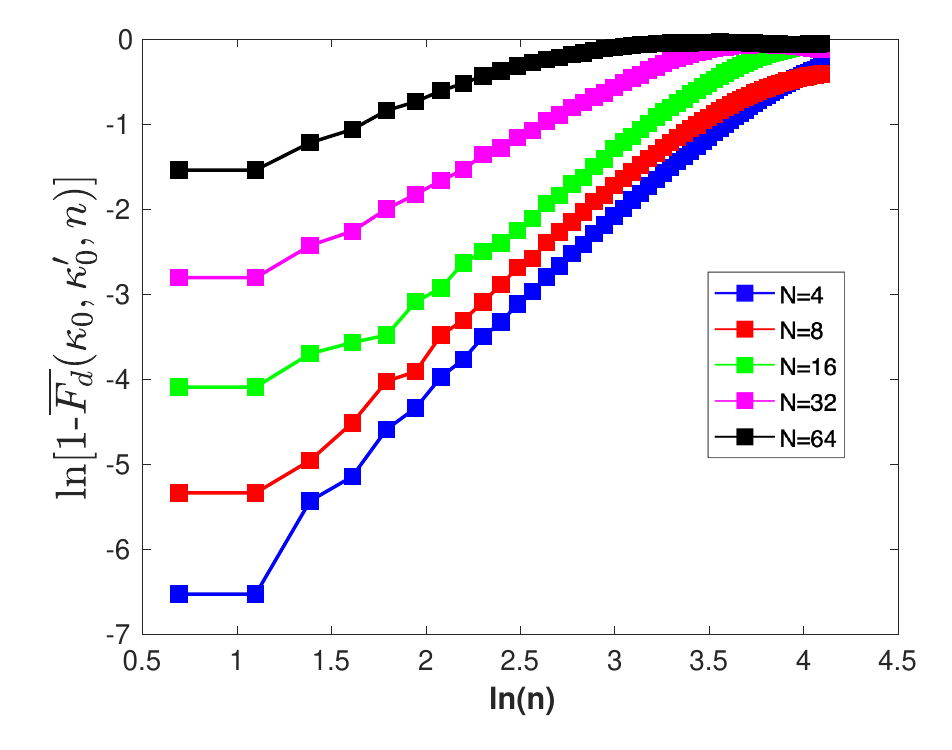}
 	\caption{The breakdown of quadratic fall for the Loschmidt echo is shown with on a  log-log scale, when $\kappa_0= 2\pi$ for a few $N$ values including $N=4$. The perturbation strength is 0.1.}
 	\label{LE4}
 \end{figure}

  \begin{figure}[!htbp]
 	\centering
 	\includegraphics[width=0.7\linewidth]{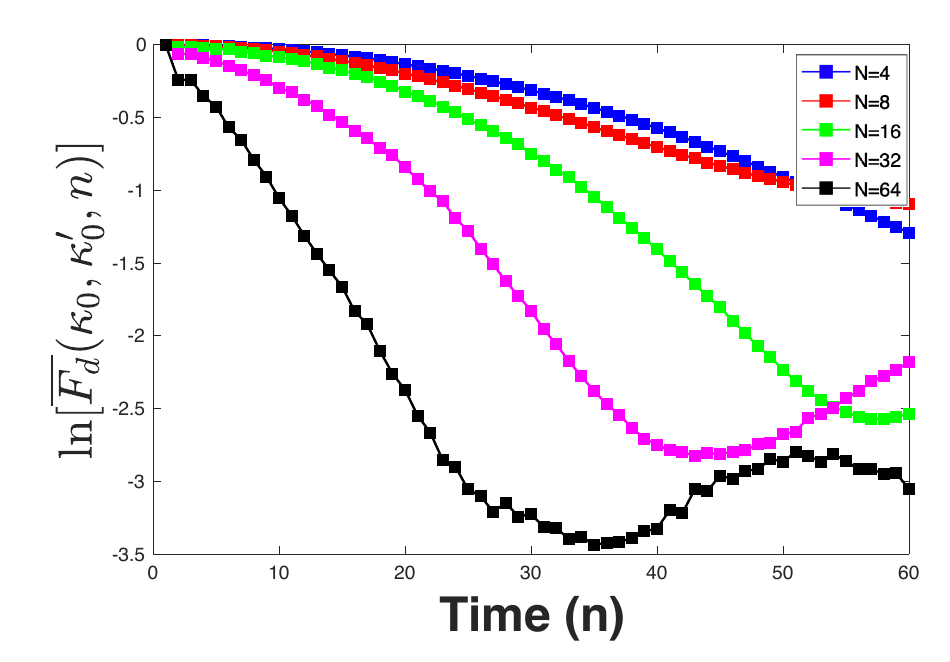}
 	\caption{Loschmidt decay on a linear log scale for some values of $N$. Perturbation strength, $\delta \kappa_0$, is 0.1 and $\kappa _0= 2\pi$. It can be seen that $N=64$ case is showing exponential decay - a forerunner of the Lyapunov decay. }
 			\label{LE5}
 \end{figure}

  \begin{figure}[!htbp]
 	\centering
 	\includegraphics[width=0.7\linewidth]{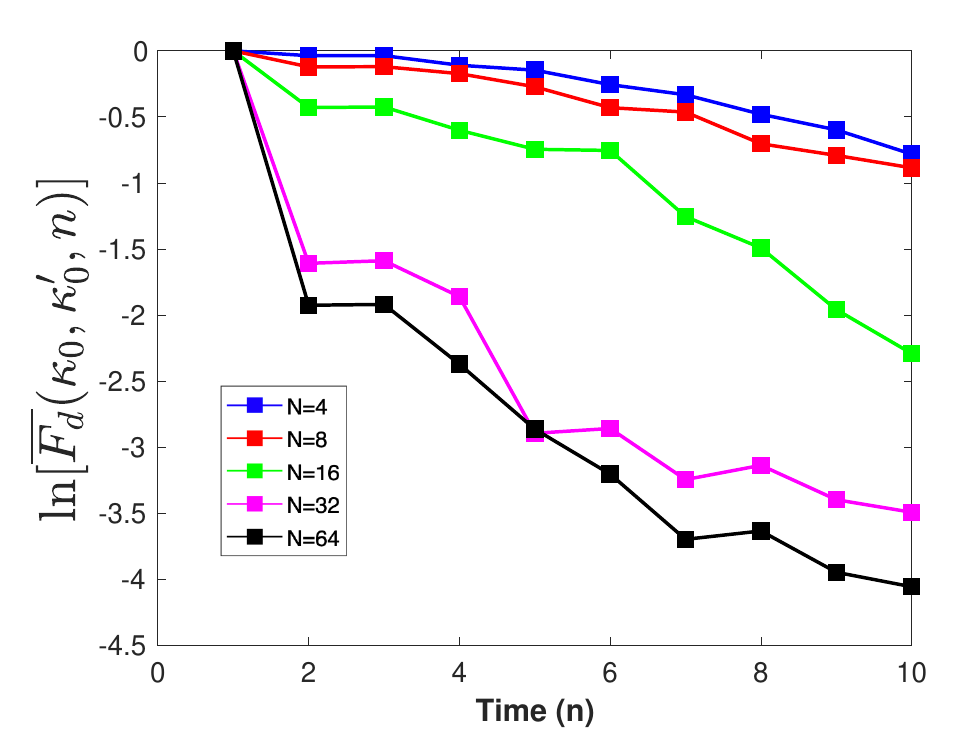}
 	\caption{Loschmidt decay on a linear log scale for some values of $N$. Perturbation strength, $\delta \kappa_0$, is $0.5$ and $\kappa_0 = 2\pi$. It can be seen that $N=32$ and  $N=64$ case is showing exponential decay - a forerunner of the Lyapunov decay. To extract the Lyapunov exponent, one needs to go for a much larger $N$.}
 	\label{LE6}
 \end{figure}
 
 \begin{figure}[!htbp]
 	\centering
 	\includegraphics[width=0.7\linewidth]{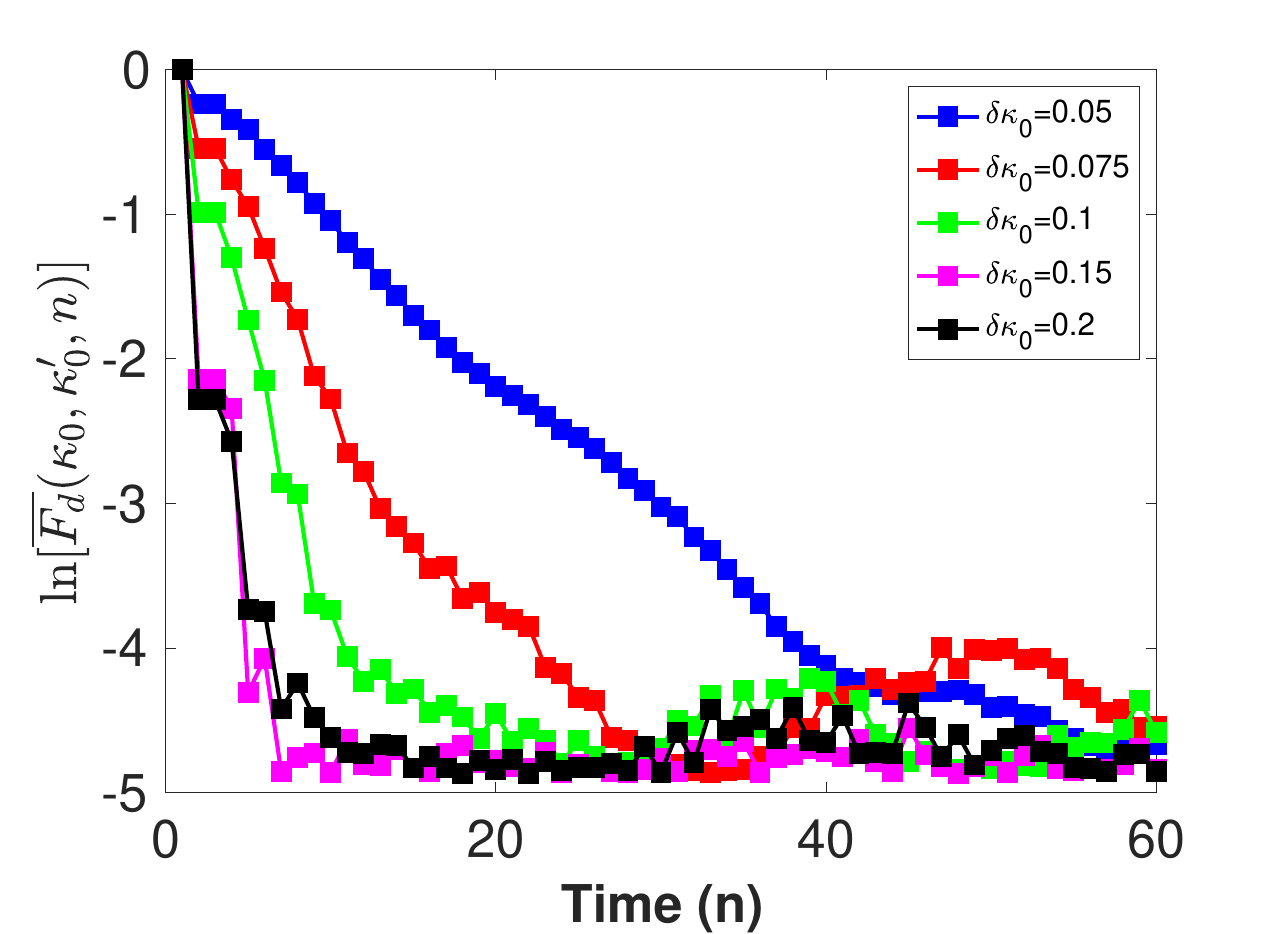}
 	\caption{Loschmidt decay on a linear log scale for some values of $\delta \kappa_0$ the perturbation strength.  $\kappa_0 = 2\pi$ and $N=128$. Increasing the perturbation strength results in saturation of the rate of exponential decay.}
 	\label{LE7}
 \end{figure}

 Figure~\ref{LE3} shows that when the perturbation strength is low (order of $10^{-2}$), we see a Gaussian (quadratic in log-log plot) decay for the 3 and 4 qubit kicked top respectively. 
 Once the size of perturbation is increased, we see a departure from the quadratic decay as is shown in Fig.~\ref{LE4}. 
 However, keeping the same perturbation strength, one observes an exponential decay in the echo as one increases the spin size for the kicked top as shown in Fig.~(\ref{LE5}). Value of $j=32$ starts showing an exponential decay as evident on a log linear scale. As one increases the perturbation strength, as in Fig.~\ref{LE6}, one sees an exponential decay for $j=16$ and $j=32$ - a forerunner for the exponential decay
 that is the hallmark of Loschmidt decay in quantum chaotic systems. For large dimensional chaotic systems, as one increases the perturbation strength, there is a transisition from quadratic to exponential decay that saturates at the value given by the classical Lyapunov exponents \citep{garcia2016lyapunov}. We do see an antecedent of this decay in Fig.~\ref{LE7}  as on increasing perturbation strength, the decay rate saturates 
 to a fixed value.  Though we are still far from the semiclassical  quantum regime of large $j$, these numerical results serve as a precursor of Lyapunov decay for higher dimensional quantized chaotic Hamiltonians.

\subsection{Fidelity decay for states}
\par
In this section, by considering the example of the 3 qubit kicked top system, we demonstrate how classical phase space features have an influence on the Loschmidt echo.  Analysis for four qubit states follows analogously.
The three-qubit  permutation symmetric 
initial states are coherent states situated at
\begin{eqnarray}
X_{0}&=&\sin\theta_0 \cos\phi_0, \nonumber \\ 
Y_{0}&=&\sin \theta_0 \sin\phi_0, \nonumber \\ 
Z_{0}&=&\cos \theta_0,
\end{eqnarray}
on the phase space, written as~\citep{Glauber,Puri},
\begin{equation}
\ket{\psi_0}=|\theta_0,\phi_0\kt = \otimes^{2j} 
(\cos(\theta_0/2) |0\kt + e^{-i \phi_0} \sin(\theta_0/2) |1\kt).
\end{equation}

We  study time evolution and fidelity decay of two completely different three-qubit states
 ((i) $|0,0\kt$ and (ii) $|\pi/2, -\pi/2\kt$), shown in Fig.~\ref{fig:classical}.
The
coherent state at $|0,0\kt$ for three qubits is  $\otimes^3|0\kt.$ It is on a period-4 orbit in the classical phase space and is represented  with a red square
in Fig.~\ref{fig:classical}.  $\otimes^3|+\kt_y$ corresponds 
to the coherent state at $|\pi/2,-\pi/2\kt$,
which is a fixed point as per regular classical phase space 
structure, and eventually becomes unstable as we move 
from regular to mixed phase space, shown by a red circle in Fig.~\ref{fig:classical}.
Let us consider the state  $\otimes^3|0\kt$. Its evolution is described by
\begin{equation}
\label{eq29}
\begin{split}
&\mathcal{U}^{n} |000\rangle \equiv |\psi_n \kt =\frac{1}{2}e^{-i n \left(\frac{3 \pi}{4}+\kappa\right)}\left\lbrace (1+i^n) \left(
\alpha_n |000\rangle  \right.  \right. \\ & \left. \left. + i \beta_n |\overline{W} \rangle
\right) + (1-i^n) \left( i \alpha_n |111\rangle - \beta_n |W \rangle
\right) \right\rbrace. 
\end{split} 
\end{equation}

Loschmidt decay can be computed by looking at the overlap of this state with another, evolved with a unitary of slightly different chaoticity parameter $\kappa_0'.$ 

\begin{align*}
F_3(\kappa_0,\kappa_0',n,\ket{\psi_0})&=|\bra{000} \mathcal{U}^{-n} (\kappa_0)\mathcal{U}^{n}(\kappa_0')  \ket{000}|^2\\
&= |\alpha_n^* \tilde{\alpha_n}+\beta_n^*\tilde{\beta_n} |^2 \numberthis
\end{align*}

Expansion in powers of $\delta \kappa_0$ at $\kappa_0=3\pi/2$ yields the quadratic term as the leading term that is non-zero for 
$n=4$ and beyond. This explains the extremely slow fall in fidelity for this state at $\kappa_0 = 3\pi/2$.  In contrast, the quadratic term in the expansion of  decay for $\kappa_0 = 0$  becomes non-zero starting with $n=1$.

\begin{figure}[!htbp]
	\centering
	\includegraphics[width=0.6\linewidth]{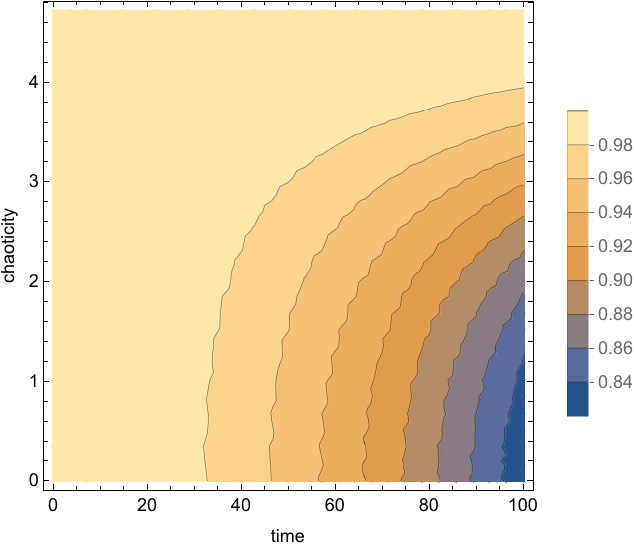}
	\caption{Loschmidt decay for $\ket{000}$ with respect to the chaoticity parameter $\kappa_0 \in [0, 3 \pi/2]$ and time $n$.  Perturbation strength is fixed at $0.005.$}
	\label{LE000}
\end{figure}
Figure~\ref{LE000} interestingly shows somewhat counter-intuitive behavior of the decay of Loschmidt echo 
with chaos. At first, it appears, more chaos leads to less echo decay for a coherent wave packet starting at 
$|0,0\kt$. However, the state $|0,0\kt$ is on a period 4 orbit and will rapidly become delocalized with support over the period 4 phase space points. Fidelity decay for delocalized states having a high participation ratio is in general inversely correlated with the degree of chaos \citep{gorin2006dynamics}. 
As a contrast, consider the three-qubit state, $|\psi_0\rangle=|+++\rangle$, corresponding to a fixed point of the map, where
$|+\rangle=\frac{1}{\sqrt{2}}(|0\rangle+i|1\rangle)$
is an eigenvector of $\sigma_y$ with eigenvalue $+1$. This state delocalizes when the fixed point loses stabiity and the echo decay increases with the increase of chaos ($\kappa_0 \in [0, 3 \pi/2]$) in the system. 
When the initial state is $\otimes^3|+\kt_y=(|\phi_1^+\kt +\sqrt{3} i |\phi_2^+\kt)/2$, the evolution is confined to the positive parity region.
We have, $\mathcal{U}^n|+++\kt_y$ equal to
\begin{align*}
|\psi_n\kt 
=\frac{1}{2} e^{-i n \left(\frac{\pi}{4}+\kappa\right)}\big[ (\alpha_n-i\sqrt{3}\beta_n^{*}) |\phi_1^{+} \rangle+(\beta_n+i\sqrt{3}\alpha_n^{*})|\phi_2^{+} \rangle \big].
\label{eq:fixedptstate}
\end{align*}
Defining $\gamma_n=(\alpha_n-i\sqrt{3}\beta_n^{*})/2$ and $\delta_n= (\beta_n+i\sqrt{3}\alpha_n^{*})/2$,
we can obtain the fidelity decay expression at time $n$ as before.

\begin{align}
F(\kappa_0,\kappa_0',n, \ket{\psi_0})&=|\bra{+++} \mathcal{U}^{-n} (\kappa_0)\mathcal{U}^{n}(\kappa_0')  \ket{+++}|^2\\
&= |\gamma_n^*\tilde{\gamma_n} +\delta_n^*\tilde{\delta_n} |^2 \numberthis
\end{align}

where $\tilde{\gamma_n}$ and  $\tilde{\delta_n}$ are $\gamma_n(\kappa'_0)$ and $\delta_n(\kappa_0')$ respectively.

\begin{figure}[!htbp]
	\centering
	\includegraphics[width=0.6\linewidth]{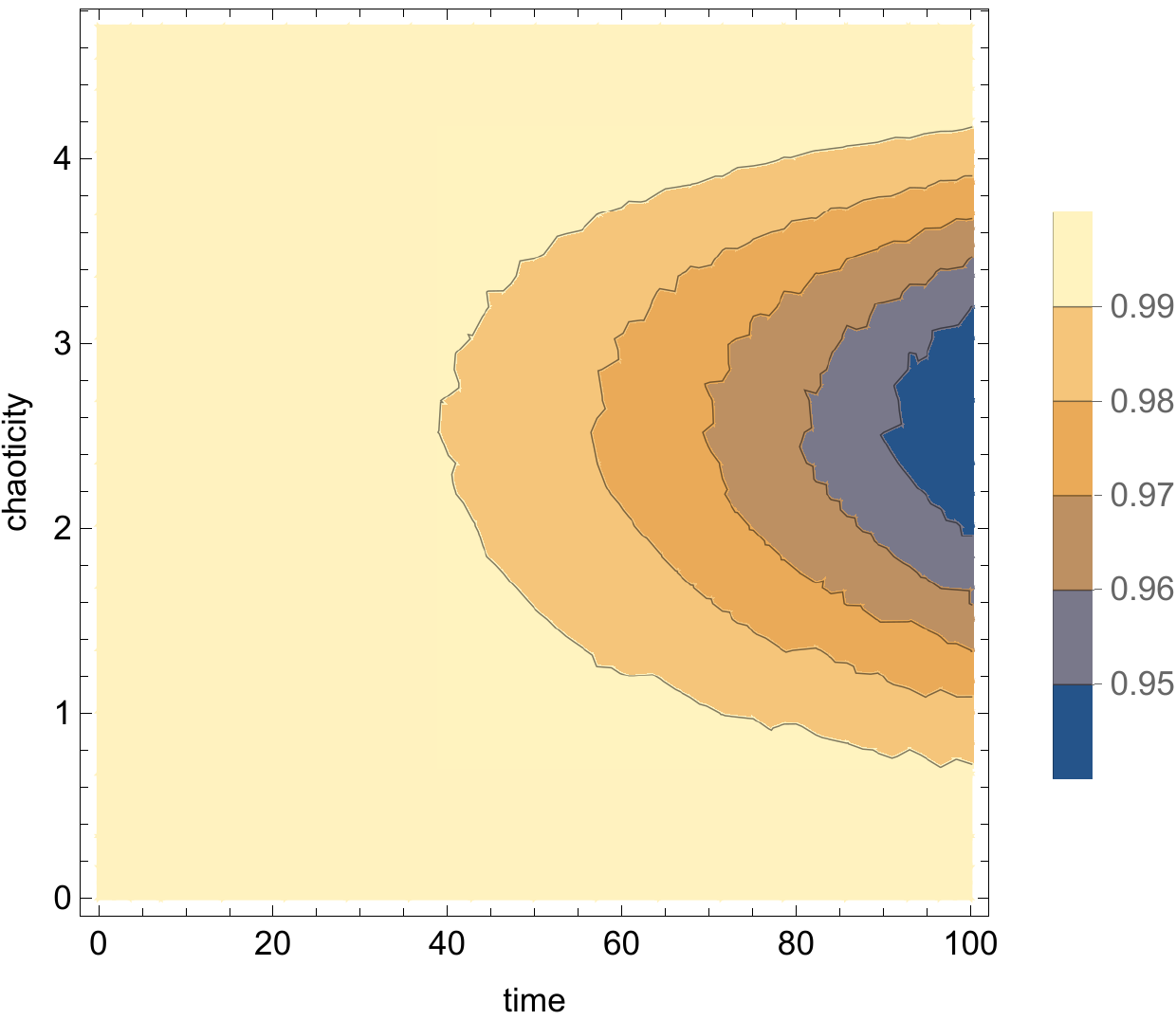}
	\caption{Loschmidt decay for $\ket{+++}$ with respect to the chaoticity parameter $\kappa_0 \in [0, 3 \pi/2]$ and time $n$.  Perturbation strength is fixed at $0.005.$}
\end{figure}
Decay of Loschmidt echo for the $\ket{+++}$ for small perturbations follows the quadratic decay and also
increases with the increase of chaos ($\kappa_0 \in [0, 3 \pi/2]$) in the system. 
\section{Summary and Discussion}
\label{sec:Conc}

Quantum chaos investigates the footprints of classical chaos in the quantum world. We posed an intriguing question - how deep in the quantum regime one can hope to find these signatures? In this chapter, we addressed this question with a provocative answer - we find signatures of classical Lyapunov exponents
as captured by OTOCs even in quantum systems consisting of as few as 3 and 4 qubits as shown in  Fig.~(\ref{fig:OTOCLogLin})  and Fig.~(\ref{fig:OTOC4Qub_pis}) respectively.
 Our results for Loschmidt echo, another quantifier of chaos based on sensitive dependence of a system to perturbations in dynamics, 
suggest a more feeble signature of chaos for the kicked top with lower values of angular momentum. Through numerical study, we have shown that one needs to go to 
sufficiently high quantum numbers to see a forerunner to the exponential Lyapunov decay in the Loschmidt echo. However, for certain initial states, we do see the effects of delocalization, periodic orbits, and chaos in the decay of the echo signal in deep quantum regime of 3 and 4 qubit kicked top. How do these states fare under environmental decoherence would be an interesting future direction to explore.

Recent studies involving a related concept, the Adiabatic Guage Potential (AGP) which is the generator of adiabatic deformations between eigenstates, serves as a probe to detect chaos in systems with large Hilbert spaces \citep{PhysRevX.10.041017}. An interesting direction for the future is to compare the effectiveness of AGP with that of Loschmidt echo in detecting chaos. 

One  interesting observation from our work was the case of $\kappa_0=\pi j$, the chaoticity parameter for the kicked top. As we saw, for the value of $j=2$ ($N = 4$ qubit), the Floequet operator in this case, has interesting decomposition in terms of 
 {\it sum} of 4
pure rotations and similar sum exists for $\kappa_0=\pi r/s$ with $r$ and $s$ relatively prime to each other. 
On the one hand, this paves way for some experiments where the nonlinear twist is replaced by a sum of rotations. On the other hand, this gives us some insights into the origin of chaos and complexity in a system
with a classical limit of just two degrees of freedom.
It is also worth noting that it is very rare that systems exhibiting signatures of chaos are exactly solvable.
A conservative system with as many constants of motion as its degrees of freedom is integrable. Such systems have regular dynamics. In the quantum world, these constants of motion become operators that commute with the Hamiltonian. Lack of sufficient constants of motion leads to non-integrability and the random matrix conjecture in the quantum domain. Exactly solvable systems give us a reference to study departure from integrability and transition to chaos upon the introduction of perturbations breaking the necessary symmetries via the KAM theorem. Our work paves way for the search for more systems that are ``chaotic" yet solvable. For example, a system of coupled kicked tops \citep{trail2008entanglement}, which consists of two spins coupled via hyperfine interactions and one of them periodically kicked can be made to have connections with a many-body model considering a large spin as a collection of spin-1/2 particles. 

Furthermore, our study has connections and applications to metrology, sensing, and study of open quantum systems 
where the environment-induced noise can potentially have an effect on future devices exploiting quantum chaos.
The connections between chaos and quantum sensing have been explored in \citep{fiderer2018quantum}, where quantum-enhanced metrology was demonstrated without the use of entangled input states and found that hypersensitivity to perturbations, quantified by Loschmidt echo was responsible for the enhancement in parameter estimation. 

Nitrogen vacancy centers in diamonds are a promising candidate for making such quantum-enhanced sensors.
These are low $N$ quantum systems and the complete Hamiltonian of NV spin system consists of terms that give rise to simple rotations as well as terms that can be interpreted as non-linear ``twists" to the spin that can potentially give rise to non-integrability and chaos in the classical limit \citep{raosuter17}.
We believe our work on signatures of chaos in low $N$ spin systems will be useful to explore non-classical and non-linear dynamics in systems like the NV-centers, to achieve better sensitivity of measurements, and also study the fundamental connection between quantum chaos, metrology, and decoherence.  Understanding the chaotic/non-integrable dynamics arising out of the non-linear terms in the NV center Hamiltonian, which has intimate connections to the decoherence mechanisms when treated as open quantum systems is useful in understanding the rate and source of errors.  Thus, our approach is useful in studying the limits on the noisy operation of quantum devices, a major obstacle in the development of quantum technologies.
We therefore hope our work is useful to the quantum chaos community as well as experimentalists.

  \chapter{Tomography with random diagonal unitary maps}
\label{chap:chap4}

\begin{figure}[H]
\centering
\includegraphics[scale=0.4]{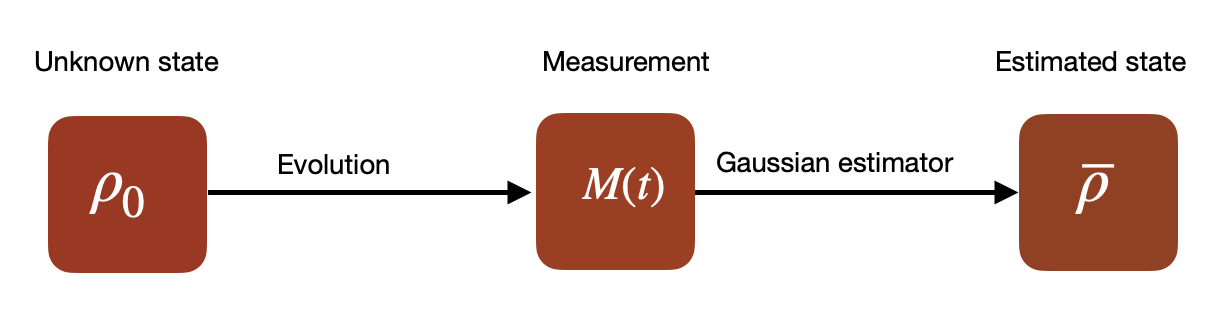}  
\caption{An overview of the state estimation procedure}
\label{tom}
\end{figure}

To determine an unknown state is a  fundamental challenge in quantum information processing.
 The process of estimating an unknown state by performing measurements on it is called quantum tomography. Since the probabilities for various outcomes during the measurements depend on the state in which we perform them, we can in principle determine the unknown density  matrix by inverting the measurement records.  \citep{RevModPhys.29.74,paris2004quantum, dariano2003quantum}. 
 The traditional method to carry out quantum state tomography is to do projective measurements. 
   Any projective measurement would collapse the wave function, deterministically evolved through the Schr\"{o}dinger's equation. Projective measurements are expensive and time-consuming and one has to repeat the process many times to get an accurate estimate of the density matrix. 
 However, one may get around this by using a weak continuous measurements protocol~ \citep{PhysRevLett.95.030402, PhysRevLett.97.180403,chaudhury2009quantum,PhysRevLett.93.163602, PhysRevA.81.032126,riofrio2011continuous,Deutsch2010QuantumCA,PhysRevLett.124.110503,smith2003faraday,PhysRevA.90.032113}.
 In this approach, a bunch of quantum states, all initialized to the same is evolved using random unitaries and measurements are made continuously ro get an ``informationally complete" measurement record. 
A set of measurement operators is called informationally complete, if they span all of the operator space.  Such a set of complete measurements has been extensively studied  \citep{renes2004symmetric,zhu2018universally,flammia2005minimal,scott2006tight,d2004informationally},  to cite a few. One does a series of measurements of several observables and obtains the outcome probabilities. Then one inverts these measurement records to obtain the original state.  An outline of the whole procedure  is shown in Fig.~\ref{tom}.
At a more fundamental level, continuous measurements provide us with a window to study the transformation of the system from following quantum laws to classical laws,
the  emergence of chaos from quantum mechanics, and information gain in tomography
under chaotic dynamics \citep{habib2006emergence,PhysRevLett.112.014102, bhattacharya2003continuous,  MADHOK2016}. 

  In this chapter, we study the connection between information gain in tomography and the randomness of quantum dynamics employed to generate the measurement record. For this purpose, we consider various families of ``random" maps.
  In particular, we consider the application of random unitaries diagonal in a fixed basis and compare this with applying a distinct Haar random map at each iteration. During the process, we obtain bounds on maximal information that can be acquired in for  state reconstruction. 
  
  Such diagonal unitaries are of natural interest in quantum computation. Experimentally diagonal gates can be implemented fault-tolerantly in superconducting and semiconducting systems \citep{aliferis2009fault, nakata2014diagonal}. Diagonal quantum circuits are  easier to realize in the lab and are robust to environmental decoherence  { \citep{buscemi2007quantum}}.{ Further, the repeated action of diagonal unitaries has been shown to achieve decoupling of two interacting quantum systems \citep{nakata2017decoupling}. This could be used in achieving environmental decoherence.} 
   Since diagonal gates commute, one does not need to worry about the order of interactions in an experimental realization of such a dynamics, and it reduces the time for implementation and makes the protocol more robust \citep{nakata2014diagonal}.
  It has been shown that diagonal gates have better computational power than classical computers \citep{PhysRevLett.112.140505, bremner2010classical}. The entangling power of such unitaries has been also of interest \citep{Arul}. Diagonal quantum circuits have been employed in the generation of random quantum states uniformly distributed according to a unitarily invariant Haar measure 
  \citep{ nakata2014generating, nakata2014diagonal}.
  Despite this, the merits of
  random diagonal unitaries are far from exhausted and little is known about their concrete applications in other quantum information processing protocols like state reconstruction and quantum control.
  
  Our findings show that random unitary maps with diagonal unitaries do not lead to information completeness as far as the task of quantum state reconstruction is concerned. However, if the state to be reconstructed lies in a lower-dimensional subspace \citep{gross2010quantum} or if we only require a lower resolution tomography that serves practical purposes \citep{aaronson2007learnability}, implementing diagonal unitaries could not only be sufficient but also efficient. In this work, we do not put any restriction on the initial state or the resolution of tomography and study the information generation in state reconstruction when the underlying dynamics are random diagonal unitaries.
  
  Our study has an intimate connection with quantum chaos since the nature of chaotic dynamics can be effectively modeled by random maps. Probing this question deeper, we further explore whether the origin of information gain lies in the spectral statistics of the quantum chaotic map or in the randomness of its eigenvectors. To address this question, we study information gain obtained when the dynamics is generated by a map whose eigenvectors are random/chaotic but spectral statistics 
  belong to a Poisonnian distribution that is characteristic of a regular system. We also study the amount of information gain when eigenvalues are chosen from a distribution characteristic of a chaotic system.

  The rest of this chapter is assembled in the following way.  
In Sec. \ref{sec:weak_continuou}, we describe a general protocol for tomography with continuous-time measurements. In Sec.~\ref{cmt_random_diagonal} we use this protocol to study state reconstruction using random diagonal unitaries. We find that despite being a restrictive case, very good fidelities can be achieved. Then we quantify the information gain using Fisher information and Shannon entropy from the measurement correlations of the estimation process in Sec.~\ref{information_gain}. We give predictions of information generation from random matrix theory, obtain bounds on information gain and introduce a new random matrix ensemble and show that our predictions agree with our numerical simulations in Sec.~\ref{statistical_bounds}. 
We discuss the connection of information gain in continuous measurement tomography and quantum chaos to spectral statistics and eigenvectors of dynamical maps in Sec.~\ref{cmt_chaos}
before we conclude with a brief discussion and overview of coherent state tomography and other results in the final two sections.
  \section{Weak Continuous Measurements} \label{sec:weak_continuou}
  The total system we work with is composed of the object system~$(S)$ and the probe/meter $(M)$. We assume that the object system and the probe start out in a product state. Their evolution is governed by the total Hamiltonian $H_\tau =H_S+ H_M +H_{int}$ \citep{svensson2013pedagogical}. Here $H_{int}$ is the interaction Hamiltonian between the system and the meter. 
  The system and the meter are initially assumed to be uncoupled.  
  They evolve together via a time evolution operator ${U},$ generated by the total Hamiltonian, $H_\tau$.
  \begin{equation}
  {U} \tau_0 {U}^\dagger= {U} \sigma_0 \otimes \mu_0 {U}^\dagger,
  \end{equation}
  where ${U}= \mathrm{exp} \left(- \frac{i}{\hbar} \int \mathrm{d}t H_\tau \right)$, $\sigma_0$ belong to the object system and $\mu_0$ to the probe.

  Let the initial state of the meter be $|m^{(0)}\rangle.$  After undergoing unitary evolution for time $t$, the state becomes $|m^{(i)}\rangle$.  The Hilbert space $H_M$ of the meter is spanned by a complete, orthogonal set of basis states $\lbrace \ket{q} \rbrace$, which are eigenvectors of an observable $\hat{Q}$, called the pointer observable/pointer variable. Expanding in terms of a continuous pointer variable $\hat{Q},$ with pointer states $|q \rangle$,
  \begin{equation}
  |m^{(0)}\rangle= \int \mathrm{d}q |q \rangle \langle q|m^{(0)}\rangle= \int \mathrm{d}q|q\rangle \psi_0(q)
  \end{equation}
  \begin{equation}
  |m^{(i)}\rangle= \int \mathrm{d}q |q \rangle \langle q|m^{(i)}\rangle= \int \mathrm{d}q|q\rangle \psi_i(q).
  \end{equation}
  Let us assume that initially, the pointer of the meter is centered around $q=0$ so that $\langle Q\rangle_0=0$. That is a natural and convenient choice because the difference in the pointer variable is what characterizes a measurement. Let us also choose the initial wave function of the meter to be a Gaussian, centered at zero. 
  \begin{equation}
  \psi_0(q)= \dfrac{1}{(2 \pi \sigma^2)^{1/4}} \mathrm{exp}\left(\frac{-q^2}{4 \sigma^2} \right),
  \end{equation}
  where $\sigma$ is the width of the Gaussian probability density. In the collective, weak measurement that we do, the collective observable say $\mathcal{O}_c= \sum \mathcal{O}^j$, where $\mathcal{O}^j$ acts on the $j^{th}$ subsystem. 
  The interaction Hamiltonian  $H_{int} = \gamma \mathcal{O}_c \otimes \hat{P}$ captures the coupling of the observable to be measured, with a  meter observable. The variable $\hat{P}$ is chosen to be the one conjugate to the pointer variable $\hat{ Q}.$ Here $\gamma$ is a  coupling constant. The measurement takes place within a very small time span $\delta t_u$, therefore
  \begin{equation}
  \int \mathrm{d}t H_{int} = \int \mathrm{d}t \gamma \mathcal{O}_c \otimes \hat{P}= \gamma \mathcal{O}_c \otimes \hat{P} \delta t_u.
  \end{equation}
  The combination $\gamma \delta t_u = g$ is an effective coupling constant.  In our case, we are driving the system using random unitaries. However,   the measurement procedure is only concerned with the interaction term in the Hamiltonian. Since our aim is is to explain the measurement process, for simplicity let us set $H_S$ and $H_M$ to zero. 
  Then
  \begin{equation}
  {U} = 
  \mathrm{exp} \left( \frac{-i}{\hbar}g \mathcal{O}_c \otimes \hat{P} \right).
  \end{equation}
  To understand how the coupled evolution changes the meter variable, assume that the system starts in a pure state  $\ket{\phi}_s=\sum_i \alpha_i|o_i\rangle$, where $\lbrace o_i \rbrace$ are the eigenstates of $\mathcal{O}_c$. Then the collective state after interaction is given by
  \begin{align}
  \sum_i \alpha_i |o_{i}\rangle \otimes |m^{(i)}\rangle &= {U} \left( \ket{\phi}_s \otimes |m^{(0)}\rangle \right)  \\&= 
  \mathrm{exp} \left( \frac{-i}{\hbar}g \mathcal{O}_c \otimes \hat{P} \right) \left(\Big(\sum_i \alpha_i|o_{i}\rangle \Big) \otimes |m^{(0)}\rangle \right) \\&= \sum_i \alpha_i|o_i\rangle \otimes \mathrm{exp} \left( \frac{-i}{\hbar}g o_i  \hat{P} \right) |m^{(0)}\rangle \\&= \sum_i \alpha_i |o_i\rangle \otimes \mathrm{exp} \left( \frac{-i}{\hbar}g o_i  \hat{P} \right) \int \mathrm{d}q |q\rangle \psi_0(q) \\&= \sum_i \alpha_i|o_i\rangle \otimes \int \mathrm{d}q |q\rangle \psi_0(q- go_i). \label{state}
  \end{align}
  The meter state after evolution is, $|m^{(i)}\rangle = \int \mathrm{d}q |q\rangle \psi_0(q- go_i)$ with probability $|\alpha_i|^2$, which implies that $\psi_i(q) = \psi_0(q-go_i)$.
  That is, the initial pointer state of the meter has been translated proportional to  an eigenvalue of the system observable. 
  Until now a measurement of the meter hasn't been performed. Now let us perform a projective   measurement of the meter.
  If the meter is projected onto a particular outcome, then the rest is an operator acting on the system Hilbert space, called a Kraus operator \citep{nielsen2002quantum}.
  \begin{align}
  M_q&= \langle q| \mathrm{exp} \left( \frac{-i}{\hbar}g \mathcal{O}_c \otimes \hat{P} \right)|m_0\rangle\\&= \mathrm{exp} \left( \frac{-i}{\hbar}g \mathcal{O}_c \otimes \hat{P} \right) \psi_0(q) \\ &= \sum_i \psi_0(q-go_i) |o_i\rangle \langle o_i| \\&= \sum_i \dfrac{1}{(2 \pi \sigma^2)^{1/4}} \mathrm{exp}\left(\frac{-(q-go_i)^2}{4 \sigma^2} \right) |o_i\rangle \langle o_i|. \label{measure}
  \end{align}
  Let $\rho_0$ denote the initial state of the object system. Then the  post measurement state is given by  
  \begin{equation}
  \rho_q'= \frac{M_q \rho_0 M_q^\dagger}{\mathrm{Prob}(q)}. \label{post}\end{equation} 
  The corresponding POVM element $ E_q = M_q ^{\dagger} M_q$ is given by
  \begin{equation}
  E_q =\sum_i \dfrac{1}{(2 \pi \sigma^2)^{1/2}} \mathrm{exp}\left(\frac{-(q-go_i)^2}{2 \sigma^2} \right) |o_i\rangle \langle o_i|.
  \end{equation}
  Note that    $\mathrm{Lim}_{\sigma \rightarrow0}E_q= |go_i =q\rangle \langle go_i=q|$. For a finite $\sigma$ the measurement has finite strength. For large $\sigma$, the measurement is very weak.
  
  The probability for a measurement  outcome $q$ is given by $\mathrm{Prob}(q)= \mathrm{Tr}(E_q \rho_0).$  For the initial  state $\ket{\phi}_s=\sum_i \alpha_i|o_i\rangle$, we get
  \begin{equation}
  \mathrm{Prob}(q) =  \dfrac{1}{(2 \pi \sigma^2)^{1/2}}\sum_i |\alpha_i|^2 \mathrm{exp}\left(\frac{-(q-go_i)^2}{2 \sigma^2} \right). \label{prob}
  \end{equation}
  In the weak-measurement regime, when $\sigma \gg o_i$ holds for all eigenvalues,  probability function in Eq.~(\ref{prob})  can be rewritten as follows.
  \begin{align}
  \mathrm{Prob}(q) &=  \dfrac{1}{(2 \pi \sigma^2)^{1/2}}\sum_i |\alpha_i|^2 \mathrm{exp}\left(\frac{-(q-go_i)^2}{2 \sigma^2} \right) \\
  &\approx \dfrac{1}{(2 \pi \sigma^2)^{1/2}} \mathrm{exp}\left(\frac{-(q-\sum_i |\alpha_i|^2 go_i)^2}{2 \sigma^2} \right) \label{approx}\\
  &= \dfrac{1}{(2 \pi \sigma^2)^{1/2}} \mathrm{exp}\left(\frac{-(q- g \langle o\rangle)^2}{2 \sigma^2} \right). 
  \end{align}
  Equation~(\ref{approx}) is obtained by Taylor expanding the exponential function up to first-order around $q=0$ \citep{vaidman1996weak}. The Gaussian spread,  $\sigma^2$ is called shot noise. There is also another noise arising due to the fundamental uncertainty in quantum measurements.  Fluctuations in the observed meter state, called the projection noise leads to variations in the system state.  However, since the shot noise $\sigma^2 $ is much larger, the effect of the projection noise can be neglected and   measurement induced back-action is insignificant \citep{smith2003faraday}.  Therefore it is possible to perform multiple measurements without needing to repeat the evolution from the start and  obtain a time-stamped series of measurement records.  
  \section{Continuous measurement tomography} \label{cmt}
     In the continuous measurement tomography protocol that we consider, one starts with a collection of $\mathcal{N}$ identically prepared states which are evolved together. The collective system is denoted by $\rho_0^{\otimes \mathcal{N}}$, where $\rho_0$ is the density matrix of a single system which is unknown.
     The ensemble is collectively controlled, coherently evolved, and continuously probed. The set of measurement operators spanning all of the operator space leads to information completeness. 
     
     We measure the sum of identical observables on all the $\mathcal{N}$ subsystems, and the measurement record at time $t$ can be written in terms of such a collective observable as
\begin{equation}
M(t)=  \langle \mathcal{O}_0 \rangle (t) + \delta M(t).
\end{equation}
Here, $\delta M(t) $ arises from the noise in the detection system, and $\langle \mathcal{O}_0 \rangle(t) = \mathrm{Tr}(\rho_0 U(t)^\dagger \mathcal{O}_0 U (t))$.
We want to determine $\rho_0$ by continuously measuring an observable $\mathcal{O}_0$ evolved in the Heisenberg picture. Such a collective measurement in principle can lead to correlations \citep{geremia2005suppression} among the states which can cause back-action. However, under the conditions of weak continuous measurements, any such quantum back-action is negligible  \citep{PhysRevLett.95.030402}. The prominent noise in the system is the intrinsic shot noise of the probe.

 We consider a discrete set of measurements separated by the time interval $\Delta t$, of observables $\mathcal{O}_n= \Big(\prod_{i=1}^n U_i^\dagger (\Delta t)\Big) \mathcal{O}_0 \Big( \prod_{i=1}^nU_i(\Delta t)\Big)$, where a different unitary governs the evolution for each $\Delta t$ interval. The unitary evolution, which would produce an informationally complete set of observables is not unique.
 The question we ask is the following - How does the performance of tomography, as quantified by the fidelities obtained, depend on the nature of the unitary or the set of unitaries employed to evolve the system.
  For example, one can choose $U( \Delta t)$s from the set of Haar random unitaries \citep{mezzadri2006generate,ozols2009generate}, and apply them to get an informationally complete measurement record that is also unbiased over time. We shall refer to this kind of dynamics as the ``Haar random" case in this work.

 One can also obtain a sequence of measurement records from repeated application of the same fixed unitary chosen at random according to Haar measure \citep{easton}, i.e., $\prod_{i=1}^nU_i(\Delta t)=U_0^n(\Delta t)$. This way,  we obtain a one-parameter family of measurement records. Although not informationally complete, this produces high fidelity reconstruction \citep{PhysRevA.81.032126}. Repeated action of a randomly chosen unitary has been studied as a paradigm to explore quantum signatures of chaos \citep{ PhysRevLett.112.014102, MADHOK2016}. { Another way of driving the operator evolution is by choosing random unitaries that are all diagonal in a particular basis. There is an extra degree of freedom of phases in this case, unlike the previously described powers of a single unitary.}

Let us now discuss the estimation procedure briefly. Considering a stroboscopic time series of measurement records, at a time $ n \Delta t$, 
\begin{equation}
M_n=  \mathrm{Tr}(\mathcal{O}_n \rho_0) +\sigma W(n), \label{00}
\end{equation}
 where we model the detector noise after a  Gaussian white noise $\delta M(t)= \sigma W(t)$. Here  $\sigma$ is the noise variance and $W(t)$ is a Wiener process {with mean zero and unit variance} \citep{durrett2019probability}.
 We can expand $\rho_0$ in a Hermitian basis consisting of  $(d^2-1)$ traceless operators $E_\alpha$ and the identity matrix \citep{MADHOK2016}.
 \begin{equation}
 \rho_0 =  \sum_{\alpha=1}^{d^2-1} r_\alpha E_\alpha + I/d. \label{01}
 \end{equation}
 Using Eq.~(\ref{01}) in Eq.~(\ref{00}), 
\begin{align}
M_n &=  \sum_{\alpha=1}^{d^2-1} r_\alpha \mathrm{Tr}( \mathcal{O}_n E_\alpha ) + \sigma W(n)\\
& =  \sum_{\alpha=1}^{d^2-1} r_\alpha \tilde{\mathcal{O}}_{n\alpha} +\sigma W(n),
\end{align}
where $\mathrm{Tr}( \mathcal{O}_n E_\alpha )=\tilde{\mathcal{O}}_{n\alpha} $.
All such measurement records $\lbrace M_n \rbrace$ together can be written in a matrix form, 
 \begin{equation}
 \bold{\tilde{M}} =  \bold{\tilde{\mathcal{O}}} \bold{r} +\sigma \bold{W}, \label{matrix}
 \end{equation}
 where $\bold{\tilde{M}}$ is a vector of measurement records.
Equation~(\ref{matrix}) says that  the conditional probability of the measurement records given the underlying parameters is a Gaussian, 
 \begin{align}
 \bold{P}({\bold{\tilde{M}/r}}) &\propto \mathrm{exp}\left(-\frac{1}{2 \sigma^2}(\bold{\tilde{M}} - \bold{\tilde{\mathcal{O}}r})^T (\bold{\tilde{M}} - \bold{\tilde{\mathcal{O}}r})\right) \label{cond} \\
 & \propto \mathrm{exp}\left(-\frac{1}{2}(\bold{r}-r^{ML})^T  \bold{C^{-1}} (\bold{r}-r^{ML}) \right). \label{cov}
 \end{align}
Equation~(\ref{cov}) can be obtained from Eq.~(\ref{cond}), look at \citep{riofrio2011continuous} for a proof.  Here  the maximum likelihood estimate  vector $r^{ML}$ of the parameters $\lbrace\alpha\rbrace$ is the one which minimizes the exponent in the Gaussian \citep{PhysRevA.55.R1561}, given by
 \begin{equation}
 r^{ML} =  \sigma^2(\bold{\tilde{\mathcal{O}}}^T \bold{\tilde{\mathcal{O}}})^{-1} \bold{\tilde{\mathcal{O}}}^T \bold{\tilde{M}},
 \end{equation}
where the quantity {$  \sigma^2(\bold{\tilde{\mathcal{O}}}^T \bold{\tilde{\mathcal{O}}})^{-1}$ is called the covariance matrix, $\bold{C}$. Therefore, $r^{ML} = \mbf{C}\mbf{\tilde{\mathcal{O}}}^{T} \mbf{\tilde{M}}$.} The eigenvalues of $\mbf{C}^{-1}$ are  the signal-to-noise ratios with which we have measured different orthogonal directions in the operator space (given by its eigenvectors). 
 
In the absence of measurement noise, and when the inverse covariance matrix $\bold{C^{-1}}= \mbf{\tilde{\mathcal{O}}}^T \mbf{\tilde{\mathcal{O}}}/\sigma^2$ is full rank, the most likelihood estimate is given by $ \rho^{ML} = \sum\limits_{\alpha=1}^{d^2-1} r^{ML}_\alpha E_\alpha + I/d $. In presence of measurement noise, or when the measurement record is incomplete, $\rho^{ML}$ can have non-physical eigenvalues. Then one has to replace $\rho^{ML}$ by its closest physical density matrix, which can be obtained by minimizing the squared distance between the new estimate $\bar{r}$ and $r^{ML}$ \citep{tarantola2005inverse,MADHOK2016}.
\begin{equation}
\left \|r^{ML}  - \bar{r} \right\|^2 = (r^{ML} -\bar{r})^T (\bold{\tilde{\mathcal{O}}}^T \bold{\tilde{\mathcal{O}}}) (r^{ML} -\bar{r}),
\end{equation}
 subject to the constraint $\sum\limits_{\alpha=1}^{d^2-1} \bar{r}_{\alpha } E_\alpha  + {I}/{d}  \geq 0$.
 \section{Continuous measurement tomography with random diagonal unitaries} \label{cmt_random_diagonal}
 We evolve the initial state using random unitaries diagonal in a fixed basis and generate a measurement record. In the Heisenberg picture, the operator evolves while the state remains the same.  After the first $\Delta t$ time interval, the  operator $\mathcal{O}_0$ changes to $U^{\dagger}(\Delta t) \mathcal{O}_0 U(\Delta t)$, where $U(\Delta t)=\sum_{j=1}^d e^{-i \phi_j} \ket{j} \bra{j} $. Since we will be indexing the unitaries as well, 
 we can rewrite this as $U_{m}(\Delta t)=\sum_{j=1}^d e^{-i \phi_{mj}} \ket{j} \bra{j} $, where $U_{m}(\Delta t)$ is the random diagonal unitary applied at time $m\Delta t$.
 
 Here the exponential phase factors $\phi_{mj}, \in [0, 2\pi]$, are chosen uniformly at random.  
 After $n$ time steps, 
 \begin{equation}
 \mathcal{O}_n = U_{n}^{\dagger}(\Delta t)U_{n-1}^{\dagger}...U_{1}^{\dagger}(\Delta t)\mathcal{O}_0U_{1}(\Delta t)U_{2}(\Delta t)...U_{n}(\Delta t), \label{002}
 \end{equation}
 which gives
 \begin{align}
 \mathcal{O}_n &= \sum_{j,k=1}^d e^{\sum_{m=1}^n {-i(\phi_{mj}-\phi_{mk})}} \bra{k}\mathcal{O}_0\ket{j} \ket{k}\bra{j} \\
&= \sum_{j,k=1}^d e^{-i(\Phi_{nj}-\Phi_{nk})} \bra{k}\mathcal{O}_0\ket{j} \ket{k}\bra{j},   \label{on}
 \end{align}
 where $\Phi_{nj}= \sum_{m=1}^n \phi_{mj}$ is the phase multiplying $j^{th}$ eigenvector after the evolution for time $n \Delta t$.
The operators $\left\lbrace \mathcal{O}_n\right\rbrace$ do not span all of the operator space.  Consider $G= \left\lbrace g \in \mathrm{su}(d) | U(t) g U^{\dagger}(t)=g\right\rbrace$, where $U(t)$ is a unitary diagonal in a particular basis considered, at time $t$.  Let $B= \left\lbrace g  \in G | \mathrm{Tr}(g \mathcal{O}_0)=0 \right\rbrace$. Then $\mathrm{Tr}(\mathcal{O}(t)g)=0, \forall g \in B$. Here $\mathcal{O}(t)$ represents the operator evolved by $U(t)$. $G$ is isomorphic to the Cartan subalgebra of $su(d)$, and the dimension of $G \geq d-1$. Therefore the dimension of the spanned space $\leq d^2-d+1$. This is very similar to arguments presented in \citep{PhysRevA.81.032126}, quantifying the dimension of the operator space spanned under repeated application of a single unitary map.  But can the random diagonal dynamics saturate this bound or do they span a strictly lower dimensional subspace? To see this, let us
 rewrite Eq.~(\ref{on}) as follows
\begin{equation}
\mathcal{O}_n=\sum_{j=1}^{d} \bra{j}\mathcal{O}_0\ket{j} \ket{j}\bra{j} +\sum_{j \neq k =1 }^d e^{-i(\Phi_{nj}-\Phi_{nk})} \bra{k}\mathcal{O}_0\ket{j} \ket{k}\bra{j}, \label{03}
\end{equation}
and use the condition
\begin{equation}
\sum_{n=0}^{d^2-d} a_n \mathcal{O}_n =0 \: \mathrm{iff}\: a_n=0 \forall n, \label{04}
\end{equation}
for linear independence of the the observables.
It has been shown that if the following  conditions are satisfied,  the set $\lbrace\mathcal{O}_n\rbrace$ is linearly independent \cite{PhysRevA.81.032126}.
\begin{enumerate}
\item $ \bra{i}\mathcal{O}_0\ket{j} \neq 0 \: \forall i,j$
\item $\Phi_{mj}-\Phi_{mk} \neq \Phi_{m'j'}-\Phi_{m'k'}, \forall (m,j,k) \neq (m',j',k')$.\label{condition2}
\end{enumerate}
In random diagonal unitaries, we pick the eigenphases uniformly at random, therefore condition  \ref{condition2} is satisfied for any typical member.     That is, the set of observables generated using diagonal in a basis unitaries span a $d^2-d+1$ operator subspace almost always.
This makes intuitive sense. Kinematically speaking, repeated application of a single unitary should do as well as a set of diagonal random unitaries with a fixed basis. Our numerical simulations and the yield of tomography give further evidence of this.

\begin{figure}
\centering
\includegraphics[width=8cm,height=5.5cm,angle=0]{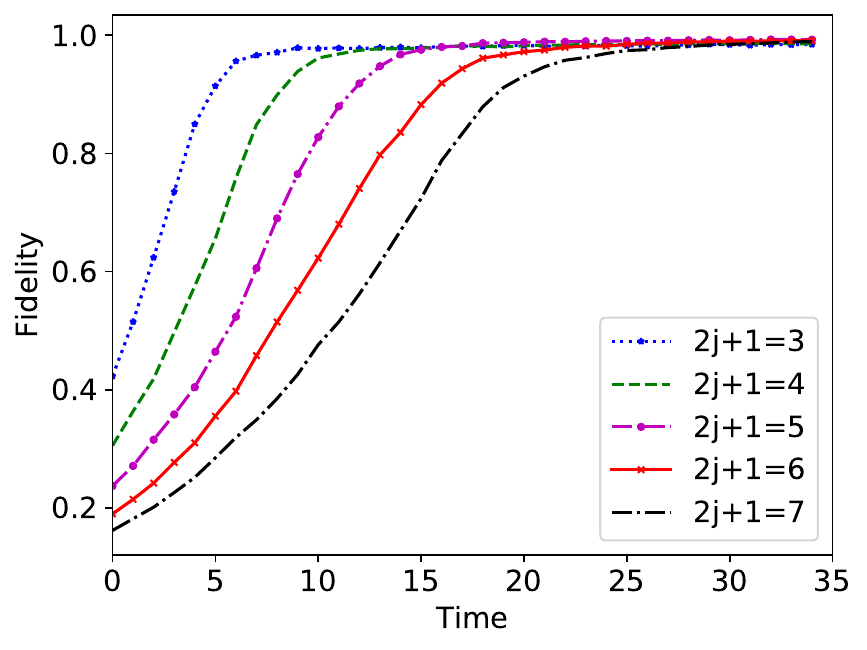} 
  \caption{Average fidelity of reconstruction with  random unitaries  diagonal in a fixed basis against time for different dimensions of Hilbert space. The X-axis represents number of applications of the unitary map. Averaging is done over reconstruction of 200 random pure states drawn according to Haar measure.  Figure shows that even for low dimensions, surprisingly high fidelity reconstruction $ >0.98$ is achieved even though measurement is not informationally complete. }
   \label{fig:1}
\end{figure}

\begin{figure*}
\subfloat[\label{sfig:testa}]{%
  \includegraphics[width=8cm,height=5.5cm,angle=0]{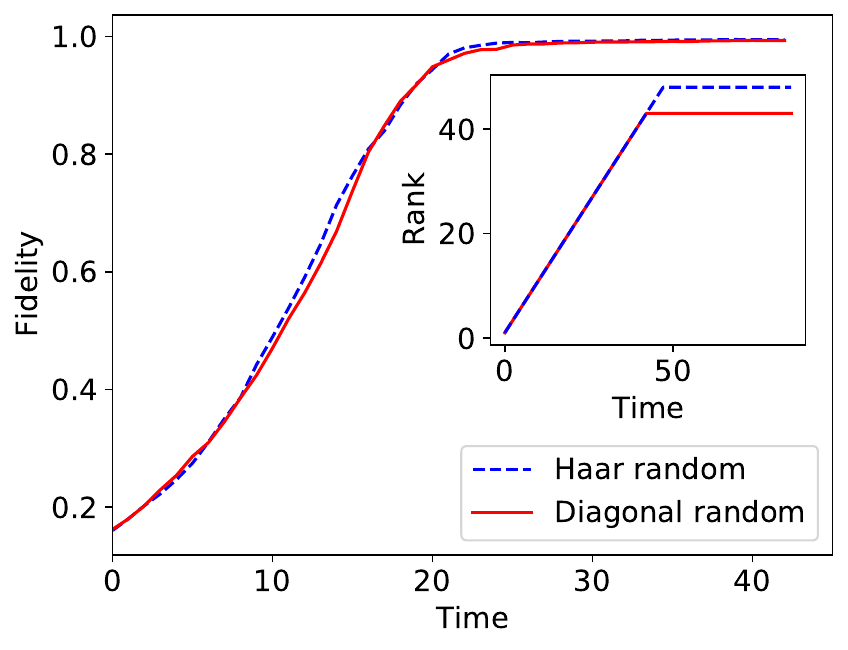}%
}\hfill
\subfloat[\label{sfig:testa}]{%
  \includegraphics[width=8cm,height=5.50cm,angle=0]{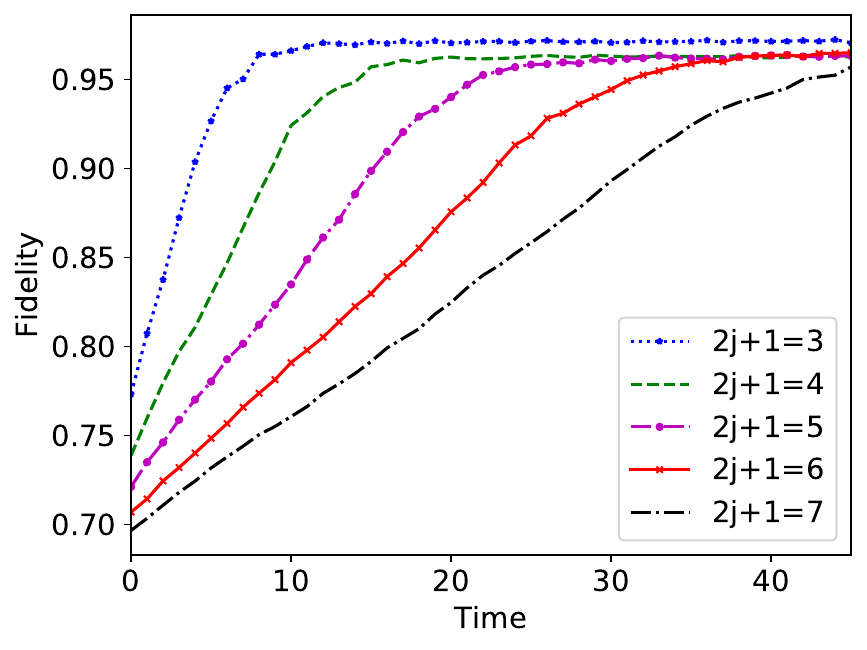}%
} 
\caption{a) Average fidelity of reconstruction  for $d=7$ with a random unitary  diagonal in a fixed basis and  completely Haar random unitary (i.e., a different Haar random unitary at each time step) against the number of applications of the unitary map. Averaging is done over the reconstruction of 100 random pure states drawn according to Haar measure. The rank of the covariance matrix against time is shown in the inset. It gives the dimension of the operator space spanned. Dynamics using   random unitary  diagonal in a fixed basis doesn't span all of the operator space, yet performance is similar. High fidelity is achieved even before rank saturates. b) Average fidelity of reconstruction against time, averaged over 200 mixed states picked according to Hilbert Schmidt measure for small dimensions. }

\label{fig:mix}
\end{figure*}

 Pure state performance when reconstructed with this algorithm for random unitary diagonal in a fixed basis evolution is shown in Fig.~\ref{fig:1}. We see that with more measurement records, the reconstruction fidelity is increasing and saturating very close to one. As the Hilbert space dimension increases, the operator subspace about which we do not have any information becomes less significant and a near-complete reconstruction is achieved. However, it is remarkable that even for small dimensions, where one would expect the effect of the subspace not spanned to be more pronounced, fidelities $>$ 0.98 is achieved. But the same process for mixed states yields a noticeable difference in the reconstruction fidelities when random diagonals are used instead of Haar random unitaries, as seen in Fig.~\ref{fig:mix}b.

\section{Quantifying Information gain in tomography} \label{information_gain}
To measure the  information acquired  during measurements,  consider the Hilbert Schmidt distance between the estimated state  $\tilde{\rho}$  and the actual state $ \rho$. 
\begin{equation}
e=
\Tr\left\lbrace(\tilde{\rho}-\rho)^2 \right\rbrace.
\end{equation} 
It is easy to see that $e$ quantifies the error in the reconstruction \citep{PhysRevLett.88.130401}.
Using the expansion in Eq. (\ref{01}), the mean error $\langle e\rangle$, obtained by repeating the reconstruction procedure, can be expressed as
\begin{equation}
\left\langle e \right\rangle= \sum_{\alpha} \langle (\Delta r_\alpha)^2 \rangle, \label{var}
\end{equation}
{where $\left\lbrace r_\alpha\right\rbrace$ are components of the state vector.}
The variances in Eq.~(\ref{var}) are  bounded from below,  called the  Cramer Rao bound \citep{cramir1946mathematical} 
\begin{equation}
\langle (\Delta r_\alpha)^2 \rangle \geq \left[\mathcal{F}^{-1}\right]_{\alpha\alpha}, \label{fisher}
\end{equation}
where $\mathcal{F}$ is the Fisher information matrix associated with the conditional distribution in Eq.~(\ref{cond}).
 When there is negligible quantum back-action, all the uncertainty in a parameter value $r_\alpha$ is due to the shot noise variance $\sigma^2$,  and the Fisher information matrix is same as the inverse covariance matrix  $\mathcal{F}=\bold{C^{-1}}$\citep{vrehavcek2002invariant}.  
Now  looking at Eq.~(\ref{fisher}),  the inverse of the total uncertainty can be written as  follows.\begin{equation} \frac{1}{\sum_\alpha \langle (\Delta r_\alpha)^2 \rangle}  = \frac{1}{\Tr(\bold{C})}. \end{equation} $1/\Tr(\bold{C})$ can be intuitively understood as
 a measure of the net information gained from measurements,  called the collective Fisher information($FI$) \citep{MADHOK2016}.  It monotonically increases with more measurements as seen in Fig.~\ref{entr}a.   Each eigenvector of $\bold{C^{-1}}$ represents an orthogonal direction in operator space  spanned during measurement process, and each eigenvalue determines the information gain or  signal to noise ratio in that direction. {If the dynamics doesn't span all of the operator space, $\bold{C^{-1}}$ is not full rank, and  $FI$ is ill-defined. To rectify this situation, a Tikhonov regularization is performed by  adding a multiple of Identity to $\bold{C^{-1}}$ before inverting \citep{ng2004feature}.
\begin{figure*}
\subfloat[\label{sfig:testa}]{%
  \includegraphics[width=8cm,height=5.5cm,angle=0]{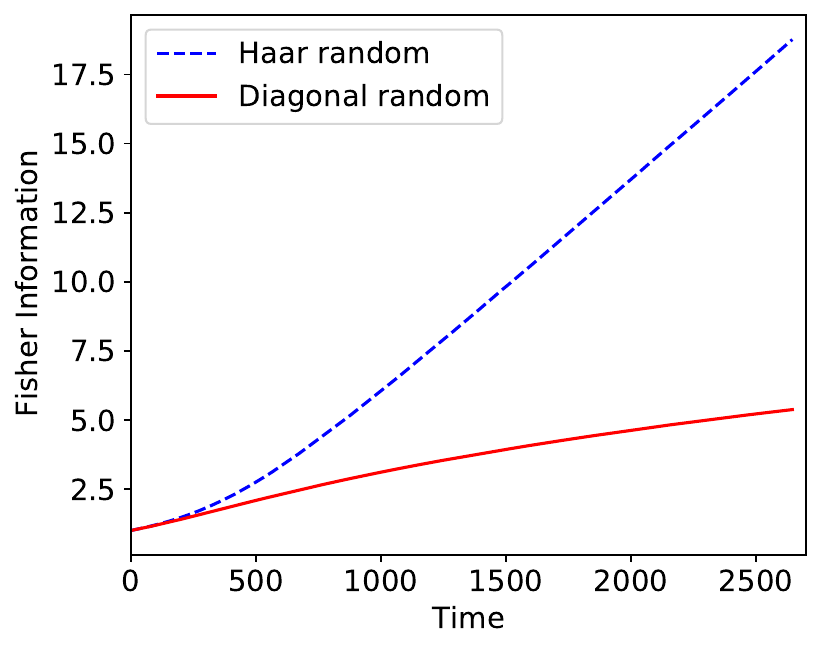}%
}\hfill
\subfloat[\label{sfig:testa}]{%
  \includegraphics[width=8cm,height=5.5cm,angle=0]{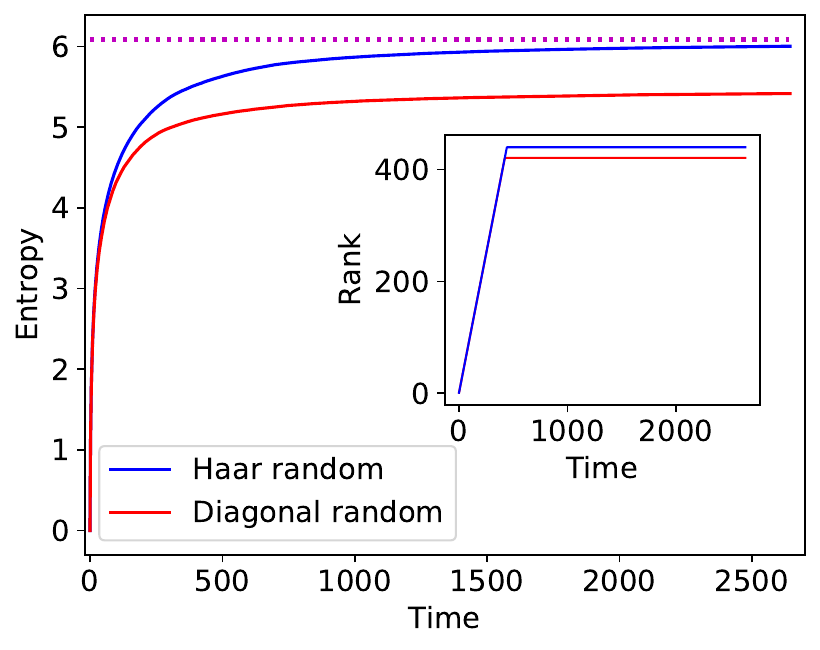}%
}
\caption{a) Comparison of collective  Fisher information of   random unitary  diagonal in a fixed basis with completely Haar random unitary (i.e., a different Haar random unitary at each time step) for $d=21$. The initial observable is $J_x$ and is evolved over time. The X-axis represents number of applications of the unitary map. The collective Fisher information, defined as inverse trace of the covariance metric ($1/\mathrm{Tr}{\bold{C}}$) quantifies the amount of information the measurement records have about the unknown parameters. b) Comparison of Shannon entropy of   random unitary,  diagonal in a fixed basis with completely Haar random unitary. As time passes,  operator space is more evenly sampled and Shannon entropy tends to saturate. The inset shows the rank of the covariance matrix. It is the dimension of the operator space spanned. Operator subspace of dimension $d-2$ is left out in the diagonal case, which is reflected in the entropy. The dotted line parallel to X-axis is $\Log(d^2-1)$, the maximum attainable entropy.}
\label{entr}
\end{figure*}
 Fisher information is closely related to other information metrics -- the mutual information $\mathcal{I}[\mbf{r};\mbf{\tilde{M}}]$ and fidelity \citep{MADHOK2016}.
 
Maximum information gain is obtained when all the eigenvalues of $\bold{C^{-1}}$ are equal \citep{MADHOK2016}. To get such an equal information gain in all the directions, the  operator dynamics needs to be unbiased. This encourages the  quantification of   the ``bias'' or ``skewness" in sampling, and Shannon entropy is a familiar metric that can achieve this.  Let us normalize  the eigenvalues of  $\mbf{C}^{-1}$  so that they become a probability distribution. As mentioned already, they represent the signal-to-noise ratios in  each direction. We can now calculate the Shannon entropy of this distribution $\mathcal{H}=- \sum \lambda_i \mathrm{log} \lambda_i$, where $\left\lbrace\lambda_i\right\rbrace$ is the set of normalized eigenvalues. With longer time  evolution, the initial observable that we started with traverses a trajectory, visiting all of the operator space that the unitary dynamics can span. If the dynamics is  unbiased, this would even out the eigenvalues which in turn maximizes the Shannon entropy.  Such an even sampling of the operator space gives high fidelity reconstruction for random pure states. This asymptotic saturation of entropy is evident in Fig.~\ref{entr}b. i.e., Random unitary dynamics maximize information gain.

\section{Statistical bounds on information gain} \label{statistical_bounds}

In this section, we study the maximum information gain that can be generated in our tomographic protocol.
We notice that the inverse covariance matrix, $\bold{C^{-1}}= \bold{\tilde{\mathcal{O}}}^T\bold{\tilde{\mathcal{O}}}/\sigma^2$ has the form similar to a matrix from the Wishart-Laguerre ensemble \citep{wishart1928generalised,livan2018introduction}, 
  obtained  from rectangular matrix of real elements. The necessary condition for the covariance matrix to have eigenvalues that behave statistically like that of a Wishart matrix
 is to have uncorrelated and identically distributed matrix elements in the constituent matrices $\bold{\tilde{\mathcal{O}}}^T$ and $\bold{\tilde{\mathcal{O}}}$. Marchenko-Pastur distribution describes the behaviour of eigenvalues of the  Wishart matrices of the form $W^TW$, where $W$ are large rectangular random matrices with independent and identically distributed entries. For a Wishart matrix constructed from $D \times N$ rectangular random matrix with  $D \leq N$, the Marchenko-Pastur density function denoted by $\rho(\lambda)$ is given by
\begin{align}
\rho(\lambda) &= \frac{N}{2 \pi  \lambda} \sqrt{(\lambda-\lambda_-) (\lambda_{+}-\lambda)} \\
\lambda_{\pm}&= \frac{1}{N}\left(1- \left(\frac{D}{N}\right)^{-1/2}\right)^2,
\end{align}
where $\lambda \in [\lambda_{-}, \lambda_{+}]$.  Note that in our protocol,  $\bold{C^{-1}}$ matrix is obtained by $\tilde{\mathcal{O}}^T\tilde{\mathcal{O}}/\sigma^2$, where
\begin{equation}
\tilde{\mathcal{O}}= \begin{pmatrix}
\tilde{\mathcal{O}}_{11} & \tilde{\mathcal{O}}_{12} &...&\tilde{\mathcal{O}}_{1d^2-1}\\
...&...&...&...\\
 \tilde{\mathcal{O}}_{N1} & \tilde{\mathcal{O}}_{N2} &...&\tilde{\mathcal{O}}_{Nd^2-1}
\end{pmatrix}.
\end{equation}
Here $N$ is the total number of time steps and $d^2-1$ is the dimension of the operator space.
An element  $\tilde{\mathcal{O}}_{n\alpha}= \mathrm{Tr}(\mathcal{O}_n E_\alpha)$ is the expectation value of operator $\mathcal{O}_n$ along the direction $E_\alpha$ in the operator Hilbert space. Since the expectation value of measurements along each direction is obtained by averaging over a large number of identically prepared systems, $\tilde{\mathcal{O}}_{n\alpha}$ follows Gaussian distribution because of central limit theorem, with variance $\sigma^2$, the shot noise. Hence each element in $\tilde{\mathcal{O}}$ is identically distributed.

Now what remains is to prove that elements of  $\tilde{\mathcal{O}}$  are independent.  The successive operators $\lbrace\mathcal{O}_n\rbrace$  are obtained  by  conjugation action on the initial operator  by a unitary map chosen uniformly at random  each time.
This makes the operators independent of each other  and hence the measurement values are uncorrelated upto one contraint, $N\lVert \mathcal{O}_0 \rVert^2=\sum_{i,\alpha} \tilde{\mathcal{O}}_{i\alpha}^2 ,$ where $\mathcal{O}_0$ is the initial operator.  However, when $N$ and $d^2-1$ are large, $\tilde{\mathcal{O}}$ behaves effectively as a random matrix with independent and identically distributed entries.  Now we numerically demonstrate that the Haar random evolution accomplishes this.
$\bold{C^{-1}}$ is a $d^2-1$ dimensional full rank matrix.  As the dimension of the Hilbert space tends to be very large, the eigenvalues  of Wishart matrices become continuous and follow the Marchenko-Pastur density function as seen  in Fig.~\ref{fig:loe}.

We estimate the collective Fisher information using the Wishart-Laguerre ensemble.  Let $\left\lbrace\lambda_i\right\rbrace_{i=1}^D $ be the eigenvalues of $\bold{C^{-1}},$ where $D=d^2-1$.  In the limit of large $N$, we can approximate the sum by an integral.
\begin{equation}
FI = \frac{1}{\sum_i^D \frac{1}{\lambda _i}} \approx \frac{1}{D \int \frac{1}{\lambda} \rho(\lambda)d \lambda} = \frac{1}{D \langle \frac{1}{\lambda}  \rangle}, \label{fi}
\end{equation}   
where $\rho(\lambda)$ is the Marchenko-Pastur density. In our numerical simulation, we evolved the system for $6d^2$ time steps and used the measurement records obtained to generate the Wishart matrix. Therefore in our simulations, the  parameters  in the density function are $D=440$ and $N=2646.$  
The collective Fisher information obtained using the integral approximation is 18.03907, in excellent agreement with the Haar random case. Using the covariance matrix of Haar random evolution,  we get $FI$=18.76207, after 2646 time steps.  The small difference in the values obtained can be attributed to the dimension being small for the eigenvalues to be continuous.


To quantify the bias in the operator space dynamics, we calculate the Shannon entropy. We normalize the eigenvalues of $\bold{C^{-1}}$  so that they form a probability distribution  and  compute the the Shannon entropy $\mathcal{H}=-\sum \ \lambda_i \Log\lambda_i$, where $\lbrace \lambda_i \rbrace$ are the normalized eigenvalues. For the  Haar random case, the { average} entropy numerically obtained for an ensemble of random states of dimension $d=21$, after 2646 iterations, which we denote by subscript ``$rs$",    is $\mathcal{H}_ {rs}=6.00415$. The Shannon entropy of Wishart-Laguerre orthogonal ensemble, denoted by subscript ``$loe$" can be calculated as  
  \begin{equation}
  \mathcal{H}_{loe}= - D\int_{\lambda_-}^{\lambda_+}\lambda \Log \lambda  \frac{N}{2 \pi  \lambda} \sqrt{(\lambda-\lambda_-) (\lambda_{+}-\lambda)}d\lambda. \label{mpdensity}
\end{equation}   
Using this integral approximation, which works better for large dimensions, we get $\mathcal{H}_{loe}=6.00363$ in remarkable agreement with the Haar random case. Fig.~\ref{fig:loe} shows the distribution of normalized eigenvalues of  $\bold{C^{-1}}$ for Haar random evolution and the Marchenko-Pastur density function. When all the  eigenvalues are equal,  the expected Shannon entropy  for  $d=21$, is $\mathcal{H}_{exp}= \log{(d^2-1)}= 6.08677$. Haar random unitaries lead to completely unbiased dynamics as reflected in the entropy values.
\begin{figure}[hbtp]
	\centering
	\includegraphics[width=8cm,height=5.5cm,angle=0]{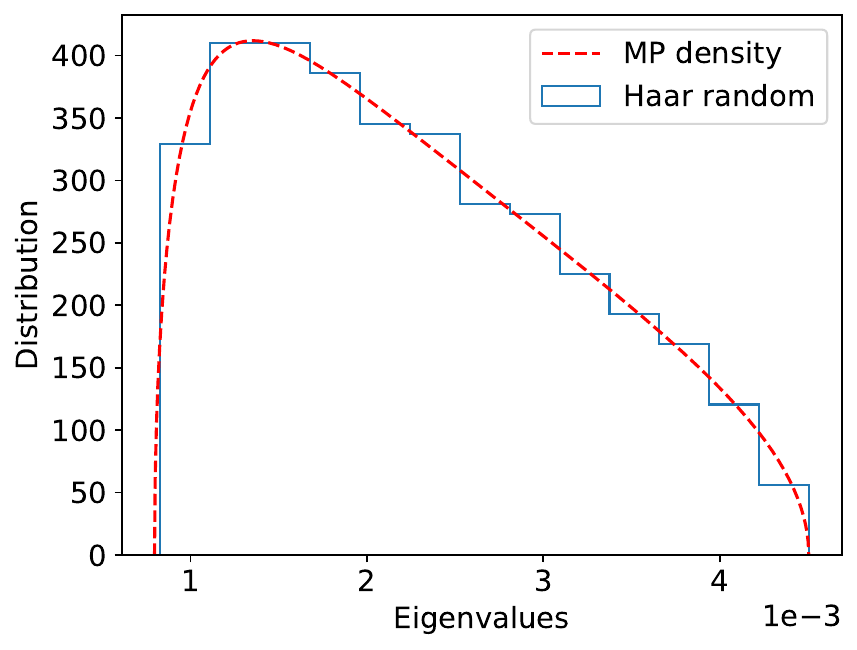}
	\caption{Histogram of eigenvalue distribution of $\bold{C^{-1}}$ of Haar random evolution and Marchenko-Pastur density function. Remarkable agreement is seen in the eigenvalue distribution.} \label{fig:loe} 
\end{figure}

\subsection*{Information gain for random diagonal unitaries}

In this section, we study the maximum information gain that can be generated in our tomographic protocol through the application of random diagonal operators. For this case, the  inverse covariance matrix does not obey the Marchenko-Pastur distribution.
 In the standard Hilbert space, the operator we apply at any time step is of the form, $U = \sum_{j=1}^d  e^{-i\phi_j}\ket{j}\bra{j}$,  where $e^{-i\phi_j}$ and $\ket{j}$ are its eigenvalues and eigenvectors, respectively. Since we will be indexing the unitaries as before,
we can rewrite this as $U_{m}=\Sigma_{j=1}^d e^{-i \phi_{mj}} \ket{j} \bra{j} $, where $U_{m}$ is the random diagonal unitary applied at time step $m$.
Here the exponential phase factors $\phi_{mj}, \in [0, 2\pi]$, is chosen uniformly at random.  
Therefore, after $n$ time steps, 
\begin{equation}
\mathcal{O}_n = U_{n}^{\dagger}(\Delta t)U_{n-1}^{\dagger}...U_{1}^{\dagger}(\Delta t)\mathcal{O}_0U_{1}(\Delta t)U_{2}(\Delta t)...U_{n}(\Delta t). \label{0003}
\end{equation}
 Now we use the superoperator picture~\citep{caves1999quantum, PhysRevLett.112.014102}. The superoperator of the measured observable after $n$ times steps, $|\mathcal{O}_n)= \bold{U_n U_{n-1} ... U_{1}} |\mathcal{O}_0)$, where the superoperator map is $\bold{U} = U^\dagger \otimes U^{T}$. Using this,
we can write our unitary superoperator map explicitly as $\bold{U} = \sum_{j, k}^d  e^{-i(\phi_k - \phi_j)}|j,k)(j,k|$, where one defines $|j,k) = \ket{j}\otimes\ket{k}^*$, with $^*$ denoting complex conjugation. Therefore in our notation, $ \bold{U_n U_{n-1} ... U_{1}} =  \sum_{j, k}^d  e^{\sum_{m=1}^n {-i(\phi_{mk} - \phi_{mj})}}|j,k)(j,k|$.

  In the superoperator representation, after $N$ time steps, the inverse of this covariance matrix is  $\bold{C^{-1}} = \sum_{n=1}^N |\mathcal{O}_n) (\mathcal{O}_n|$. We can write this as
\begin{equation}
\label{eq:InvCExp}
\bold{C^{-1}} = \sum_{j,k=1}^d\sum_{j',k'=1}^df_{j,k}^{j',k'}(j',k'|\mathcal{O}_0)(\mathcal{O}_0|j,k)|j',k')(j,k|,
\end{equation}
where,
 \begin{equation}
f_{j,k}^{j',k'}=\sum_{n=1}^N e^{\sum_{m=1}^n {-i(\phi_{mj} - \phi_{mj'}-\phi_{mk}+\phi_{mk'})}}, 
\end{equation}
which can be simplified as in Eq.~(\ref{on})
 \begin{equation}
f_{j,k}^{j',k'}=\sum_{n=1}^N e^{ {-i(\Phi_{nj} - \Phi_{nj'}-\Phi_{nk}+\Phi_{nk'})}}. 
\end{equation}
{Notice that if $\Phi_{nj}-\Phi_{nj'} - \Phi_{nk}+\Phi_{nk'}=0$, $\forall n$, for a particular choice of $j,k,j',k'$, the quantity $f_{j,k}^{j',k'}=N$. For an arbitrary unitary map $U$, we will assume that the only way this can happen is if $(j=k) \wedge (j'=k')$ or $(j=j')\wedge (k=k')$, which is certainly true for a random unitary. }

With this assumption, we can approximate $\bold{C^{-1}}$ by terms that scale with $N$ in the large $N$ limit. The inverse covariance matrix is

\begin{equation}
\label{inverseC}
\bold{C^{-1}} \approx N\left[\sum_{j,k=1}^d|(j,k|\mathcal{O}_0)|^2|j,k)(j,k|+ \sum_{j\ne k=1}^d (j,j|\mathcal{O}_0)(\mathcal{O}_0|k,k)|j,j)(k,k| \right] .
\end{equation}

{In this superoperator representation, $\bold{C^{-1}}$ is a $d^2 \times d^2$ dimensional matrix with $d^4$ elements in total. Notice that in  Eq.~(\ref{inverseC}), the total number of terms are only of order $d^2$.  Therefore the matrix $\bold{C^{-1}}$ is sparse for large $d$.  The degree of sparsity increases with growing dimensions of the Hilbert space.
The first sum in Eq.~(\ref{inverseC}) contains the diagonal elements.  In the limit of large $d$, eigenvalues of $\bold{C^{-1}}$ are very close to the diagonal terms, because of the limited interaction with other elements in the matrix.  Since the inverse covariance matrix is a superoperator in a real vector space of $d^2$ dimensions, let us compare the normalized eigenvalues with the Porter-Thomas distribution \citep{wootters1990random}. The motivation for this comparison is that $\bold{C^{-1}}$ can be thought of as being picked from a unitarily invariant measure by its construction, with real eigenvalues. }}
\begin{figure}
\centering
\includegraphics[width=8cm,height=5.5cm,angle=0]{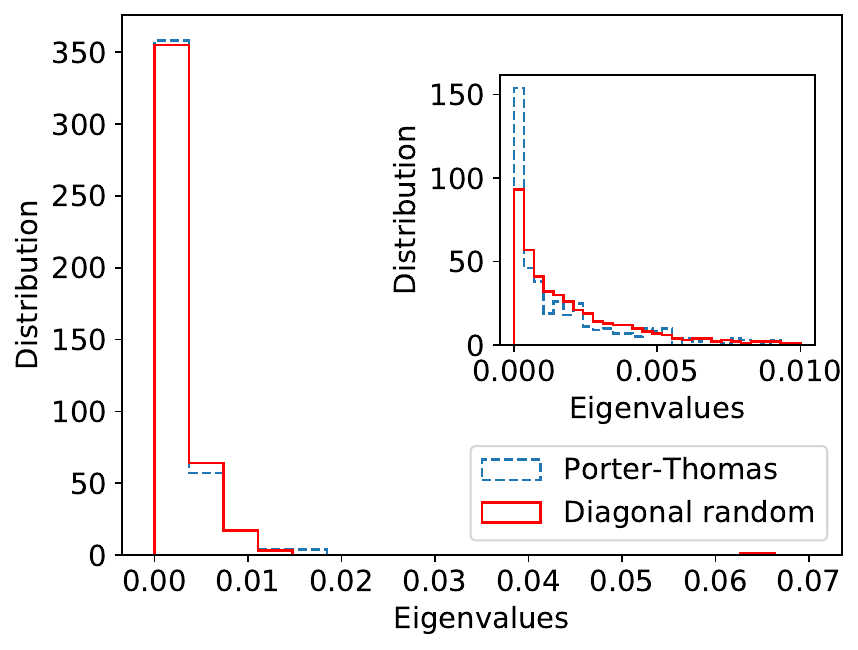} 
\caption{Distribution of eigenvalues of $\bold{C^{-1}}$ for the diagonal random case, along with the distribution of   random numbers generated from  Porter-Thomas distribution. The inset shows the zoomed in distribution for very small eigenvalues.}
\label{ptfig}
\end{figure}
{The Porter-Thomas distribution given below in Eq.~(\ref{pt}) represents frequency distribution of components of a pure unit vector, chosen uniformly at random in a $d^2$- dimensional real Hilbert space. Let $a_i$ be the $i^{th}$ component of the random real pure state, then  probability for obtaining the $i^{th}$ outcome $p_i= a_i^2$. When the dimension of the Hilbert space $d^2$ is large,  the  $i^{th}$ outcome occurs  $\lambda_i= d^2p_i$ times. The distribution of these frequencies follow the Porter-Thomas distribution
\begin{equation}
\rho(\lambda)= \frac{1}{\sqrt{2\pi \lambda}} e^{-\lambda/2}. \label{pt}
\end{equation}
  We denote the Shannon entropy obtained from Porter-Thomas distribution by subscript $``pt"$.
\begin{equation}
\mathcal{H}_{pt}=-d^2 \int_0^\infty \frac{\lambda}{d^2} \Log \left(\frac{\lambda}{d^2}\right) \frac{1}{\sqrt{2\pi \lambda}} e^{-\lambda/2} dx=5.35941.\label{enpt}
\end{equation}
Also using properties of random states in a real vector space \citep{wootters1990random}, the expected entropy of pure states with real coefficients,  $\mathcal{H}_{exp}=\Log(d^2) -0.729637 $, gives $5.35941$. Both these values obtained for entropy are very similar, and  in very good agreement with $\mathcal{H}_{rs} = 5.41684$ as obtained by our numerical simulations using random diagonal unitaries.  Further evidence of this is given in Fig.~\ref{ptfig}  which compares the eigenvalues of $\bold{C^{-1}}$ with the Porter-Thomas distribution. In very high dimensions, sparsity of the matrix is so high that  all the correlations die, and eigenvalues of $\bold{C^{-1}}$ form a truly random vector. In that asymptotic limit, eigenvalues  follow the Porter-Thomas distribution.
}

{Now  let us look at the rate of information generation.  The Fisher information obtained for the diagonal random case in our numerical simulation is $5.37541$ after $N=2646$ time steps. Using the constraint $\Tr(\bold{C^{-1}})= N ||\mathcal{O}_0||^2$ to re-scale the numbers generated according to Porter-Thomas distribution, so that they are in same footing with the numerical case,
\begin{equation}
\lambda_i \rightarrow \frac{\lambda_i}{d^2} \Tr(\bold{C^{-1}})+ d^2,
\end{equation} 
where $d^2$, has been added as a constant regularization factor which we also had in our simulations to avoid infinities while finding the Fisher information. 
\begin{equation}
FI= \frac{1}{\left[d^2 \int_0^{\infty} \left(\frac{d^2}{\lambda \Tr(\bold{C^{-1}})+ d^4 } \right)\frac{1}{\sqrt{2 \pi \lambda}}e^{-\lambda/2} d\lambda\right] },  \label{fipt}
\end{equation}
which yields $4.31174$, a low value compared to the one we obtained numerically. This is because Porter-Thomas distribution is heavily populated by very small numbers.  $\bold{C^{-1}}$ has a lesser number of very small eigenvalues as seen in the inset of Fig.~\ref{ptfig}. Therefore the eigenvalues of $\bold{C}$, which is the set of inverted eigenvalues $\left\lbrace 1/\lambda_i\right\rbrace$ has more smaller numbers than the  corresponding Porter-Thomas set.  Hence  the trace of the covariance matrix is smaller and  the $FI$ larger, compared to the Porter-Thomas case.} 

 \section{Continuous measurement tomography and its connection to quantum chaos and spectral statistics} \label{cmt_chaos}

Our results of the previous section indicate a connection between random diagonal unitaries and quantum chaos. One can characterizes quantum chaos dynamically, by ``ergodic mixing", i.e., something that takes a localized state in phase space and maps it to a random state, smeared across phase space.  As shown in previous literature, a quantum chaotic map takes a localized state to a pseudorandom state in Hilbert space.  This is characterized by the entropy production of the probability distribution with respect to the standard basis \citep{BandyopadhyayArul2002, Lakshminarayan, Bandyopadhyay04, Zyczkowski1990, ScottCaves2003, trail2008entanglement}. Intuitively,  information gain in quantum tomography is a closely related phenomenon where ergodic mixing due to chaos can be viewed in the Hiesenbeg picture and interpreted
as the rate of obtaining information in different directions of the operator space.

In contrast, a common approach is to characterize chaos using static properties.  In this approach, the signature of chaos in a quantum system is in the energy level statistics of the Hamiltonian (or phases of a Floquet map).  Depending on the symmetries, quantum chaotic systems are classified as Gaussian or Circular Orthogonal (GOE/COE), Gaussian or Circular unitary GUE/CUE and Gaussian or Circular Symplectic Ensembles (GSE/CSE) \citep{Haake}. 

The question then is, does ergodic mixing depend sensitively on the eigenvalues of the Hamiltonian $H$ or Floquet map $\mathcal{U},$ or just on the eigenvectors?  
Is the power of $\mathcal{U}$ to generate randomness related to its eigenvectors, and not eigenvalues?   Are any two Hamiltonians or Floquet maps with the same eigenvalue spectrum the same?    The physical Hamiltonians for regular systems will have nonrandom eigenvectors as well as level statistics corresponding to a Poissonian distribution. 

To decouple the role played by eigenvalues and eigenvectors of the dynamics in the rate of information gain in tomography, we construct quantum maps that have an eigenspectrum corresponding to regular systems and eigenvectors that are random with respect to a standard basis. This is obtained by doing a unitary transformation to a given eigenbasis. The other possibility of regular eigenvectors but eigenspectrum exhibiting level repulsion- a signature of chaos, is also considered.

\begin{figure}[hb]
	\centering	\includegraphics[width=8cm,height=5.5cm,angle=0]{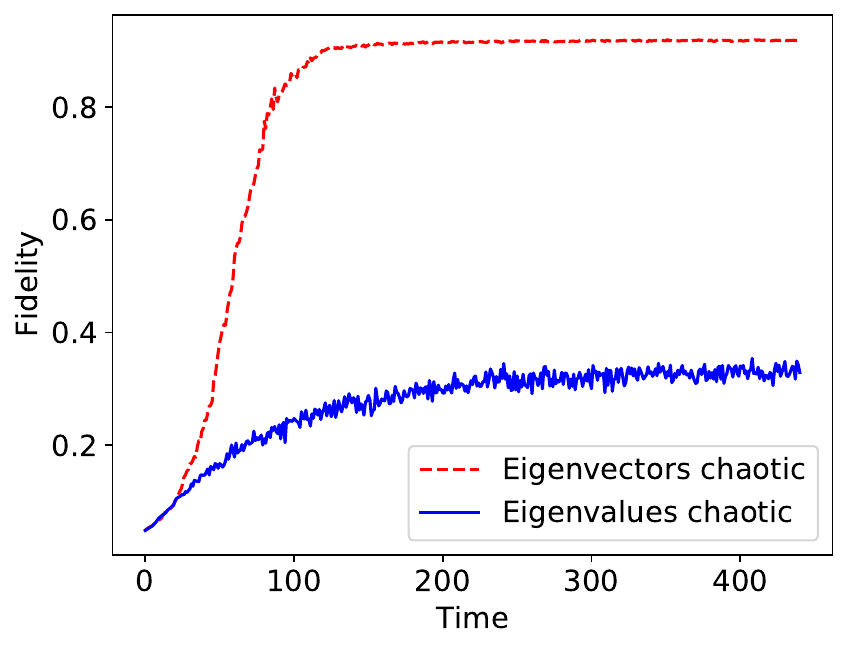}
	\caption{Average reconstruction fidelity over ten random states of $d=21$ using repeated application of a kicked top unitary, $\mathcal{U}= e^{-i1.4 J_x} e^{\frac{-ik_0}{(n-1)} J_z^2}$.  $k_0$ is the chaoticity parameter of the map. The $X$ axis shows the number of applications of the Floquet map. The `eigenvalues chaotic' case is when eigenvalues are picked from the floquet in the chaotic regime with chaoticity 7 and eigenvectors are picked from the floquet with chaoticity 0.5. The other case follows similarly. } \label{fig:fisher} 
\end{figure}

\begin{figure*}[!htb]
\centering
	\subfloat[\label{sfig:testa}]{%
		\includegraphics[width=8cm,height=5.5cm,angle=0]{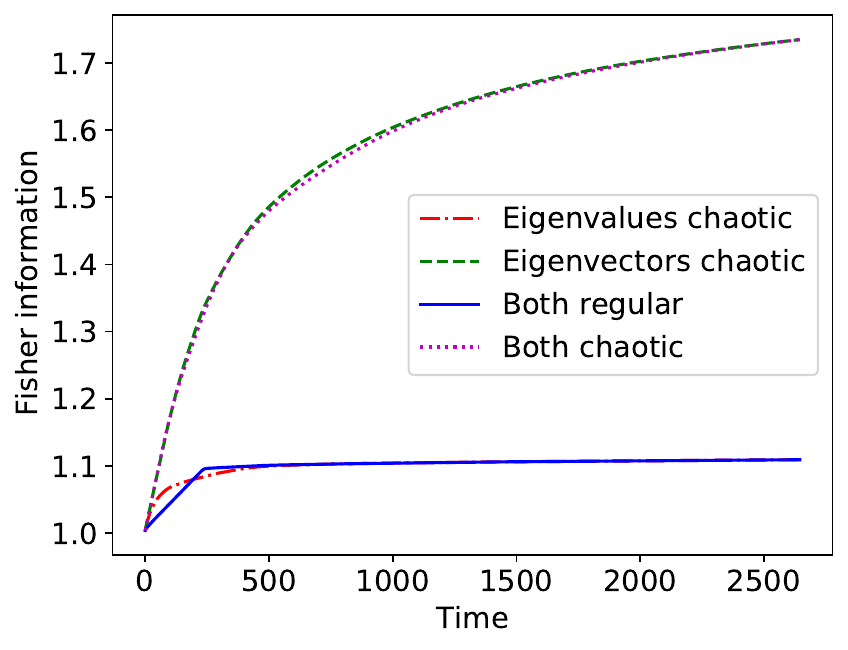}%
	}\hfill
	\subfloat[\label{sfig:testa}]{%
		\includegraphics[width=8cm,height=5.5cm,angle=0]{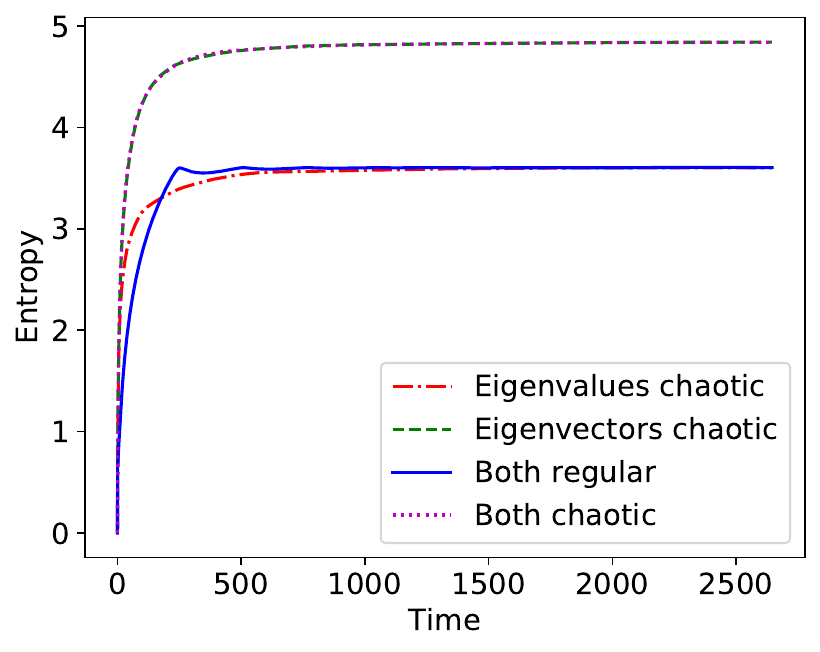}%
	}
	\caption{{The initial observable is $J_z$ and is evolved over time using repeated application of kicked top flouquet operator, $\mathcal{U}= e^{-i1.4 J_x} e^{\frac{-ik_0}{(n-1)} J_z^2}$ for $J=10$.  $k_0$ is the chaoticity parameter. The $X$ axis shows the number of applications of the Floquet map.  The `eigenvalues chaotic' case is when eigenvalues are picked from the floquet in the chaotic regime with chaoticity 7 and eigenvectors are picked from the floquet with chaoticity 0.5. Other cases follow similarly. We observe that when eigenvectors are spread out with more support in the Hilbert space, information gain is  more.  b) Comparison of Shannon entropy with  kicked top evolution for various cases described in part a.  The difference in the saturation value means that when eigenvectors are picked from a  chaotic unitary, they span more operator space.}}
	\label{floquet}
\end{figure*}

\begin{figure*}[!htb]
	\subfloat[\label{sfig:testa}]{%
		\includegraphics[width=8cm,height=5.5cm,angle=0]{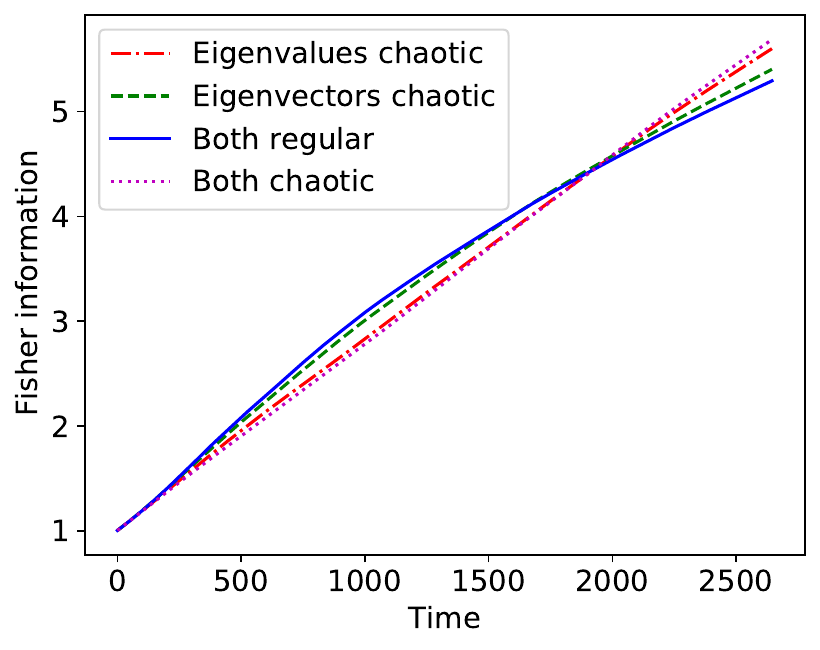}%
	}\hfill
	\subfloat[\label{sfig:testa}]{%
		\includegraphics[width=8cm,height=5.5cm,angle=0]{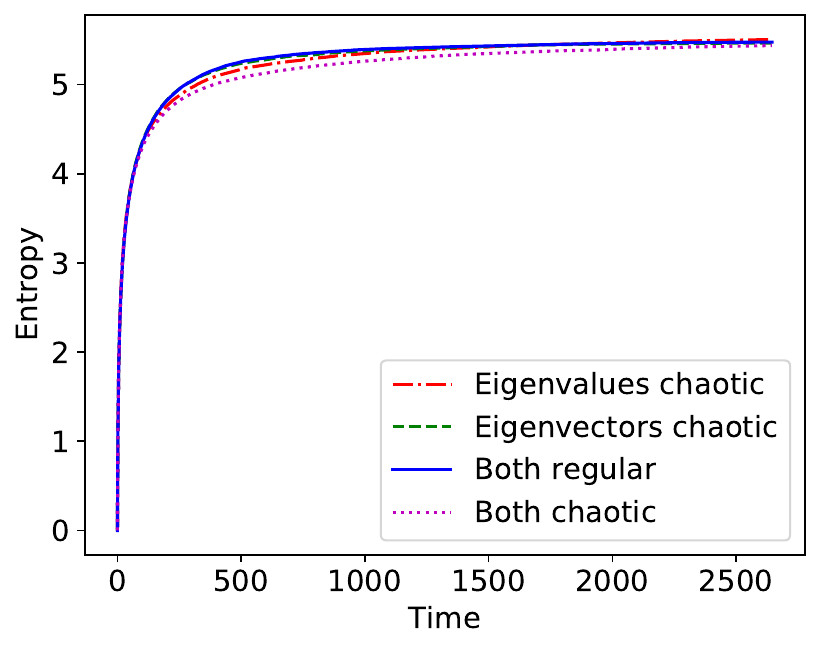}%
	}
	\caption{{ When the initial operator $J_z$ is rotated by a random unitary to $U^{\dagger} J_z U$, the differences seen in the information generation and entropy in the previous figure disappear. The $X$ axis shows the number of applications of  the Floquet map. A larger saturation value of Shannon entropy shows that more of the operator space is spanned, much more than the previous case. }}
	\label{floquet1}
\end{figure*}

 {To this end, we use the repeated application of the kicked floquet to evolve the initial operator $J_z.$  Results are shown in Fig.~\ref{fig:fisher} and Fig.~\ref{floquet}. There is a stark increase in the achieved fidelity when eigenvectors are chosen from a chaotic kicked top unitary.  Figure \ref{floquet} shows that when eigenvalues of the unitary are non-degenerate, the rate of information generation and the amount of operator space spanned during the evolution are solely dependent on the nature of eigenvectors.   Choosing the eigenphases from a chaotic unitary doesn't give any advantage in this case. Figure~\ref{floquet1} shows the evolution of the same system with a rotated initial operator.}

From these observations, it is clear that the eigenvalue statistics of  $\mathcal{U}$, which is a basis-independent criterion of quantum chaos, is not necessary for the information gain in tomography. This suggests that it is the RMT statistics of the eigenvectors of $\mathcal{U}$ that is responsible for faster quantum state reconstruction.
Randomizing the eigenvectors of the initial operator has resulted in washing away the differences seen in the rate of information generation and entropy, in Fig.~\ref{floquet}.  This again demonstrates the basis dependence  of randomness generation.
The message our study imparts is that in general,  the dynamical signatures of chaos, like the generation of near maximally entangled random states and information gain in tomography, are a basis-dependent feature of a system. Either we need initially random operators that might be hard to implement, or one needs a dynamics
that gives rise to pseudorandomness in operator space that generates observables that has support over 
almost the entire $d^2-1$ dimensions.

 Figure~\ref{entr} shows a faster growth of Fisher Information 
 as compared to the case to Fig.~\ref{floquet} for the kicked top. As discussed above,
 different Haar random unitaries saturate the full $d^2 -1$ dimensional operator space, whereas 
a repeated application of a single Haar random unitary misses $d-2$ dimensions. The Kicked top dynamics therefore misses a $d-2$ dimensional subspace as well.  This manifests in the values of Fisher Information.
The information gain when one employs different Haar random unitaries is naturally more rapid as compared to being restricted to repeated application of a single kicked top which is further restricted by additional constraints that we describe below.

In addition, the kicked top has a parity symmetry {given by $R = \exp(-i\pi j_x)$}. In the basis {in which} the parity operator {is diagonal}, the Floquet map has a block diagonal structure {corresponding to the $+1$ and $-1$} parity eigenvalues.
The parity operator $R = e^{-i \pi J_x}$ commutes with the kicked top unitary, $\mathcal{U}$(supplementary section in \citep{PhysRevLett.112.014102}. Therefore, one can find a basis which diagonalizes  both $J_x$ and $\mathcal{U}$.
In the case of global chaos, the Floquet unitary acts as if it is chosen from the circular orthogonal ensemble(COE)~\citep{Haake}.
{Because of the additional parity symmetry, we must choose a block diagonal matrix whose blocks are sampled from the COE in the basis in which the parity operator $R$ is diagonal, thus having the same block structure as the Floquet map}.  This is the reason that the saturation value of Shannon entropy in Fig.~\ref{floquet} ($4.8398$) is lower than that of the random diagonal case Fig.~\ref{entr} (where Shannon entropy reached 5.41684).  That means the Floquet dynamics spans a smaller subspace of the operator space.  
\section{The role of chaos in coherent state tomography} \label{chaos_tomo}
\begin{figure*}[!htb]
	\subfloat[\label{sfig:testa}]{%
		\includegraphics[width=8cm,height=5.5cm,angle=0]{ 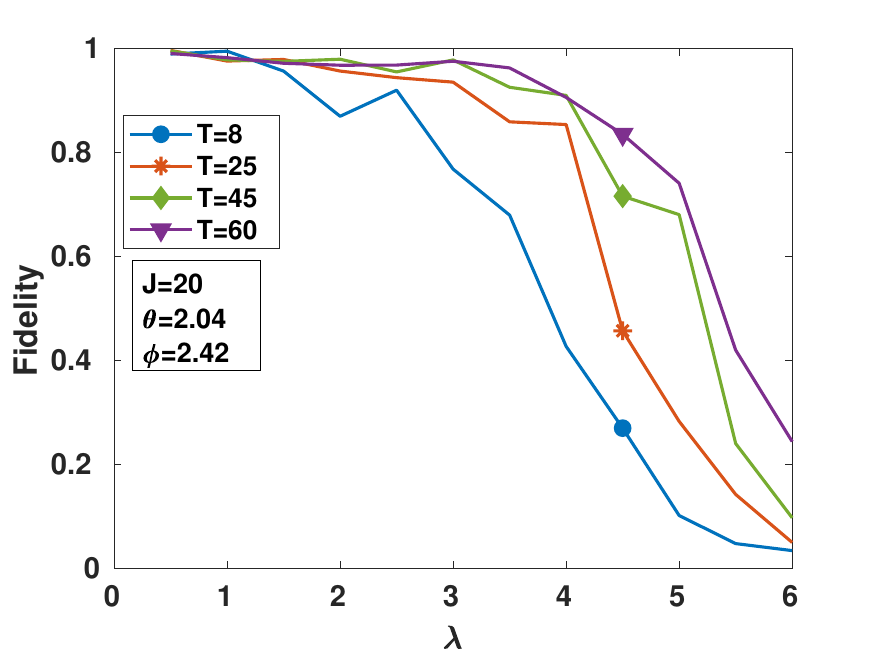 }%
	}\hfill
	\subfloat[\label{sfig:testa}]{%
		\includegraphics[width=8cm,height=5.5cm,angle=0]{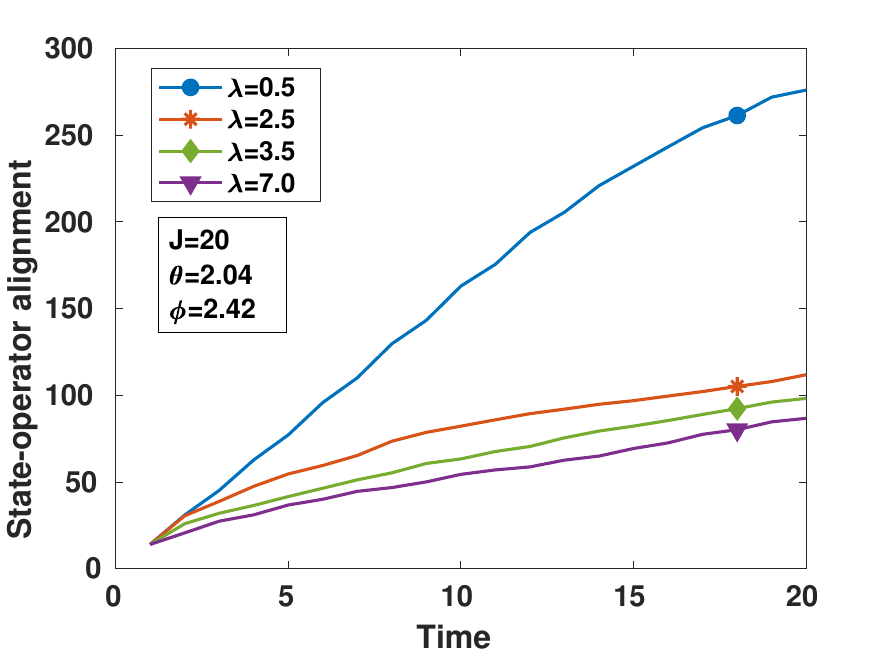}%
	}
	\caption{{(a) Fidelity of a random coherent state with chaoticity at different times. (b) State-operator alignment for different chaoticities against time. }}
	\label{fig:coherent}
\end{figure*}

A fascinating question about chaos and tomography is how chaos and randomness generation affect spin coherent state reconstruction. Coherent states have strikingly different behavior from random states in that they are localized in phase space as shown in Fig.~\ref{fig:coherent}(a). The observable measured is more scrambled in the operator space with more chaos in the dynamics. As a result,  more states make up the operator in the coherent state basis. The amount learned from a measurement about any coherent state of interest is low. Therefore, coherent states show the opposite trend of random states with respect to chaos.

We can explain the behavior of these localized states more quantitatively using alignment of the measured observable with the state. We define the \textit{alignment matrix} as follows
\begin{equation}
\tilde{\mathcal{S}}=
\begin{pmatrix}
r_{1}\tilde{\mathcal{O}}_{11} & r_{2}\tilde{\mathcal{O}}_{12} & .. & .. & r_{d^2-1}\tilde{\mathcal{O}}_{1d^2-1}\\
r_{1}\tilde{\mathcal{O}}_{21} & r_{2}\tilde{\mathcal{O}}_{22} & .. & .. & r_{d^2-1}\tilde{\mathcal{O}}_{2d^2-1}\\
.. & .. & .. & .. & ..\\
.. & .. & .. & .. & ..\\
r_{1}\tilde{\mathcal{O}}_{n1} & r_{2}\tilde{\mathcal{O}}_{n2} & .. & .. & r_{d^2-1}\tilde{\mathcal{O}}_{nd^2-1}
\end{pmatrix} 
\end{equation}
{where $\tilde{\mathcal{S}}_{n\alpha}=r_{\alpha} \tilde{\mathcal{O}}_{n\alpha} = r_{\alpha} \mathrm{Tr}[\mathcal{O}_{n}E_{\alpha}]$, and $\mathcal{O}_n=U^{\dagger n}\mathcal{O}U^{n}$. }
Then the extent of the operator alignment along the state of interest is given by $\mathrm{Tr}[\mathcal{T}]$, where $\mathcal{T}=\tilde{\mathcal{S}}^{T}\tilde{\mathcal{S}}$.

 Figure~\ref{fig:coherent}(b) shows that as chaotcity increases, the operators are less aligned along the state, and measuring them give less information about the state. This results in lower fidelity reconstruction for high chaoticity dynamics.
Details of this investigation are in \citep{sahu2022revisiting}. 
\section{Discussion}

Quantum tomography is a resource-intensive process of fundamental importance in quantum information theory. The challenge is to accomplish this in an efficient manner and research is focused on optimizing  protocols. Many techniques like compressed sensing \citep{gross2010quantum} or the taking advantage of the positivity constraint \citep{kalev2015power} focus on the prior information available in state estimation. In this chapter, we took a different approach and showed that with a dynamics that is not informationally complete, we can still get very high fidelities in quantum state reconstruction.
In particular, what we have seen in this chapter is that random unitaries diagonal in a particular basis do almost as good as Haar random unitaries in terms of fidelity of  state reconstruction. This is despite missing out on the information from $d-1$ dimensional subspace of the operator space. We quantified the rate of information gain using collective Fisher information and used Shannon entropy to quantify uniformity in operator sampling. We gave statistical bounds on information gain and also discussed how close diagonal random unitary dynamics come in saturating these bounds.
Finally, we saw that asymptotic evolution using  Haar random unitaries is modeled remarkably well by the Wishart-Laguerre orthogonal ensemble. We also obtained an intuitive understanding of the vector space visited by the random unitary maps considered. Thus our work is an important contribution towards the applications of random matrix theory in quantum information.

One interesting question that arises from our study is the performance of quantum process tomography using states generated by random diagonal unitaries as inputs. Quantum process tomography is the process of determining the trace-preserving completely positive map that is applied on the system and therefore $d^4-d^2$ real numbers are required to completely characterize it.
Using continuous measurement quantum process tomography, how close does a random dynamics or random dynamics diagonal in a fixed basis come in getting an accurate description of the map?

The flip side of quantum tomography is quantum control. One requires an informationally complete set for perfect state reconstruction. Similarly, such an informationally complete dynamics
will be able to steer an initial state to any target state in the Hilbert space.  
 The ability of random diagonal unitaries to generate information in $d^2-d+1$ dimensions tells us as to what target states are achievable starting from a fiducial state.  
 
 Randomized
benchmarking is used to 
estimate the fidelity between the applied map
and the target unitary, in the presence of errors. How well do randomized benchmarking protocols work when one only has random diagonal unitaries at disposal? Is there a way to perform randomized benchmarking with a restricted set of unitaries? These are questions we like to address in the future.

\chapter{Exponential speedup in measuring OTOCs}
\label{chap:chap5}
Connections between non-integrability, many-body physics, complexity, ergodicity, and entropy generation are the cornerstones of statistical mechanics. The aim of quantum chaos is to extend these questions in the quantum domain. Foundational works in this aspect include semiclassical methods connecting classical periodic orbits to {the} density of states
 level statistics \citep{berry1977level}, characteristics of Wigner functions \citep{Berry77a},  quantum scars in chaotic phase spaces \citep{heller1984bound} and connections to random matrix theory.
Search for these footprints of chaos, and characterization of ``true" quantum chaos, independent of any classical limit, has important consequences both from a foundational point of view as well for quantum information processing. 
For example, such studies address complexity in quantum systems and play a potentially crucial role in information processing protocols like quantum simulations that are superior to their classical counterparts. 

Characterization of chaos in the quantum domain has been much contested since, unlike its classical counterpart, unitary quantum evolution preserves the overlap between two initial state vectors and hence rules out hypersensitivity to initial conditions. However, a deeper study reveals chaos in quantum systems. 

These issues have been extensively studied in the last few decades and several quantum signatures of classical chaos have been discovered. This interestingly coincides with exquisite control of individual quantum systems in the laboratory and the ability to coherently drive these systems with non-integrable/chaotic Hamiltonians. Recent trends include studies involving connections of quantum chaos to {out-of-time-ordered} correlators (OTOC) and the rate of scrambling of quantum information in many-body systems with consequences ranging from the foundations of quantum statistical mechanics, quantum phase transitions, and thermalization on the one hand to information scrambling inside a black hole on the other hand \citep{nicole2,manybody1, manybody2, manybody3, manybody4, qgravity1, chaos1,chaos2,SivaAL-2019,arul2,arul3,arul_vaibhav,santos, shock1, pawan, shenker2014black, kitaev1, kitaev2}.

OTOCs have been much talked about in the quantum information circle recently and a number of ways to measure OTOCs have been proposed including a protocol employing an interferometric scheme in cold atoms \citep{swingle2016measuring}.  An alternative method involving two-point projective measurements was proposed \citep{campisi}, giving a scheme for the measurement of OTOCs using the two-point measurement scheme, developed in the field of non-equilibrium quantum thermodynamics  elucidating the connections between information scrambling and thermodynamics.
While various experimental schemes are reported in \citep{meas1,meas2,meas3,nicole1,lata_group}, the focus of this chapter is to give an exponentially fast quantum algorithm to measure OTOCs. Measurement of OTOCs for an Ising spin chain in an NMR simulator has been reported \citep{PhysRevX.7.031011,wei}. A many-body time-reversal protocol using trapped ions has been proposed and demonstrated \citep{garttner2017measuring} which though universal is not scalable.
The focus of the literature is on infinite temperature OTOCs, an observation that will be important for our algorithm.

 { In order to explore any quantum signatures of chaos, one has to numerically process data structures whose computational complexity scale 
	exponentially with the number of qubits required to simulate the system. 
	In this chapter, we give a quantum algorithm that gives an exponential {speedup} in measuring OTOCs provided that the number of gates, $K$, required in the decomposition of the times evolution operator of the system scales \textit{polynomially} with $n$, where $n$ is the number of qubits used in the implementation and, $N$, the dimension of the Hilbert space with $N = 2^n$. This implies that the algorithm measures the OTOCs in a time that scales as \textit{poly(n)}, which is exponentially faster than any classical algorithm. Furthermore, we give a method for efficient estimation of gate fidelities.
	Our algorithm is based on the Deterministic Quantum Computation with one pure qubit (DQC1) algorithm, which is the first mixed state scheme of quantum computation. Therefore, this can be naturally implemented by a high-temperature NMR based quantum information processor.  It involves a deterministic quantum control of one qubit model, using scattering circuit \citep{knill,scattering}. This algorithm is also called the `power of one qubit' as the main primary resource required for this algorithm is one pure qubit. Moreover, the essential part of simulations, state initialization, and readout, that are often quite involved in certain models of quantum computation \citep{van2001powerful}.  We give a quantum circuit to evaluate OTOCs—which bypasses the need to prepare a complex initial state and can be accomplished by a very simple measurement.
	Applications include estimation of fidelity decay and density of states in quantum chaos \citep{exponential, PhysRevA.68.022302}, computing Jones {polynomials} from knot theory \citep{shor2007estimating, jones2004nuclear} and phase estimation in quantum metrology \citep{PhysRevA.77.052320}. Although the DQC1 model of quantum information processing (QIP) is believed to be less powerful than a universal quantum computer, its natural implementation in high-temperature NMR makes it an ideal candidate for probing OTOCs and mixed state quantum computation protocols.}

\section{Out-of-time-ordered correlators (OTOCs)}
OTOC was first proposed by Larkin and Ovchinnikov while studying superconductivity in the semiclassical limit \citep{larkin}.
 They later reemerged in the study of many-body systems \citep{manybody1,manybody2, manybody3, manybody4} quantum gravity \citep{qgravity1} and  quantum chaos \citep{chaos1,chaos2, shock1, pawan, shenker2014black, shenker3, kitaev1, kitaev2}. In quantum information literature, OTOC is used as a  probe to study the dynamics of information. One can probe the macroscopic irreversibility of the dynamics,  the spread of quantum information from a localized point to the rest of the system via entanglement and correlations, and also the aspects of thermalization \citep{Swingle-2018,scrambling1, scrambling2,thermal1, del_campo}. Consider a chain of interacting spins. Then a correlator of two operators acting at two different sites can be defined as
\begin{equation}\label{eqq}
C_{W,V}(\tau)=\dfrac{1}{2}\langle[W(x, \tau), V(y, 0)]^{\dagger}[W(x, \tau), V(y, 0)]\rangle, 
\end{equation}
where the local operators $W$ and $V$ are unitary and/or Hermitian that act on sites $x$ and $y$ respectively and $W(x, \tau)=U^{\dagger}(\tau)W(x, 0)U(\tau)$ is the Heisenberg evolution of operator $W$ under time evolving operator $U(\tau)$. The average is taken with respect to the thermal state at some temperature which we take to be infinite. In particular, if the operators $W$ and $V$ are unitary, the above equation becomes,
\begin{equation}
 C_{W, V}(\tau)=1-\mathtt{Re}\langle W(x,\tau)^{\dagger}V(y, 0)^{\dagger}W(x, \tau)V(y, 0)\rangle.
\end{equation}

In classical physics, the chaos is defined as the sensitive dependence on initial conditions. If we replace $W$ and $V$ in the Eq.~(\ref{eqq}) with position($Q$) and momentum ($P$) operators, and taking a semi-classical limit, we notice that $\hbar^2\{Q(\tau), P(0) \}^2=\left(\hbar\frac{\delta Q(\tau)}{\delta Q(0)}\right)^{2} \approx \mathrm{exp}(2\lambda \tau)$. The quantum-classical correspondence principle implies that the quantity $C_{W, V}(\tau)$ grows exponentially till the Ehrenfest time ($\tau_{Eh}$). However, unlike the classical systems, the lyapunov exponent($\lambda $) calculated from OTOC is bounded by $\frac{2\pi}{\beta}$ \citep{chaos1}. Beyond the $\tau_{Eh}$, the quantum corrections start dominating and the quantum-classical correspondence breaks down. 

An interesting feature of OTOC is that it measures the growth on support of an initially localized operator  over the system as it evolves in Heisenberg fashion \citep{pawan,ope1, ope2, ope3, ope4, ope5}. Consider a pair of local operators $W$ and $V$ that act on different subspaces of total Hilbert space($\mathcal{H}$) under a chaotic time evolution $U(\tau)=\exp(-iH\tau)$. We assume that the Hamiltonian is generic with local interactions. Under this evolution, the operator $W$ will evolve in time and it can be expanded in Taylor series around $\tau=0$ as 
\begin{eqnarray}\label{eqq2}
 W(\tau)&=&\sum_{n}\dfrac{\tau^n}{n!}\dfrac{d^n W}{d\tau^n}\nonumber\\
&=& W(0)+i\tau[H, W]+(i\tau)^2[H,[H, W]]+...
\end{eqnarray}
This implies that the operators $W(\tau)$ and $V$ in general do not commute for time $\tau \neq 0$. For example, consider one dimensional Ising spin chain with nearest-neighbor interactions. Let $W(i, \tau=0)=\sigma_{z}^i$ acts on site $i$ at time $\tau =0$. On substituting $W$ in second line of the series in the Eq.~(\ref{eqq2}), the first order commutator will give us the sum of products of local operators acting on the sites $i-1, i$ and $i+1$ \textit{i.e.,} $[H, \sigma_{z}^{i}]=f(i-1, i, i+1)$. As time flows, the higher ordered nested commutators also will contribute to the expansion of $W(\tau)$ thus making the quantity $[W(\tau), V]\neq 0$ \citep{shock1}.  

Lieb and Robbinson \citep{lieb1972finite} showed that for short range interacting Hamiltonians, the quantity $C_{W, V}$ is bounded i.e $C_{W,V}(\tau)\leq ce^{-a(i-v\tau)}$. Where $a$ and $c$ are constants and $v$ is called Lieb-Robbinson velocity. This bound on OTOC imply a light cone like structure in quantum lattice models. 
 it is worthwhile to note that the growth of the OTOC
is a  quantum measure, can be used in systems with no 
obvious classical limits.
\section{Determinstic Quantum Computation with one pure qubit (DQC1)}
{ Single qubit quantum computation, although limited in applicability is interesting from a fundamental point of view. Despite involving minimal entanglement, DQC1 gives an advantage over classical computing. It has been shown that none of the classical models simulate DQC1 efficiently \citep{animesh}}.  In this model, we start with a known state of an ancilla or probe qubit and couple it to the system. If the system state is known, we can perform spectroscopy of the controlled operation acting  on the system. Else if the operation is known, one can do tomography with the same circuit \citep{scattering}.  In both cases, a measurement performed on the ancilla qubit after the interaction reveals information about the system or the operation. 
{The circuit diagram for DQC1 is shown below.}
	\begin{figure}[!ht]
		\centering
	\begin{quantikz}
		\lstick{$\ket{0}$} & \gate{H} &  \ctrl{1}   & \gate{H} & \meter{} \\
		\lstick{$\ket{\psi_0}$ or $\mathbb{I}/2^{n}$}  & \qwbundle[alternate]{} & \gate{U}  \qwbundle[alternate]{}& \qwbundle[alternate]{} & \qwbundle[alternate]{} 
	\end{quantikz}
\caption{Quantum circuit for the DQC1 protocol (when the input is $\mathbb{I}/2^{n}$). The circuit gives an efficient algorithm for trace estimation of a unitary with only one qubit of quantum information.}
\end{figure}
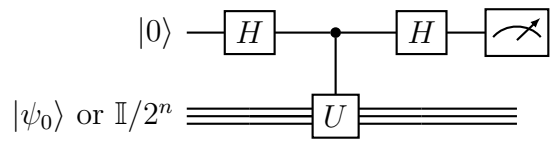

{The top qubit (the pure qubit that is also the control qubit) is acted upon by a Hadamard gate. This transforms state $\ket{0}$ to $\frac{(\ket{0} +\ket{1})}{\sqrt{2}}$. Then a controlled unitary $U$ is applied followed by another Hadamard gate. It is to be noted that the controlled unitary $U$, and the state 
 $\ket{\psi_0}$ can belong to an arbitrarily large Hilbert space.
Measuring the control qubit, we observe  $\ket{0}$ and  $\ket{1}$ with probabilities}
\begin{align}
P(0)= \frac{1}{2}(1+ \mathtt{Re} \bra{\psi_0} U \ket {\psi_0})  \nonumber \\
P(1)= \frac{1}{2}(1 - \mathtt{Re} \bra{\psi_0} U \ket {\psi_0}) .
\end{align}

 {Instead of a pure state  $\ket{0}$, if the lower set of qubits are in a completely mixed state, with density matrix, $\rho =  \mathbb{I}/2^{n}$, we get }
\begin{align}
P(0)= \frac{1}{2}(1+ \frac{1}{2^n}\mathtt{Re}  (\Tr U))  \nonumber \\
P(0)= \frac{1}{2}(1 - \frac{1}{2^n}\mathtt{Re} (\Tr U)).   
\end{align}

{By a trivial modification of this scheme, one can make these probabilities depend on $\mathtt{Im} (\Tr U)$
and therefore, this gives a quantum algorithm to estimate the trace of a unitary matrix. $L$ measurement of the top qubit will give us an estimate of the trace with fluctuations of size $1/\sqrt{L}$. Therefore, to achieve an accuracy $\epsilon$ one requires $L \sim 1/\epsilon^2$ implementations of the circuit. If $P_e$ is the probability that the estimate departs from the actual value by an amount $\epsilon$, then one needs to run the experiment $L \sim \log(1/P_e)/\epsilon^2$ times.
This accuracy in the estimate does not scale with the size of the unitary matrix and hence provides an exponential {speedup} over  classical algorithms, provided the unitary admits an efficient gate decomposition. It is known that if the gate decomposition scales as \textit{poly(n)}, the controlled version of these gates also scales \textit{polynomially} in $n$.
Moreover, the result is obtained by a mesaurement of only the top qubit and hence independent of the size of the readout register.}
As a last remark, it is worthwhile to note that, while we have assumed the probe qubit to be in a pure state, this is not necessary. With the probe qubit in a state, $\alpha \ket{0}\bra{0} + \frac{(1- \alpha)}{2}\mathbb{I} $, the model with a tiniest fraction of a qubit is computationally equivalent to the DQC1 circuit described above.
More specifically, the number of runs of the trace estimation algorithm goes as  $L \sim \log(1/P_e)/\alpha^2\epsilon^2$. Therefore, as long as $\alpha$ is non-zero, the circuit provides an efficient estimate of the trace.


\section{Using DQC1 to calculate OTOC}
\begin{figure*}
\centering
  \includegraphics[width=12cm,height=2.5cm,angle=0]{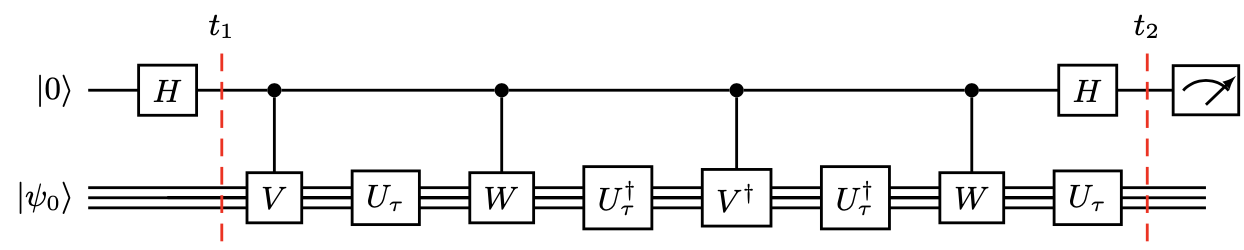}%
  
\caption{  Evaluates the expectation value of OTOC with respect to $\ket{\psi_0}.$ Time progresses along the horizontal line.  The top register is the single-qubit ancilla or probe. The bottom register is the system on which controlled gates act. When the probe qubit is $\ket{0}$, the system is left unchanged, whereas when the probe is $\ket{1}$, controlled operations take place. Measurement of $\sigma_z$ or $\sigma_y$ is performed on the probe qubit, in the end, revealing the value of OTOC.   }

\label{fig1}
\end{figure*}

We now adapt the DQC1 algorithm to measure OTOCs. This is shown in the circuit in Fig.~\ref{fig1}.
Here we initialize the probe to $\ket{0}$ and for simplicity let us say the system state is prepared in a pure state  $\ket{\psi_0}$. The controlled gates act on the system only when the control qubit is $\ket{1}$. $H$ is the Hadamard gate, and $U_\tau$ is the unitary determined by a Hamiltonian which evolves the system up to time $\tau$.  The state of the probe $+$ system at time $t_1$ is $\frac{(\ket{0} +\ket{1})}{\sqrt{2}} \otimes \ket{\psi_0}.$ After the interaction, at time $t_2$, the combined state is $\frac{1}{2}\ket{0} \otimes (1+\mathcal{U}) \ket{\psi_0}+\frac{1}{2}\ket{1} \otimes (1-\mathcal{U}) \ket{\psi_0} $  where $\mathcal{U}= W_\tau^\dagger V^\dagger W_\tau V.$  After the action of the second Hadamard on the probe qubit, measurement of $\sigma_z \otimes \mathbb{I}$, with $\sigma_z$ on the probe qubit yields $\mathtt{Re} \bra{\psi_0}W_\tau^\dagger V^\dagger W_\tau V \ket{\psi_0}$ and measurement of $\sigma_y$ on the probe yields $\mathtt{Im}\bra{\psi_0}W_\tau^\dagger V^\dagger W_\tau V \ket{\psi_0}$. If we perform the circuit sufficiently many times, then we get
\begin{align}
    \langle \sigma_z \rangle &= \mathtt{Re} \bra{\psi_0}W_\tau^\dagger V^\dagger W_\tau V \ket{\psi_0} \nonumber \\ \langle \sigma_y \rangle &= \mathtt{Im} \bra{\psi_0}W_\tau^\dagger V^\dagger W_\tau V \ket{\psi_0}.
\end{align}
Thus we have obtained the OTOC values. As mentioned previously, assuming we have an efficient gate decomposition and fix the size of fluctuations in our answer, the complexity of this algorithm does not scale with the dimension of Hilbert space of the physical system under consideration. This is not an unreasonable assumption as efficient decomposition of some quantized chaotic systems is known \citep{benenti2001efficient, emerson2003pseudo, PhysRevA.57.1634} and used in quantum simulations \citep{PhysRevA.68.022302, emerson2002fidelity}.
In the above, the inherent assumption is that the system state at the beginning  is perfectly known. By taking the initial state $\ket{\psi_0}\bra{\psi_0}$ to be completely mixed, that is proportional to $\mathbb{I}$, we get the trace of OTOC, which is the measurement with respect to a thermal state at infinite temperature. Therefore, OTOCs with respect to the thermal state at infinite temperature is a perfect candidate for the implementation with DQC1, that employs only 1 qubit of quantum information, and hence a happy accident.

\section{Estimating the eigenvlaue spectrum of OTOC}
\begin{figure}
    \centering
    \includegraphics[width=11cm,height=3cm,angle=0]{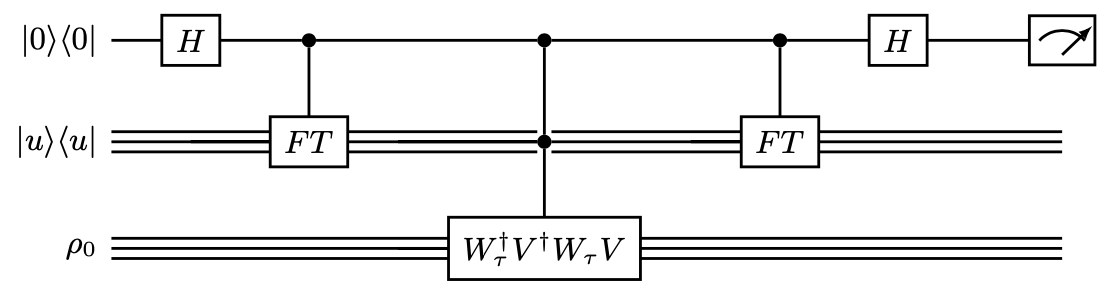}%
  
    \caption{The circuit for obtaining the spectral density of OTOC. Now there are two ancillas. Controlled Fourier transform is applied twice on the second ancilla. The operation $W_\tau^\dagger V^\dagger W_\tau V$ which acts on the system is written in a condensed form and should be implemented by decomposing into constituent gates as in  Fig.~\ref{fig1}. Only the single-qubit probe/ancilla is measured in the end as before.}
    \label{fig_2}
\end{figure}

Not only the expectation value of OTOCs, the eigenvalue spectrum of OTOCs is also of interest. Just like  energy eigenvalue spacing for integrable and chaotic systems form distinct distribution,  the level spacing of OTOCs also shows marked difference \citep{spectrum2, spectrum1}. One can obtain the eigenvalue density of OTOCs using a DQC1 algorithm. The circuit is similar to the previous one. But now, apart from the $n$-qubit register for the system,  we also need an extra $n_2$-qubit ancilla  and  perform discrete  Fourier transforms. The circuit is shown in Fig.~\ref{fig_2}.


In this circuit, $\ket{u}$ is the initialized state of the second ancilla register of $n_2$ qubits, with the expectation value of OTOC equal to $u$. { The OTOC, $W_\tau^\dagger V^\dagger W_\tau V$, which can be implemented as before, along with two quantum Fourier transforms collectively form the net unitary operation, $\mathcal{U'}.$ After the first Fourier transform, $\ket{u} \rightarrow \sum_{s=0}^{N_2-1} \mathrm{exp}(i2 \pi u s/N_2) \ket{s} $, where $N_2=2^{n_2}$ and$ \lbrace\ket{s}\rbrace$ are the transformed basis states. The controlled OTOC unitary action yields $\ket{s}_2\ket{\psi_0}_3 \rightarrow \ket{s}_2 (W_\tau^\dagger V^\dagger W_\tau V)^s\ket{\psi_0}_3$ when the first register is $\ket{1}$. No change takes place when probe qubit is $\ket{0}$.} Second Fourier transform completes the circuit.
Measuring $\sigma_z$ and $\sigma_y$ on the probe qubit as before, we get
{
\begin{align}
f(u)=& \Tr(\mathcal{U'}\rho_0)\\=& \frac{1}{N_2} \sum_{s=0}^{N_2-1} \mathrm{exp}(i4 \pi u s/N_2) \Tr[(W_\tau^\dagger V^\dagger W_\tau V)^s\rho_0].
\end{align}
}
Spectral information is now contained in the phases, and can be estimated \citep{scattering}. Normalized $f(u)$ can be directly mapped to the spectral density of eigenvalues in a region around $u$, with the resolution and range determined by the number of ancilla qubits $n_2$ and the time scale $\delta$ \citep{scattering}. As in the previous case, the DQC1 implementation provides an exponential speed up in obtaining spectral density over any known classical algorithm.
\section{Estimation of gate fidelity}
Another important application of the single-qubit computation is in determining the fidelity of unitary gates in quantum circuits. { Quantum process tomography is an exact way to quantify the errors \citep{nielsen2002quantum}. But it is computationally demanding, and becomes infeasible to perform, once the system size gets large. Here we implement a scalable algorithm to characterize the errors originally introduced in \citep{benchmarking1}. Characterization of errors in implementing the gates is important from the perspective of quantum control.}

Let us say the target unitary we want to implement is $U,$ such that after its action, the state $\rho$ gets mapped to $U \rho U^\dagger.$ However, because of the errors that creep in, the actual outcome becomes $U' \rho U'^\dagger.$  In real life implementations, there are always errors due to environmental decoherence and other external noises. They can be modeled by a completely positive trace-preserving map \citep{benchmarking1}. 
\begin{figure}
\centering
  \includegraphics[width=12cm,height=4cm,angle=0]{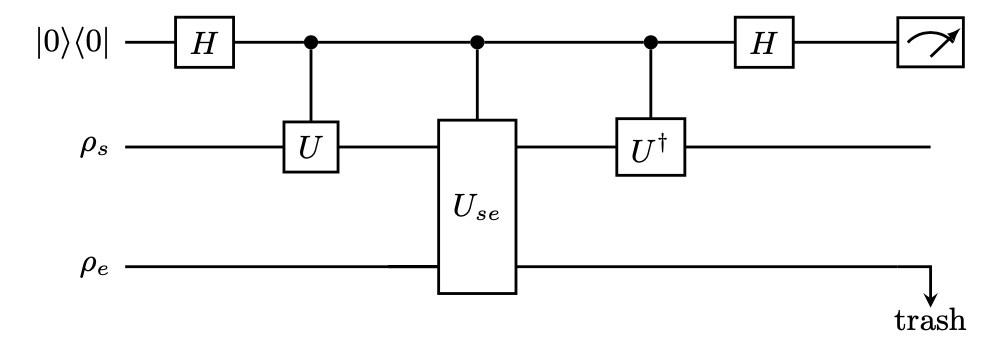}%
\caption{ The circuit for obtaining the fidelity of a unitary gate $U$. The errors are represented by a quantum channel. The system state is initialized to $\rho_s= \ketbra{\psi}{\psi}.$ This circuit finds the overlap between the state after the actual implementation of the gate with the ideal output. The state of the lower ancilla after the combined evolution with the system according to $U_{se}$ is discarded.}
\label{fig4}
\end{figure}
\begin{figure}
    \centering
    \includegraphics[width=11cm,height=5.5cm,angle=0]{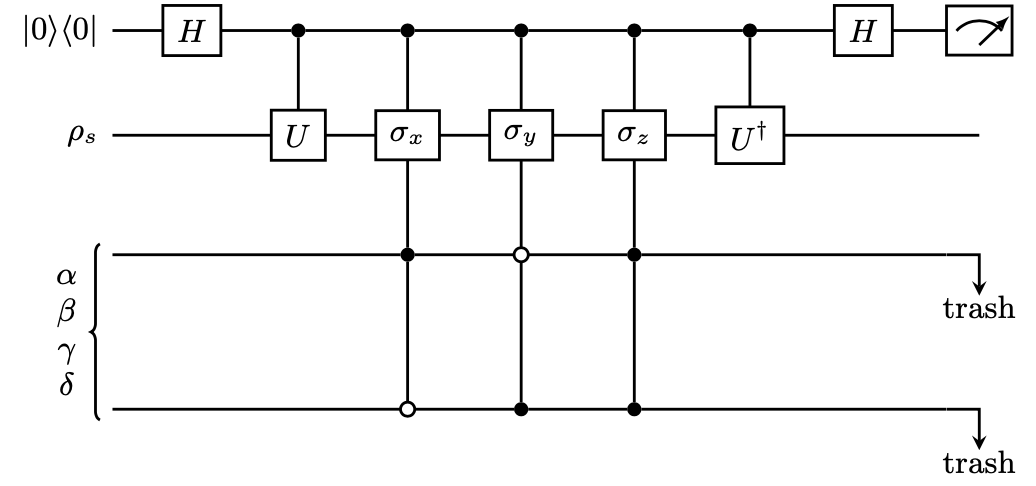}%
  
    \caption{ Computing Haar averaged gate fidelity of $U.$ Here a single qubit system is initialized to $\rho_s= \ketbra{\psi}{\psi}$. The second ancilla, co-evolved with the system, consists of two qubits initialized to $\alpha \ket{00}+\beta \ket{01}+ \gamma \ket{10}+ \delta \ket{11}, $ with $|\alpha|^2= p$ and $|\beta|^2=|\gamma|^2=|\delta|^2=\frac{(1-p)}{3} $. This achieves the standard depolarizing channel \citep{depolarizing}. Final measurements in the single qubit ancilla gives Haar averaged gate fidelity.}
    \label{fig5}
\end{figure}
Let us denote the CPTP map characterizing the errors by $\Lambda$. Then the fidelity of the gate $U,$ with respect to the state $\rho_s= \ketbra{\psi}{\psi}$  is given by 
\begin{equation}
   F_g(U,\Lambda,\psi)= \bra{\psi} U^\dagger \Lambda(U\ketbra{\psi}{\psi}U^\dagger) U \ket{\psi},
\end{equation}
 where $\Lambda(\rho)= \sum_k A_k \rho A_k^\dagger.$ Figure~\ref{fig4} demonstrates how to obtain $F_g$ using DQC1. We have an extra ancilla initialized to $\rho_e.$ Errors are represented by  CPTP maps, implemented by a collective evolution of the system and ancilla and then discarding the environment. 
 To obtain the average gate fidelity, we need to average over all pure states according to Haar measure.
 \begin{equation}
     \mathbb{E}_\psi(F_g)= \int d\psi \bra{\psi} U^\dagger \Lambda(U\ketbra{\psi}{\psi}U^\dagger) U \ket{\psi}.
 \end{equation}
 Equivalently, one can Haar average over all the CPTP maps instead, and appeal to the identity \citep{benchmarking1}
 \begin{equation}
     \mathbb{E}_\psi(F_g)=\mathbb{E}_U(F_g),
 \end{equation}
 where $\mathbb{E}_U(F_g)= \Tr \left[\rho \Lambda_{avg}(\rho)\right].$  
 The Haar averaged CPTP map $\Lambda_{avg}$ is a depolarizing channel and can be implemented experimentally.  Its action on a state $\rho$ is given by
 \begin{equation}
     \Lambda_{avg}(\rho)= p\rho+ (1-p) \frac{\mathbb{I}}{d},
 \end{equation}
 where the noise strength parameter $p$ is completely characterized by the set of Kraus operators $\lbrace A_k \rbrace$ of the map. { When the dimension of the Hilbert space is large, the phenomenon of concentration of measure along with Levy's lemma gives a rather nice result. If one chooses the target unitary to be implemented uniformly at random, then its fidelity is exponentially close to the average fidelity $\mathbb{E}_U(F_g)$ \citep{benchmarking1}.
 \begin{equation}
    F_g(U,\Lambda,\psi)= \mathbb{E}_U(F_g)+ O\left(\frac{1}{\sqrt{d}}\right),
 \end{equation}
 where the initial system state $\ket{\psi}$ can be fixed arbitrarily.}

 To achieve the Haar averaged gate fidelity using DQC1 circuit, we need to simply replace the general quantum operation in Fig.~\ref{fig4} by a depolarizing channel. 
 A sample circuit for a single qubit system is illustrated in Fig.~\ref{fig5}. The second ancilla consists of two qubits initialized to $\sqrt{p}\ket{00}+ \sqrt{\frac{1-p}{3}}\left(\ket{01}+\ket{10}+\ket{11}\right).$
 Measurement of the single-qubit probe  at the end of the circuit reveals $\mathbb{E}_U(F_g)$, and hence the average gate fidelity $\mathbb{E}_\psi(F_g).$ 


\section{Conclusions}
We have shown that using a single bit of quantum information, one can estimate OTOCs  exponentially faster than classical methods.
In the spirit of the slogan, ``classical chaos generates classical information, as captured by classical Lyapunov exponents and the classical Kolmogorov-Sinai entropy, quantum chaos generates quantum information", leading to the growth of OTOCs (till the Ehrenfest time), {which are} popular quantifiers for this. In this chapter, we have given an efficient quantum algorithm for estimating OTOCs and capturing the growth of quantum complexity. {We also have constructed a simple quantum circuit to find out fidelity of unitary gates. Both these problems of estimating OTOCs and  benchmarking  quantum gates are well sought after in quantum information science.}
One possible avenue is to estimate the semiclassical 
formulas, like the Gutzwiller trace formula on a quantum computer. 
There are existing algorithms for this \citep{georgeot08} that give a 
polynomial speedup over similar implementations on a classical 
computer. 

We aim to explore the possibility of such computations using the DQC1 model of quantum computation, which can even operate on highly mixed initial states. One can also consider a perturbed OTOC where the operator $W_\tau^\dagger$ that occurs in  $W_\tau^\dagger V^\dagger W_\tau V$,
 undergoes time evolution with a slightly perturbed Hamiltonian as compared to  $W_\tau$ and therefore provides a direct analogue to classically chaotic systems under stochastic noise. {
Moreover, understanding the power behind DQC1 is still an open question.
 Future directions include determining the nature of resources 
quantum mechanics provides for information processing tasks that are 
superior to their classical counterparts as well as other avenues where mixed state quantum computation can be applied.}

\chapter{Concentration of measure and applications}
\label{chap:chap6}
The laws of physics and also several other natural phenomena often play out in very high dimensional spaces. Classical and quantum statistical mechanics, dynamical systems, 
classical and quantum information processing \citep{cover2012elements, nielsen2002quantum}, biological evolution and adaptive speciation in genotypic spaces and even viral and RNA evolution \citep{eigen1988molecular, nowak1992quasispecies}
are all interestingly characterized by a  higher dimensional geometry. Hilbert spaces in quantum information and computation, phase space in classical statistical mechanics, genomic spaces in evolutionary biology, and sequence spaces in viral/RNA quasispecies evolution and typical sequences in Shannon's information theory are some examples. On the one hand, this leads to an increase in complexity and, therefore, an exponential amount of resources for an accurate and complete description of such phenomena. On the other hand, such large dimensional spaces are perfect candidates to make statistical arguments and describe the system using certain average variables that attain near-equilibrium values for time scales of interest. Foundations of statistical mechanics, ergodic theory, random matrix theory in nuclear physics, and analysis of higher dimensional dynamical systems all rely on statistical arguments where fluctuations from equilibrium values become vanishingly small in the limit of higher dimensions of underlying space.

The concentration of measure is a general phenomenon applied in statistical mechanics, probability theory, and measure theory and deals with 
quantifying how close is a random variable to its mean.
Consider a large number of scalar random variables, $X_1, ..., X_n$, with mean and variance of order one $(O(1))$. Then how much  does the sum, 
$S_n = X_1 + ... + X_n$, deviate from the mean? If each component  varies with order $O(1)$, the sum varies in the interval $O(n)$.
However, the phenomenon of concentration of measure says that, given a sufficient amount of independence between constituent random variables,  $X_1,X_2, ..., X_n$, the deviation of the sum is typically in an interval of size $O(\sqrt{n})$ and 
sharply concentrates in a narrow range about the mean. 
 Hoeffding's inequality \citep{hoeffding1994probability}, expresses this in a mathematical form as 
 \begin{equation}
\textrm{Prob}\left\lbrace  |S_n- \mathbb{E}[{S_n}] |\geq \epsilon \right\rbrace \leq  \mathrm{exp} \left({- 2n \epsilon^2}\right),
\label{Hoeff}
\end{equation}
where $ \mathbb{E}[{S_n}]$ stands for expectation value.
 The phenomenon of concentration of measure applies not only to linear expressions such as the sum but to more general functions, $F(X_{1}, ..., X_{n})$, that might be a non-linear combination of the component random variables. The basic intuition behind the concentration of measure is that the independence of constituent random variables makes it hard for all of them to pull together to drive the function $F(X_{1}, ..., X_{n})$ significantly away from its mean. 

As stated above, the phenomenon of concentration of measure is quantified in terms of large deviation inequalities that give upper bounds on the probability that a random variable deviates by a certain amount from its mean. This formulation, in its general form was proposed by {Talagrand} \citep{talagrand1996new, talagrand1996new1}, who observed that if a function depends on several independent random variables in a balanced and smooth way, the effect of randomness evens out, rendering the function to be essentially a constant with weak fluctuations about its mean.

Another way of looking at the measure concentration phenomenon is from the perspective of higher dimensional geometry, especially an $n$- dimensional hypersphere \citep{bengtsson2017geometry}. It is well known that a random point in an $n$-dimensional ball, $\textbf{B}^n$ is likely to be situated close to the boundary, ${S}^{n-1}$. Therefore, the skin of a higher dimensional fruit contains most of its mass. A closely related phenomenon, known as the Levy's Lemma, states that if one chooses any point on a higher dimensional sphere as the north pole and then selects a point
 randomly according to uniform measure on the sphere, it is likely to be close to the equator \citep{ledoux2001concentration, levy1951problemes}. 
 In another form, the lemma gives a lower bound on the likelihood that the value of a well behaved function on the surface of the hypersphere at a point picked uniformly at random, lies away from its expected value. 
 
While the connection between higher dimensional geometry over a hypersphere and the related concentration of measure phenomenon and the statistical description of higher dimensional systems has been well studied in 
quantum information theory \citep{Rigol16, bengtsson2017geometry}, random matrix theory and permutations, probability and statistical physics \citep{ledoux2001concentration}, they rely on a uniform spherically symmetric distribution over the hypersphere. 
 For example, Levy's lemma deals with the property of concentration of measure over the surface of a hypersphere when the upper bounds on the probability of deviation from the mean value are calculated for points picked uniformly at random on the hypersphere. Intuitively, Levy's lemma is a consequence of central limit theorem, and it has applications in statistical mechanics \citep{Rigol16} and entanglement theory \citep{page1993average, lubkin1978entropy} among others.

 In this chapter, we generalize Levy's Lemma to get measure concentration inequalities for points picked 
   according to any Lipschitz distribution, including uniformly random distribution as its special case.
  Thus, our work liberates us from uniform measure over the hypersphere and the requirement of spherical symmetry.

A reasonable question to ask is, where in nature can one find such distributions over the hypersphere? We demonstrate applications of our results to  quantum information theory when higher dimensional spaces are encountered. The effects of concentration of measure in evolutionary biology is discussed in appendix \ref{appendixA}.
   
\section{Measure concentration on a higher dimensional sphere}
\label{MC}
In this section, we discuss the concentration of measure on a higher dimensional hypersphere. The concentration function denoted by $\alpha_X$ captures the concentration of measure on a metric measure space $X$. For an $\epsilon > 0$, 
\begin{equation}
\alpha_X(\epsilon) := \mathrm{sup} \left\lbrace \mu(X \setminus N_\epsilon(S)) \: | \:\mu(S)=1/2 \right\rbrace, 
\end{equation}
where \begin{equation}N_\epsilon(S) = \left\lbrace  x \in X \: | \: \exists s \in S : d(s,x) < \epsilon \right\rbrace.
\end{equation}
When $\alpha_X(\epsilon)$ is small, the measure on $X$ is said to be highly concentrated. Levy's lemma is a statement on the concentration of measure in higher dimensional metric measure spaces.
We start by asking a question - what is the probability that a point taken at random on a higher dimensional sphere  lies in a narrow belt surrounding any particular equator? According to Levy's Lemma \citep{ledoux2001concentration, levy1951problemes, milman2009asymptotic}, as the dimension of the sphere increases this probability approaches one. That is, the concentration function $\alpha_X$ approaches zero, called the concentration of measure phenomenon. A surprising result is when the measure of the higher dimensional sphere is uniform, almost all of its surface area is concentrated around its equator! 

Levy's lemma is useful in the study of entanglement in large bipartite systems. When the dimension of the Hilbert space is large, almost all pure bipartite states are shown to be maximally entangled. Such properties of typical states follow Levy's lemma in higher dimensional spaces. 
\\  
\\
\textbf{Levy's lemma}: Given a Lipschitz continuous function $f: S^{n-1} \rightarrow \mathbb{R}$ defined on a large dimensional hypersphere $S^{n-1}$, and a point $x \in S^{n-1}$ chosen uniformly at random. Then
\begin{equation}
\mu\left\lbrace x \in S^{n-1}: |f(x)- \mathbb{E}{f(x)} |\geq \epsilon \right\rbrace \leq 2 \mathrm{exp} \left(\frac{- kn \epsilon^2}{\eta^2} \right), \label{one}
\end{equation}
where $\mu$ denotes the measure of points on the hypersphere under consideration.  $\mathbb{E}{f(x)}$ denotes the average value of $f(x)$, $\eta$ is the Lipschitz constant of $f$, given by $\eta= \mathrm{sup}|\bigtriangledown f|$ and $k$ is a positive constant. We assume a uniform, normalized probability measure on the space. i.e., $\mu(S^{n-1})=1$.  When the hypersphere is of higher dimensions, the probability that $f$ differs from its expectation value by more than an $\epsilon$ for a point picked uniformly at random is close to zero \citep{gerken2013measure,madhok2019typicality, popescu2006entanglement}. We consider a slowly varying measure instead of a uniform measure on the sphere. In the following, we see that a similar result holds even when we pick a point on the sphere not uniformly at random but according to a Lipschitz density function. We analytically find the bounds on corresponding probability.  

\subsection{Generalisation of Levy's Lemma}

\textbf{Theorem 1}:
Given a Lipschitz continuous function $f: S^{n-1} \rightarrow \mathbb{R}$ defined on a large dimensional hypersphere $S^{n-1}$, and a point $x \in S^{n-1}$ chosen at random from another Lipschitz continuous probability distribution  $\rho(x)$. Then,

\begin{align}
\mu\left\lbrace x \in S^{n-1}:|f(x)- \mathbb{E}{f(x)}| \leq \epsilon \right\rbrace   & \geq (1-\epsilon)\left[ 1- \left( 2\textrm{exp}  \left(- \frac{kn \epsilon^2}{\eta^2}\right) + 2\textrm{exp}  \left(- \frac{kn \epsilon^2}{\eta'^2}\right)\right) \right], \label{eq:6} 
\end{align}
where $\mathbb{E}{f(x)}$ denotes the average $f(x)$, $\eta$ is the Lipschitz constant of $f$, given by $\eta= \mathrm{sup}|\bigtriangledown f|$. Similarly, $\eta'$  is the Lipschitz constant of $\rho(x)$ and $k$ is a positive constant. Intuitively speaking, this means that, for a high enough dimension,  concentration of measure results still hold good.   
When $n$ is large, and $ \epsilon$ tends to zero, $(1- \epsilon) \rightarrow 1$ and $\left[ 1- \left( 2\textrm{exp}  \left(- \frac{kn \epsilon^2}{\eta^2}\right) + 2\textrm{exp}  \left(- \frac{kn \epsilon^2}{\eta'^2}\right)\right) \right]   \rightarrow1$. Hence the RHS of   Eq.~(\ref{eq:6}) tends to 1.
For instance, when $n \epsilon^2$ is 1000, and $\eta=\eta'=1$, the exponential term comes out to be of the order of $10^{-8}$, which is really small.

\textbf{Proof:} The proof of this theorem appears in an appendix of \citep{madhok2019typicality}, which we reproduce here.
 $f(x)$ is a  Lipschitz continuous function and $\rho(x)$ is a Lipschitz continuous probability density function on the surface of $S^{n-1}$. The expectation value $\mathbb{E} \rho(x) $;
\begin{equation}
	\mathbb{E} \rho(x) =\frac{ \int_{S^{(n-1)}} \rho(x) d\mu}{\int_{S^{(n-1)}} d\mu} =1. \label{q}
	\end{equation}
	
	Here, $d\mu$ is a differential area element on the hypersphere. We consider the normalized metric measure space, i.e., $\int_{S^{(n-1)}} d\mu=1$. The numerator in Eq.~(\ref{q}) is also equal to 1, since it is an integral of a probability density over the whole space. If the argument of the function, $x$ is chosen uniformly at random,  Levy's lemma applies.
Let S denote $\left\lbrace x \in S^{n-1} : |f(x)- \mathbb{E}{f(x)}| \leq \epsilon, \:  \epsilon \rightarrow0 \right \rbrace $, where $\mathbb{E}{f(x)}$ is the expectation value of $f(x)$ when $x$ is chosen uniformly at random.

Then by Levy's lemma, measure of all such points 
\begin{equation}
\mu\left\lbrace x \in S^{n-1} :|f(x)- \mathbb{E}{f(x)}| \leq \epsilon \right\rbrace \geq 1-2\textrm{exp}  \left(- \frac{kn \epsilon^2}{\eta^2}\right). \label{2}
\end{equation}
Similarly, let T denote  the set of points$\left\lbrace x \in S^{n-1} : |\rho(x)- \mathbb{E}{\rho(x)}| \leq \epsilon,\:  \epsilon \rightarrow0 \right \rbrace $.
Since $\rho$ is Lipschitz,
\begin{equation}
\mu\left\lbrace x \in S^{n-1} :|\rho(x)- 1| \leq \epsilon \right\rbrace \geq 1-2\textrm{exp}  \left(- \frac{ kn \epsilon^2}{\eta'^2}\right),\label{3}
\end{equation}
where we noted that $\mathbb{E}{\rho(x)} =1$.
From set theory, $\overline{(S \cap T)} = \overline{S} \cup \overline{T}$. Therefore,
\begin{equation}
\mu(S \cap T)=  1- \mu\overline{(S \cap T)} =1- \mu(\overline{S} \cup \overline{T}). \label{4}
\end{equation}
\begin{figure}[hbtp]
\centering
\includegraphics[scale=0.6]{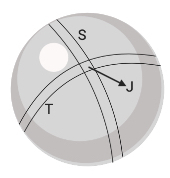}
\caption{By Levy's lemma, the sets $S$ and $T$ are regions around the equator. They contain almost all of the domain in a higher dimensional sphere. Their intersection is the set $J$, marked in the figure. } 
\end{figure}
Combining equations (\ref{2}),(\ref{3}) and (\ref{4}), 
\begin{align}
\mu(S \cap T)& > 1- \left( 2\textrm{exp}  \left(- \frac{kn \epsilon^2}{\eta^2}\right) + 2\textrm{exp}  \left(- \frac{kn \epsilon^2}{\eta'^2}\right)\right) \\ & \approx1, \textrm{when} \:n \epsilon^2\: \textrm{is large}
\end{align}
Let $ J= S \cap T$, then $\mu(J) \approx1$ when $n$ is very large and  $\epsilon$ is very small such that $n\epsilon^2$ is large.  Now, if $x$ is chosen according to a Lipschitz distribution, the measure of points at which the value of the function lies inside an $\epsilon$ neighbourhood around the mean of $f$ is the integral of the density function over the region $S$.

\begin{align}
\mu\left\lbrace x \in S^{n-1}:|f(x)- \mathbb{E}{f(x)}| \leq \epsilon \right\rbrace \label{aa}  &= \int_S \rho(x) d \mu \\ &\geq\int_J \rho(x) d \mu \\
& \geq \int_J (1-\epsilon) d\mu \label{5}\\&= \mu(J) - \epsilon \mu(J) \\& \geq (1-\epsilon)\left[ 1- \left( 2\textrm{exp}  \left(- \frac{kn \epsilon^2}{\eta^2}\right) + 2\textrm{exp}  \left(- \frac{kn \epsilon^2}{\eta'^2}\right)\right) \right]
 \label{6} 
\end{align}
In Eq.~(\ref{5}), we have used the fact that $(1-\epsilon)$ is the lower bound of  $\rho(x)$ in the region $J$.  When $n$ is large, and $ \epsilon$ tends to zero, $(1- \epsilon) \rightarrow 1$ and $\left[ 1- \left( 2\textrm{exp}  \left(- \frac{kn \epsilon^2}{\eta^2}\right) + 2\textrm{exp}  \left(- \frac{kn \epsilon^2}{\eta'^2}\right)\right) \right]   \rightarrow1$. Hence the RHS of Eq.~(\ref{6}) tends to one. In this limit, we see that $ \left\lbrace \mu(S) , \mu(T)  \right\rbrace\approx1$  and so is the measure of their intersection, the region  $J$.   
	
	A stronger result can be obtained in which the expectation value of the function, $\mathbb{E}{f(x)}$ in the left hand side of Eq.~(\ref{aa}) is replaced with $\mathbb{E}_{liptz}{f(x)}$, the Lipschitz expectation value defined as follows.
	If $f(x)$ is a Lipschitz continuous function from $S^{(n-1)} \rightarrow \mathbb{R}$, the
expectation value of $f(x)$ when $x$ is chosen at random according to a Lipschitz continuous density function $\rho(x)$ is
\begin{equation}
\mathbb{E}_{liptz} f(x) = \int_{S^{(n-1)}} f(x) \rho(x) d\mu. \label{0}
\end{equation}

\textbf{Theorem 2:}
\begin{equation}
\mu\left\lbrace x:|f(x_{liptz})- \mathbb{E}_{liptz}f(x)|\leq \epsilon(1+|f_{max}|)\right\rbrace \geq  (1-\epsilon),
\end{equation}  
where $|f_{max}|$ denotes the maximum value attained by $f(x)$ in the space $S^{(n-1)}$.

\textbf{Proof:}
Note that $\int_{S^{(n-1)}} \rho(x) d\mu=1$, since we consider a normalized metric space. Using this Lipschitz expectation value essentially means the measure of a given open set in the space $S^{(n-1)}$ is no more proportional to its area, but is weighted by a density function $\rho(x)$ which is very well behaved. In order to reach the desired result,
we need to find how  $\mathbb{E}_{liptz} f(x)$ is related to $\mathbb{E}{ f(x)}$,  the expectation with uniform measure, given by
\begin{equation}
\mathbb{E}{ f(x)} = \int_{S^{(n-1)}} f(x)  d\mu.
\end{equation}
 Note that $\rho(x)$ is a probability density function, and it is always positive. In the region $J$, $\rho(x) \in (1-\epsilon, 1+ \epsilon)$. Thus $(1+ \epsilon)$ could be regarded as an upper bound of $\rho(x)$ in this region. In the rest of the metric space, i.e., in $ S^{(n-1)} \setminus J$, $\rho(x)$ varies more than $\epsilon$ from the uniform density 1.  One can in principle find the least upper bound of this density function, and denote it by $(1+\delta)$, where $\delta> \epsilon >0$. 
Intuitively, as the dimension of the sphere gets large, we expect both the averages to be close to each other, which is what we see if we calculate the modulus of their differences.


\begin{align}
 |\mathbb{E}_{liptz}f(x) - \mathbb{E}f(x)| &=\left\vert   \int_{s^{(n-1)}}f(x) \rho(x) d\mu - \int_{s^{(n-1)}} f(x) d\mu  \right\vert 
 \\&=\left\vert   \int_{s^{(n-1)}}f(x) (\rho(x)-1) d\mu \right\vert
 \\&\leq \left\vert   \int_Jf(x) (\rho(x)-1) d\mu \right\vert + \left\vert \int_{s^{(n-1)} \setminus J} f(x) (\rho(x)-1)d\mu  \right\vert 
 \\& \leq   \int_J |f_{max}| \epsilon d\mu  + \int_{s^{(n-1)} \setminus J}  |f_{max}| \delta d\mu  \label{c}
 \\&=   \int_{J}|f_{max}| \epsilon d\mu  + \int_{s^{(n-1)} \setminus J}|f_{max}| \epsilon d\mu + \int_{s^{(n-1)} \setminus J} |f_{max}| (\delta -\epsilon) d\mu  
 \\ & =    \int_{s^{(n-1)}}|f_{max}| \epsilon d\mu  + \int_{s^{n-1} \setminus J}  |f_{max}| (\delta -\epsilon) d\mu   \label{a} 
\\& \leq \epsilon |f_{max}|  + 2|f_{max}| (\delta- \epsilon)\left[ \textrm{exp}  \left(- \frac{kn \epsilon^2}{\eta^2}\right) + \textrm{exp}  \left(- \frac{kn \epsilon^2}{\eta'^2}\right)\right]. \label{b}
\end{align}

\begin{figure}
\begin{subfigure}{.5\textwidth}
\centering
\includegraphics[scale=0.5]{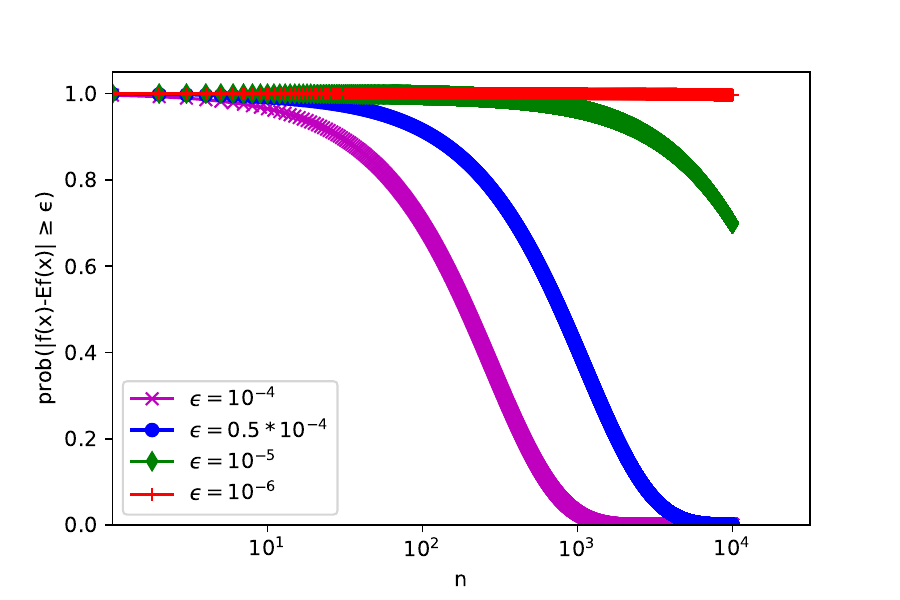} 
\caption{}
\label{fig:sub-first}
\end{subfigure}
\begin{subfigure}{.5\textwidth}
\centering
\includegraphics[scale=0.5]{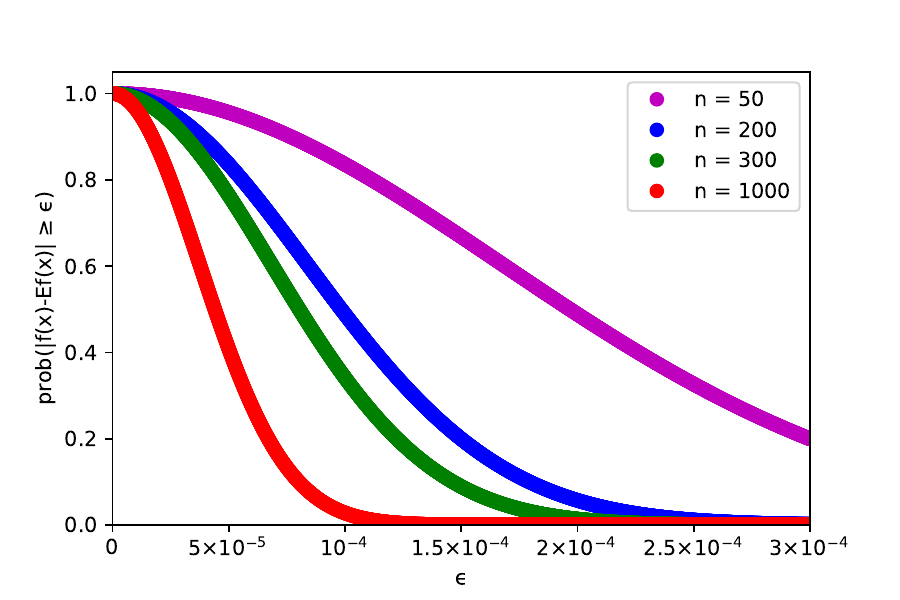} 
\caption{}
\label{fig:sub-second}
\end{subfigure}
\caption{Sample numerical simulations using Eq.~(\ref{d}) showing the behaviour of the probability, taking  $\eta=\eta'=10^{-4}$. Figure~\ref{fig:sub-first} is a semi-log plot which shows how the probability bound falls for various values of $\epsilon$, with the  dimension of the Hilbert space. Figure~\ref{fig:sub-second} shows probability bound against $\epsilon$ for fixed Hilbert space dimensions}
\end{figure}
 $|f_{max}|$ denotes the maximum value attained by $f(x)$ in the space $S^{(n-1)}$.
 In Eq.~(\ref{c}), we  use the fact that replacing $(\rho(x)-1) $ and $ f(x)$ by the maximum value they can attain, produces an upper bound of the integrals. The fact that these functions are Lipschitz, guarantees that the maximum value attained by these functions on the hypersphere are finite. For instance, consider $\rho(x)$. By Lipschitz condition, $| \rho_{max} - \rho_{min}| \leq \eta'  |x_{max} -x_{min}|$. Since the hypersphere we consider is of finite radius, $\rho_{max}$ is finite.    Note that the right hand side of Eq.~(\ref{b}) tends to zero when $n \epsilon^2$ is large. 
 Now we can rewrite Eq.~(\ref{6}) using the Lipschitz average $\mathbb{E}_{liptz}f(x)$ instead of the uniform $\mathbb{E}f(x)$.

\begin{multline}
\mu\left\lbrace x:|f(x_{liptz})- \mathbb{E}_{liptz}f(x)|\leq \epsilon + \epsilon |f_{max}|  + 2|f_{max}| (\delta- \epsilon)\left[ \textrm{exp}  \left(- \frac{kn \epsilon^2}{\eta^2}\right) + \textrm{exp}  \left(- \frac{kn \epsilon^2}{\eta'^2}\right)\right] \right\rbrace \\ \geq (1-\epsilon)\left[ 1- \left( 2\textrm{exp}  \left(- \frac{kn \epsilon^2}{\eta^2}\right) + 2\textrm{exp}  \left(- \frac{kn \epsilon^2}{\eta'^2}\right)\right) \right].\label{d}
\end{multline}

In the large $n$ limit, $\mu\left\lbrace S^{(n-1)}\setminus J\right\rbrace \rightarrow 0$, and Eq.~(\ref{d}) simplifies as
\begin{equation} 
\mu\left\lbrace x:|f(x_{liptz})- \mathbb{E}_{liptz}f(x)|\leq \epsilon(1+|f_{max}|)\right\rbrace \geq  (1-\epsilon).
\end{equation} 

Hence Levy's lemma holds in this extended form when $x$ is chosen from a Lipschitz distribution on a higher-dimensional sphere.
Why should nature care about Lipschitz distributions over a sphere? It turns out that the dynamics on a sphere finds fundamental applications in  quantum information science.

\section{Application to entanglement theory}
\label{app}

Most applications in quantum information theory require a very high dimensional Hilbert space. Such spaces give us a forum to use concentration measure results that yield simple but surprising results
\citep{page1993average, popescu2006entanglement,hayden2006aspects,hayden2010concentration1}. 
 For example, entanglement properties of random bipartite states are very important
from the perspective of quantum information theory and quantum chaos \citep{cover2012elements, nielsen2002quantum,hamma2012quantum,
refael2009criticality,Enr_quez_2018,vznidarivc2008entanglement,
giraud2007distribution,gross2009most,white2009minimally}. 
 From a quantum information perspective, there is an intimate connection between entanglement properties of random states and random subspaces and quantum communication protocols like the superdense coding \citep{hayden2010concentration1}. Random subspaces in higher dimensions yield very good error correcting codes due to high entanglement. Randomly picked subspaces can be used, for example, for noisy quantum communication, saturating the highest
 systematically achievable rate \citep{PhysRevA.55.1613, devetak2005private}.
Random entangled states are also very important from a foundational point of view in statistical mechanics \citep{popescu2006entanglement} and quantum chaos  \citep{Lakshminarayan, trail2008entanglement, ScottCaves2003, Zyczkowski1990, Madhok2018_corr, Bandyopadhyay04, MillerSarkar, Ghose2004, Wang2004}. 
Chaotic maps take localized states to highly entangled random states,  smeared all over the Hilbert space.
This is closely related to the connections between ergodicity, integrability, and chaos, as well as eigenstate thermalization in closed quantum systems \citep{Rigol16, Deutsch91,  Srednicki94}.

The phenomenon of concentration of measure and Levy's Lemma, therefore, have numerous applications in quantum information theory \citep{page1993average,popescu2006entanglement,hayden2006aspects,hayden2010concentration1}.
We can use our results to get generalized bounds on all of these applications. In this section, we provide a few illustrative examples.

Let us consider a bipartite quantum system consisting of subsystems $A$  and $B$ of dimensions $d_{A}$ and $d_{B}$ respectively.  Consider a pure state, chosen uniformly at random from this combined system $A\otimes B$.
The average entanglement of such a state picked at random from a $d_{A} \otimes d_{B}$ dimensional tensor product Hilbert space is given by the average Von Neumann entropy $S(\rho_A)$ of the reduced density matrix of one of the subsystems \citep{page1993average, lubkin1978entropy, hayden2006aspects, Page_proof,PhysRevLett.77.1,lloyd1988complexity} 
\begin{equation}
\mathbb{E}S(\rho_A) = \sum^{d_Ad_B}_{k=d_A+1} \frac{1}{k}-\frac{d_A-1}{2d_B}, d_B\geq d_A.
\end{equation}
For large dimensions, $\mathbb{E}S(\rho_A)  \approx \log d_A - d_A/(2 d_B)$, which is nearly maximally entangled, but saturates slightly below.

Levy's lemma quantifies the likelihood for the entanglement of a random state to deviate from the
mean value of entanglement. Using the result from \citep{hayden2010concentration1}

 \begin{equation}
\textrm{Prob}\left\lbrace  |S(\rho_A)- \mathbb{E}[{S(\rho_A)}] |\geq \epsilon \right\rbrace \leq  \mathrm{exp} \left({- \frac{(d_{A}d_{B} -1)C \epsilon^2}{(\log d_A)^2}}\right),
\label{Hoeff}
\end{equation}  
for some $C>0$ and $d_B \geq d_A \geq 3$.

This result can be easily generalized using Eq.~(\ref{6}) for the probability that the entanglement of a state selected randomly according to a Lipschitz distribution will deviate significantly from the
mean value of entanglement.
Therefore, it allows us to calculate the value of entanglement for a ``typical" state in a bipartite tensor product Hilbert space.


An interesting case of the above is the following.
Let us consider a quantum system consisting of subsystems $A$ and $B$, dimension of $A$ is very small compared to $B$, i.e., $d_{A} \ll d_{B}$. We can think of $A$ as our system and $B$ as its environment. Consider a pure state, chosen uniformly at random from this combined system $A\otimes B$. 
We can find $\mathrm{Tr}( \rho_{A}^{2})$ by tracing out system $B$. Then the probability for the chosen state to have a local trace distance of more than a small $\epsilon$ from the maximally mixed state \citep{muller2012random} is given by  

\begin{equation}
\textrm{Prob} \left\lbrace \left\Vert \rho_A - \frac{\mathbb{I}_A}{d_{A}}\right\Vert_1 \geq  \sqrt{\frac{d_{A}}{d_{B}}} +\epsilon \right\rbrace \leq \\ 2 \textrm{exp} \left(-\frac{d_{A}.d_{B} \epsilon^2}{18 \pi^3 \eta^2} \right).
\end{equation}
If one chooses a point from $A \otimes B$ at random according to a Lipschitz continuous probability function with Lipschitz constant $\eta'$,
Then by Eq.~(\ref{6})
\begin{multline}
\textrm{Prob} \left\lbrace \left\Vert \rho_A - \frac{\mathbb{I}_A}{d_{A}}\right\Vert_1 \geq \sqrt{\frac{d_{A}}{d_{B}}} +\epsilon \right\rbrace \leq  \\ \epsilon + 2(1-\epsilon) \left[ \textrm{exp}  \left(-\frac{d_{A}.d_{B} \epsilon^2 }{18 \pi^3 \eta^2} \right)+ \textrm{exp}  \left(-\frac{d_{A}.d_{B} \epsilon^2}{18 \pi^3 \eta'^2} \right)\right]. \label{7}
\end{multline}
We can see that in the limit $d_{A}d_{B}$ getting very large and $\epsilon$ tending to zero, the right-hand side of Eq.~(\ref{7}) also tends to zero. Also remember that we had $d_{A} \ll d_{B}$. Both these conditions together tell us that with high probability, a state $\rho_A$ is close to the maximally mixed state.
Another way of seeing this entanglement property is by using the continuity of entropy. Von Neumann entropy is a continuous function everywhere in its domain \citep{watroustheory}. Fannes' inequality is a quantification regarding continuity of entropy.
 It gives a bound on the change in Von Neumann entropy between two nearby states as a function of their trace distance \citep{nielsen2002quantum,fannes1973continuity}. If $\rho$ and $\sigma$ are two states, close together, then
\begin{equation}
\left\vert S(\rho)-S(\sigma) \right\vert \leq T(\rho,\sigma) \left(\textrm{log}\:n- \textrm{log}\:T(\rho,\sigma)\right),
\end{equation}
where $S(\rho)$ represents Von Neumann entropy of the state $\rho$ and $T(\rho,\sigma)$ is the trace distance between the two states in the argument, $\left\Vert \rho - \sigma\right\Vert_1$. Now consider a bipartite Hilbert space $A \otimes B$ where $d_{A}  \ll d_{B}$, with a Lipschitz continuous measure. We can apply Fannes' inequality to a state $|\psi\rangle$ picked from this bipartite Hilbert space. 
\begin{equation}
\left\vert S(\rho_A) -S\left( \frac{\mathbb{I}_A}{d_{A}}\right)\right\vert \leq \left(\epsilon +\sqrt{\frac{d_{A}}{d_{B}}} \right) \mathrm{log}\:d_{A} + \alpha 
\label{fannes}
\end{equation}
Eq.~(\ref{fannes}) holds with high probability as one can see from Eq.~(\ref{7}). Here 
$\alpha=\left(\epsilon+\sqrt{\frac{d_A}{d_B}} \right) \mathrm{log}\left(\epsilon+\sqrt{\frac{d_A}{d_B}} \right) $. When $\epsilon \rightarrow0$ and since $d_A  \ll d_B$, $\alpha$ tends to zero as well.  Therefore, entropy of any state $\rho_A$ chosen according to a Lipschitz continuous probability distribution is close to the maximal entropy with a lower bound given by
\begin{equation}
S(\rho_A) \geq \mathrm{log}\: d_A\left[1 - \left(\epsilon + \sqrt{\frac{d_A}{d_B}} \right) \right]  - \alpha,
\end{equation}
 with probability given by ($1\:-$ RHS) of Eq.~(\ref{7}).
 
 The results on bipartite entanglement can be generalized to the multipartite scenario.
 Consider a random state of $n$ qudits from a Lipschitz probability distribution, such that $|\psi\rangle \in (\mathcal{C}^{d})^{\otimes n} $. The entanglement across any bipartite
 cut will be nearly maximal with a high probability as the dimensionality increases (assuming $n$ is fixed and $d$ is allowed to increase).

 \section{Conclusion and Discussion - random subspaces, quantum communication and quantum chaos} 
\label{disc}

  What we have seen in this chapter is that if the probability distribution on the space is smooth enough,  a relation similar to Levy's lemma holds. That is, if you pick a random point on the space according to a smooth enough probability density, you are very likely to find the value the function assumes at that point to be in a small neighborhood around the expectation value of the function. This likelihood increases with an increase in the dimension of the space. In other words, Levy's lemma can be seen as a particular case of a more general law, where the Lipschitz continuous probability distribution happens to be uniform.

We also discussed how the relations we derived for the measure concentration could be used to calculate the probability of significant departure of entaglement of a given bipartite quantum state, picked according to any Lipschitz probability distribution, from the average value of  entanglement 
for such states over the entire Hilbert space. As discussed, this helps us get generalized bounds and study special cases 
when one of the subsystems can be considered the ``environment" or the bath. 

Since the convergence in Eq.~(\ref{4}) is exponentially rapid, we can even make statements about random subspaces in addition to random states \citep{hayden2010concentration1}. Levy's lemma also guarantees the existence of random subspaces where all states
are nearly maximally entangled. Random subspaces find applications in quantum error correction as they yield remarkably good error-correcting codes when the dimension of the Hilbert space is high. Our results suggest that random subspaces of high entanglement also exist when the subspaces are chosen according to a Lipschitz probability distribution. This is indeed remarkable as the entanglement properties of random subspaces also find applications in extending quantum communication protocols from bits to qubits.

Another possible application of such ideas is to find the connection between the generation of pseudo-random states in the Hilbert space associated with a quantum map, whose classical counterpart exhibits a mixed phase space instead of global chaos.  Typical entanglement of a Haar random state predicts the entanglement reached by eigenstates and projected coherent states via evolution in the case of global chaos~\citep{Lakshminarayan, trail2008entanglement, ScottCaves2003, Bandyopadhyay04}. However, the question of entanglement generation in the mixed phase space is yet unexplored. A simple symmetry cannot describe the structure of the chaotic sea in the mixed phase space. Hence 
studies are limited to numerical estimation of entanglement \citep{trail2008entanglement}. However, our results that generalize Levy's Lemma can help get useful
bounds on entanglement generation in mixed phase space. Classically, it is well known that generic Hamiltonian dynamical systems are neither integrable nor ergodic \citep{markus1974generic}, i.e., they exhibit a mixed phase space. Therefore, studies in mixed phase space, which is the dynamics that takes a fiducial state to a pseudo-random state according to an arbitrary measure, becomes important from a foundational point of view. This is especially the case when we study thermalization, and the origins of statistical mechanics~\citep{popescu2006entanglement}. We believe our results can shed some light on these issues. All this will be the subject of future
exploration.

\chapter{Conclusions and Outlook}
\label{chap:chap7}
This thesis has investigated some important problems in quantum mechanics and information theory. The main theme of the thesis was studying vestiges of chaos in quantum systems. In the third chapter, we have investigated the signatures of chaos in small quantum systems. The origin of chaos in classical physics is a fundamental area of research. We considered three and four qubit systems in this analysis. We see that OTOCs give us the Lyapunov exponent using the dynamical signatures up to principal order. The Loschmidt echo showed a  weaker signature of chaos. It takes a higher $J$ for the Lyapunov decay to manifest. However, specific states we considered showed correlations with classical phase space.

It is rare to find quantum systems showing chaotic signatures that can be exactly solved. Here we could obtain analytical relations for OTOCs and Loschmidt echo, which is remarkable. Exactly solvable systems like the kicked top give us a reference to study departure from integrability and transition to chaos upon introducing perturbations breaking the necessary symmetries via the KAM theorem. Recent studies involving a related concept, the Adiabatic Guage Potential (AGP) serves as a probe to detect chaos in systems with large Hilbert spaces \citep{PhysRevX.10.041017}. An interesting direction for the future is to compare the effectiveness of AGP with that of Loschmidt echo in detecting chaos. 

 Another advantage of few qubit kicked tops is that they are experimentally implementable using superconducting qubits. This renders them important to study fundamental questions like the mechanism of ergodicity and internal equilibration in closed quantum systems. In 2016, \citep{Neill16} experimentally studied ergodic behavior in a three-qubit kicked top. 

Quantum kicked top systems are also useful for quantum chaotic sensing. Quantum chaotic sensing uses extreme sensitivity on parameter values of quantum chaotic systems to improve measurements of physical quantities. In the small perturbation limit, the Loschmidt echo is related to the Fisher information, the smallest variance achievable for an estimator. For example, the measurement precision of a classical magnetic field can be substantially improved by using quantum kicked top as a sensor \citep{fiderer2018quantum}. 

The fourth chapter addressed quantum state tomography using a reduced set of random unitaries, which does not lead to information completeness of measurements. Despite this, we see that good fidelity reconstruction is achieved. Information contained in the $d-1$ dimensional subspace is missed in this case. The state reconstruction protocol using weak measurements gave us an advantage in the amount of resources required. We quantified the rate of information gain using collective Fisher information and used Shannon entropy to quantify the bias in operator sampling. 

Tomography is another domain where the effects of chaos can be studied. The rate of information gain is closely related to ergodic mixing due to chaos. Exploring the quantum kicked top as a model system,  we characterized the effects of eigenvalues and eigenvectors of the Floquet in the rate of information gain and fidelity. We found that ergodic mixing is basis-dependent, with the rate of information gain depending on eigenvector distribution. Random matrix connections of the covariance matrix were also discerned.

One of the possible extensions of this study is to perform quantum process tomography using the diagonal-in-a-basis unitaries and continuous weak measurements. Another viable application is in randomized benchmarking. How well do randomized benchmarking protocols work when one only has random diagonal unitaries at disposal? Is there a way to perform randomized benchmarking with a restricted set of unitaries? 

A quantum control process where one drives an unknown initial state into a target state is closely related to the ergodicity in the dynamics. Because of the information completeness,  Haar random unitary dynamics would be well equipped to steer the state into any target state in the Hilbert space. A restricted dynamics might not lead to good fidelity control if the target state has significant support in the unexplored part of the Hilbert space. 

Achievable control of a quantum system is also highly correlated with chaos. In a recent article, {Nicolás Mirkin} \textit{et al.} found that amount of chaos in the system is negatively correlated with the control of a subsystem \citep{mirkin2021quantum}. Remarkably, this study was concerned with extremely short spin chains. Another way to establish the effect of chaos on control would be in a state steering experiment towards a  time-evolving target state under a chaotic dynamics. By smoothly varying the chaoticity parameter, one should be able to observe the effect on the achieved fidelity. A recent paper on quantum state steering achieved control without ``classical at heart" measurements. They used sequential interactions with a set of quantum control systems to steer the state. This problem can be recast in a realistic setting, including qubit decay. An extension to a Hamiltonian control protocol where an initial unitary is steered to a fixed target unitary is also worth pursuing.

Another significant contribution of the thesis is in designing a DQC1 algorithm for computing OTOCs. This circuit can be implemented in NMR systems, making it more interesting from a practical perspective. The highlight of our method is that it evaluates OTOCs exponentially faster than any known classical algorithm. In performing any quantum computation faithfully, one needs information about the interactions of quantum gates with the environment and the resulting loss of fidelity of the gates used. Quantifying their imperfections thus becomes an important problem. We have given a DQC1 algorithm that measures average unitary gate fidelity. Producing desired results in computations requires us to understand how well the gates perform. We think that the DQC1 algorithm can be used to estimate some of the semiclassical formulas, like the Gutzwiller trace formula, which can achieve speedup over the existing algorithms. Having seen that DQC1 achieves great speedup, exploring other avenues where mixed-state quantum computation can be applied becomes interesting. 

The final chapter discussed the concentration of measure phenomenon in Hilbert spaces. We presented a generalization of Levy's lemma and its application in the entanglement of pure states. Typicality relations appear at many places in quantum information. Most of the pure bipartite states have entanglement very close to the average value of entanglement. Similarly, quantum coherence also shows such typical value for pure states. Naturally, one can ask whether such a typicality relation exists for quantum discord. 

There are consequences for the measure concentration phenomena in evolutionary biology as well, which we discussed in the appendix. In a recent work, \citep{madhok2019typicality} established the robustness of the fitness function to mutations on a high dimensional genetic sequence space. One possible way to extend this work could be to estimate the fitness function itself. Genetic properties are encoded by a sequence of nucleotides which form a large combinatorial space. The fitness function maps each sequence to a real value. Experimentally measuring the fitness of all  the sequences is impossible. There are estimation protocols to learn the fitness function by measuring a subset of them \citep{jin2002framework, brookes2022sparsity}. Suppose the fitness function is Lipschitz continuous, which would mean that the function is very well behaved over its domain. Can we get a better fidelity estimate for the fitness with fewer measurements? When it comes to the geometry of higher dimensional spaces, can we find the concentration of measure phenomenon other than on a hypersphere? These are questions we would like to answer in the future.
\appendix
\chapter{Application to evolutionary dynamics}
\label{appendixA}

\textbf{Theorem 3}: Any system of equations for elements,  $x_{1},x_{2},\dots,x_{n}$, on an $n$ dimensional simplex can be recast into a dynamics taking place on an $n$ dimensional hypersphere. 
 As $n$ becomes large,  and if $x_{1},x_{2},\dots,x_{n}$, 
 are taken to be IID (independent and identically distributed) exponential random variables, they correspond to a random probability distribution according to the uniform measure on the probability space \citep{wootters1990random}.  Then such a probability distribution corresponds to a Lipschitz probability distribution on the sphere. 
 By probability space, one means the $n-1$ dimensional space of normalized probability distributions.

\textbf{Proof:} This theorem is proved in \citep{madhok2019typicality} and here we give an outline for completeness.
 It is given that each $x_i$ is chosen at random according to the exponential distribution, $f(x_i)=e^{-x_i}$. The $x$'s  are all independent, and their collective distribution is given by $f(x_1,x_2,...x_n)= e^{-(x_1+x_2+...x_n)}.$ When ${x}$'s are normalized to unity, the joint distribution is independent of the frequency values and hence form a uniform distribution in the probability space \citep{wootters1990random}.
 In the case of normalized frequencies $(x_1,x_2,..x_n)$,  the point $Y= \sqrt{X_{eq}}=(\sqrt{x_{1}},\sqrt{x_{2}},\dots,\sqrt{x_{n}}) \in S^{(n-1)},$ the unit sphere.  For $n$  large, the numbers $p_i=x_i/n$ form an automatically normalized set \cite{wootters1990random}. Now $Y=(y_1,y_2,..y_n)= (\sqrt{x_{1}},\sqrt{x_{2}},\dots,\sqrt{x_{n}})$ with $f(y_i)\approx 2y_ie^{-y_i^{2}}.$ This distribution is Lipschitz continuous on the sphere $S^{n-1}$ \cite{madhok2019typicality}.



\subsection{Evolutionary Game Dynamics} 
Evolutionary game dynamics describes the natural selection of strategies in evolutionary games.
The essential point is that the dynamics is frequency-dependent. This means that the fitness of a particular individual/strategy 
depends on its phenotypic composition and also on its interaction with the environment. The environment is composed of both the abiotic environment like temperature, pressure, resources, etc., and the biotic environment. The biotic environment can be, for example, the types of all other individuals/strategies with which it interacts with regards to resources such as nutrition, mating preferences, etc. Another example of the biotic environment is predators in the population. 
In the case of evolutionary dynamics, the evolution of a set of genetic sequences in a population is given by 

\begin{equation}
\frac{dX}{dt} = W X- f(X)X.
\label{dxdt}  \end{equation}

The vector $X$ is composed of the population densities of the individual sequences,

\begin{equation}
X = (x_{1},x_{2},...,x_{n}), 
\label{xs}  \end{equation}

$f$ is the fitness function and the matrix $W$ consists of individual replication rates, $a_{i}(X), i=1,2,...n$, along with the mutation rates for transition between individual sequences, $i$ and $j$, given by $Q_{ij}$.
$f$ is the fitness function and $W$ is a matrix containing replication rates of each sequence denoted by $a_{i}(X),$ where $ i=1,2,...n$. The mutation frequency of each  sequence is captured in $Q_{ij}.$ A mutation leads to a transformation of one sequence to another.
\[
W = 
\begin{bmatrix}
    a_{1}(X)Q_{11} & a_{2}(X)Q_{12} & \dots  & a_{n}(X)Q_{1n} \\
    a_{1}(X)Q_{21} & a_{2}(X)Q_{22} & \dots  & a_{n}(X)Q_{2n} \\
    \dots & \dots & \dots & \dots \\
    a_{1}(X)Q_{n1} & a_{2}(X)Q_{n2} & \dots  & a_{n}(X)Q_{nn}
\end{bmatrix}
\]

The fitness function for the sequence of population densities is defined as
\begin{equation}
    {f}(X) = \sum_{i=1}^{n} a_{i}x_{i}/\sum_{i}^{n} x_{i},
\label{fx}\end{equation}
Equilibrium condition  for the system undergoing dynamics described in  Eq. (\ref{dxdt}) is obtained by solving the eigenvalue equation
\begin{equation}
 WX = \lambda X.
\label{WX}  \end{equation}
Since $W$ has positive entries, Frobenius-Perron theorem \citep{perron1907theorie} guarantees a unique largest real eigenvalue.
The equilibrium frequency distribution is represented by the eigenvector corresponding to the largest eigenvalue, $\lambda_{max}=\overline{\big(\sum_{i=1}^{n} a_{i}x_{i}/\sum_{i}^{n} x_{i}\big)}$, which turns out to be the mean replication rate.
Because of Levy's lemma, any random point in the frequency space of population densities have fitness value very close to $\lambda_{max}$, which is equal to $\bar{f}(X),$ the average fitness function. In other words, a random point on the  sequence space of  population densities  is very close to the equilibrium point of the system~\citep{madhok2019typicality}.
  \bibliographystyle{abbrvnat}
\bibliography{refs}

\end{document}